\documentclass[epsf]{iopart}
\eqnobysec
\usepackage{iopams,amsfonts,rotating,graphicx,fancybox,fancyhdr}
\newcommand{\mkbox}[3]{\hbox{\vrule
      \vbox to  #1{\hrule \vss
                  \hbox to #2{\hss#3\hss}\vss
                  \hrule}\vrule}}

\begin{document}

\title[Integrable structure of GinOE and a Pfaffian integration theorem]
        {Integrable structure of Ginibre's ensemble of real random matrices
        and a Pfaffian integration theorem}

\author{Gernot Akemann$^1$ and Eugene Kanzieper$^2$}

\address{
    $^1$ Department of Mathematical Sciences and BURSt Research Centre, Brunel
    University West London, Uxbridge UB8 3PH, United
    Kingdom\\
    $^2$ Department of Applied Mathematics, H.I.T. -- Holon Institute of
    Technology, Holon 58102,
    Israel} \eads{\mailto{Gernot.Akemann@brunel.ac.uk},
    \mailto{Eugene.Kanzieper@weizmann.ac.il}}

\begin{abstract}
In the recent publication [E. Kanzieper and G. Akemann, Phys. Rev.
Lett. {\bf 95}, 230201 (2005)], an exact solution was reported for
the probability $p_{n,k}$ to find exactly $k$ real eigenvalues in
the spectrum of an $n\times n$ real asymmetric matrix drawn at
random from Ginibre's Orthogonal Ensemble (GinOE). In the present
paper, we offer a detailed derivation of the above result by
concentrating on the proof of the Pfaffian integration theorem, the
key ingredient of our analysis of the statistics of real eigenvalues
in the GinOE. We also initiate a study of the correlations of complex eigenvalues and derive a formula for the
joint probability density function of all complex eigenvalues of a
GinOE matrix restricted to have exactly $k$ real eigenvalues. In the
particular case of $k=0$, all correlation functions of complex
eigenvalues are determined.
\end{abstract}

\pacs{02.10.Yn, 02.50.-r, 05.40.-a, 75.10.Nr\\
Dated: March 05, 2007} \vspace{1pc}

\tableofcontents
\newpage
\section{Introduction} \label{Sec1}

\subsection{Motivation}
This study grew out of our attempt to answer the question raised by
A. Edelman (1997): ``What is the probability $p_{n,k}$ that an
$n\times n$ random real Gaussian matrix has exactly $k$ real
eigenvalues?'' In the physics literature, an ensemble of such random
matrices is known as GinOE -- Ginibre's Orthogonal Ensemble (Ginibre
1965). Looking into this particular problem, we have realised that
no comprehensive solution for the probability $p_{n,k}$ can be found
without undertaking an in-depth study (Kanzieper and Akemann 2005)
of the integrable structure of GinOE. The results of our
investigation are reported in the present paper.

\subsection{Main results}
\label{SubSec12}

For the benefit of the readers, we collect our main results into
this easy to read subsection with pointers to the sections
containing detailed derivations of each statement.
\subsubsection{Real part of GinOE spectrum}
\textsc{\newline\newline (I)~Probability of Exactly $k$ Real
Eigenvalues.} Let ${\boldsymbol {\cal H}}$ be an $n\times n$ random
real matrix whose entries are statistically independent random
variables picked from a normal distribution $\textsf{N}(0,1)$. Then,
for $n-k=2\ell$ even, the probability $p_{n,k}$ of exactly $k$ real
eigenvalues \footnote[1]{The number of complex eigenvalues
$n-k=2\ell$ is always even since the complex part of the spectrum
consists of $\ell$ pairs of complex conjugated eigenvalues.}
occurring is \numparts
\begin{eqnarray}
\label{St-1} p_{n,k}=p_{n,n-2\ell} = \frac{p_{n,n}}{\ell!}  \,
Z_{(1^\ell)}(p_1,\cdots,p_\ell),
\end{eqnarray}
where $p_{n,n}$ is the probability $p_{n,n} =2^{-n(n-1)/4}$ of
having {\it all} $n$ eigenvalues real (Edelman 1997). The {\it
universal} multivariate functions $Z_{(1^\ell)}$, solely determined
by the number $\ell$ of pairs of complex conjugated eigenvalues, are
so-called {\it zonal polynomials} (Macdonald 1998) that can be
written as a sum over all partitions \footnote[2]{The notation
$\blambda = (\ell_1^{\sigma_1},\cdots,
    \ell_g^{\sigma_g})$ is known as the frequency representation of
    the partition $\blambda$ of the size $|\blambda|=\ell$. It implies
    that the part $\ell_j$ appears
    $\sigma_j$ times so that $\ell = \sum_{j=1}^g \ell_j\, \sigma_j$,
    where $g$ is the number of inequivalent parts of
the partition. In particular, the partition $\blambda=(1^\ell)$
equals $(\underbrace{1,\cdots, 1}_{\ell {\rm \; times}})$.}
$\blambda = (\ell_1^{\sigma_1},\cdots, \ell_g^{\sigma_g})$ of the
size $|\blambda|=\ell$
\begin{eqnarray}
\label{SymPol}
     Z_{(1^\ell)}(p_1,\cdots,p_\ell) =
    (-1)^\ell  \ell! \sum_{|\blambda|=\ell} \, \prod_{j=1}^g
    \frac{1}{\sigma_j!}
\left(
    - \frac{p_{\ell_j}}{\ell_j}
\right)^{\sigma_j}.
\end{eqnarray}
A few first zonal polynomials are displayed in Table
\ref{zonal-table}. The arguments $p_j$'s of the zonal polynomials
are {\it nonuniversal} \footnote[3]{The notation ${\rm tr}_{(0,a)\,}
{\boldsymbol {\hat M}}$ denotes the trace of an $(a+1)\times (a+1)$
matrix ${\boldsymbol {\hat M}}$ such that ${\rm tr}_{(0,a)\,}
{\boldsymbol {\hat M}} = \sum_{j=0}^{a} {\hat M}_{jj}$. Also,
$\lfloor x \rfloor$ stands for the floor function. In what follows,
the ceiling function $\lceil x \rceil$ will be used as well. }
\begin{eqnarray} \label{traces}
    p_j = {\rm tr}_{(0,\lfloor n/2 \rfloor-1)} \, {\boldsymbol {\hat
    \varrho}}\,^j,
\end{eqnarray}
for they depend on a {\it nonuniversal} matrix ${\boldsymbol {\hat
\varrho}}$. For $n=2m$ even, its entries are
\begin{eqnarray}
    \label{rho-even}
    \hspace{-0.2cm}
    {\hat \varrho}_{\alpha,\, \beta}^{\rm even} &=&
    \int_0^\infty dy\, y^{2(\beta-\alpha)-1} \, e^{y^2}\, {\rm
    erfc}(y \sqrt{2})
    \Big[
        (2\alpha+1) \,\nonumber \\
    &\times& L_{2\alpha+1}^{2(\beta-\alpha)-1}(-2y^2)
        + 2y^2 \,
        L_{2\alpha-1}^{2(\beta-\alpha)+1}(-2y^2)
    \Big]
\end{eqnarray}
\begin{table}
\caption{\label{zonal-table} Explicit examples of zonal polynomials
$Z_{(1^\ell)}(p_1,\cdots,p_\ell)$ as defined by the equation
(\ref{SymPol}). Another way to compute $Z_{(1^\ell)}$ is based on
the recursion equation (Macdonald 1998)
\begin{eqnarray}
\fl Z_{(1^\ell)}(p_1,\cdots,p_\ell)=(\ell-1)!\, \sum_{r=0}^{\ell-1}
\frac{(-1)^{\ell-r-1}}{r!} \,p_{\ell-r}\,
Z_{(1^{r})}(p_1,\cdots,p_{r}) \nonumber
\end{eqnarray}
supplemented by the formal ``boundary'' condition $Z_{(1^0)}=1$.
They are also tabulated in the manuscript by H.~Jack (1976).}

\begin{indented}
\lineup
\item[]
\begin{tabular}{@{}*{2}{l}} \br
$\0\0\blambda$&$Z_{\blambda}(p_1,\cdots,p_\ell)$ \cr \mr

$\0\0(1^1)$&$p_1$ \vspace{0.1cm} \cr

$\0\0(1^2)$&$p_1^2-p_2$ \vspace{0.1cm}\cr

$\0\0(1^3)$&$p_1^3-3p_1p_2+2p_3$\vspace{0.1cm} \cr

$\0\0(1^4)$&$p_1^4+8p_1p_3-6p_1^2p_2+3p_2^2-6p_4$ \vspace{0.1cm}\cr

$\0\0(1^5)$&$p_1^5-10p_1^3p_2+20p_1^2p_3+15p_1p_2^2-30p_1p_4-20p_2p_3+24p_5$\vspace{0.1cm}\cr

\br
\end{tabular}
\end{indented}
\end{table}
while for $n=2m+1$ odd,
\begin{eqnarray}
   \label{rho-odd}
    {\hat \varrho}_{\alpha,\,\beta}^{\rm odd} =
    {\hat \varrho}_{\alpha,\,\beta}^{\rm even} - (-4)^{m-\beta}
    \frac{m!}{(2m)!}\frac{(2\beta)!}{\beta!}\,
    {\hat \varrho}_{\alpha,\,m}^{\rm even}.
\end{eqnarray}
\endnumparts
Here, the notation ${\rm erfc}(\phi)$ stands for the complementary
error function,
\begin{eqnarray}
\label{cef} {\rm erfc} (\phi) = \frac{2}{\sqrt{\pi}} \,
\int_\phi^{\infty} dt\, e^{-t^2},  \nonumber
\end{eqnarray}
while $L_j^\alpha(\phi)$ denote the generalised Laguerre polynomials
\begin{eqnarray}
    L_j^\alpha(\phi) = \frac{1}{j!}\phi^{-\alpha}\, e^{\phi}\,
    \frac{d^{\,j}}{d\phi^{\,j}} \left( \phi^{\,j+\alpha}\, e^{-\phi} \right).
    \nonumber
\end{eqnarray}
\newline
The above result (\ref{St-1}) will be derived in Section \ref{Sec6}.

\vspace{0.2cm}\textsc{\newline (II)~Generating Function For
Probabilities $p_{n,k}$.} The generating function $G_n(z)$ for the
probabilities $p_{n,k}$ is
\begin{eqnarray}
 \label{gf=1}
    G_n(z) = \sum_{\ell=0}^{\lfloor n/2 \rfloor} z^\ell p_{n,n-2\ell} = p_{n,n} \, {\rm
    det}\,
    \big[{\boldsymbol {\hat 1}} + z\, {\boldsymbol {\hat
    \varrho}}\big]_{\lfloor n/2 \rfloor \times\lfloor n/2 \rfloor}.
\end{eqnarray}
Equation (\ref{gf=1}) with the ${\boldsymbol {\hat \varrho}}$ of
needed parity provides us with yet another way of computing the
entire set of $p_{n,k}$'s at once! Table \ref{prob-table} contains a
comparison of our analytic predictions with numeric simulations. The
result (\ref{gf=1}) will be proven in Section \ref{Sec6}.
\newline
\vspace{0.2cm}\textsc{\newline (III)~Probability $p_{n,n-2}$ of
Exactly One Pair of Complex Conjugated Eigen\-values.} For $k=n-2$,
the probability function $p_{n,k}$ reduces to \numparts
\begin{eqnarray}
    \label{probn-2a}
    p_{n,n-2} = 2\, p_{n,n}\,
    \int_0^\infty dy\, y \, e^{y^2}\, {\rm
    erfc}(y \sqrt{2}) \,
     L_{n-2}^2(-2y^2).
\end{eqnarray}
An alternative expression reads:
\begin{eqnarray}
    \label{probn-2b}
    p_{n,n-2} = p_{n,n} \,\left[ \sqrt{2}\, \sum_{j=0}^{\lfloor n/2 \rfloor -1}
    3^{j+\alpha_n/2}\, P_{2j+\alpha_n}\left( \frac{2}{\sqrt{3}}
    \right) - \lfloor n/2 \rfloor\right].
\end{eqnarray}
Here, $\alpha_n = \lceil n/2 \rceil - \lfloor n/2 \rfloor$, and
$P_n(\phi)$ stands for the Legendre polynomials
\begin{eqnarray}
    P_j(\phi) = \frac{(-1)^j}{2^j j!} \, \frac{d^j}{d\phi^j}\left[
        (1-\phi^2)^j
    \right].
    \nonumber
\end{eqnarray}
The leading large-$n$ behaviour of the probability $p_{n,n-2}$ is
given by
\begin{eqnarray}
\label{probn-2c}
    p_{n,n-2} \approx \frac{3^{n+1/2}}{8\sqrt{\pi\,n}}\, p_{n,n}.
\end{eqnarray}
\endnumparts
The above three results will be derived in Section \ref{SubSec41},
Section \ref{SubSec71} and Section \ref{SubSec72}, respectively.

\subsubsection{Complex part of GinOE spectrum}
\textsc{\newline\newline (IV)~Joint Probability Density Function of
All Complex Eigenvalues Given There Are $k$ Real Eigenvalues.} Let
${\boldsymbol {\cal H}_k}$ be an $n\times n$ random real matrix with
$k$ real eigenvalues such that its entries are statistically
independent random variables picked from a normal distribution
$\textsf{N}(0,1)$. Then, the joint probability density function
(j.p.d.f.) of its $2\ell=n-k$ complex eigenvalues is

\Table{\label{prob-table}Exact solution for $p_{12,k}$ (first and
second column) compared to numeric simulations (third column)
performed by direct diagonalisation of $1,000,000$ of $12 \times 12$
matrices.} \br
&\centre{2}{Analytic solution} &\\
\ns
&\crule{2}&Numeric\\
\centre{1}{$k$}&\centre{1}{Exact}&Approximate&simulation\vspace{0.2cm}\\
\mr $\00$&\centre{1}{\large $\,\frac{29930323227453 -
    20772686238032\sqrt{2}}{17592186044416}\,$}&\centre{1}{0.031452}&\centre{1}{0.031683}\vspace{0.3cm}\\
$\02$&\centre{1}{\large $\,\frac{3(1899624551312\sqrt{2} -
2060941421503)}{4398046511104}$}&\centre{1}{0.426689}&\centre{1}{0.427670}\vspace{0.3cm}\\
$\04$&\centre{1}{\large $\,\frac{3(2079282320189 -
505722262348\sqrt{2})}{8796093022208}$}&\centre{1}{0.465235}&\centre{1}{0.464098}\vspace{0.3cm}\\
$\06$&\centre{1}{\large $\,\frac{252911550974\sqrt{2}-27511352125}{4398046511104}$}&\centre{1}{0.075070}&\centre{1}{0.075021}\vspace{0.3cm}\\
$\08$&\centre{1}{\large
$\,\frac{15(1834091507-10083960\sqrt{2})}{17592186044416}$}&\centre{1}{0.001552}&\centre{1}{0.001526}\vspace{0.3cm}\\
$10$&\centre{1}{\large $\,\frac{3(1260495\sqrt{2}-512)}{2199023255552}$}&\centre{1}{0.000002}&\centre{1}{0.000002}\vspace{0.3cm}\\
$12$&\centre{1}{\large
$\,\frac{1}{8589934592}$}&\centre{1}{0.000000}&\centre{1}{0.000000}\vspace{0.2cm}
\\
\br
\end{tabular}
\end{indented}
\end{table}

\begin{eqnarray}
    \label{St-3}
    \fl
    P_{{\boldsymbol {\cal H}_k}} (z_1,\cdots, z_\ell) &=   \frac{p_{n,n}}{\ell!}\,\left(\frac{2}{i}\right)^\ell
    \, \nonumber \\
    &\times \prod_{j=1}^\ell \, {\rm erfc}
    \left( \frac{z_j - {\bar z}_j}{i\sqrt{2} } \right)
    \,
    {\rm pf\,} \left[
    \begin{array}{cc}
      {\cal D}_{n}(z_i, z_j) & {\cal D}_{n}(z_i, {\bar z}_j) \\
      {\cal D}_{n}({\bar z}_i, z_j) & {\cal D}_{n}({\bar z}_i, {\bar z}_j) \\
    \end{array}
    \right]_{2\ell \times 2\ell}.
\end{eqnarray}
Here, ${\rm pf}$ denotes the Pfaffian. The above j.p.d.f. is
supported for $({\rm Re\,}z_1,\cdots, {\rm Re\,} z_\ell)\in {\mathbb
R}^\ell$, and $({\rm Im\,}z_1,\cdots, {\rm Im\,} z_\ell)\in
({\mathbb R}^+)^\ell$. The antisymmetric kernel ${\cal
D}_{n}(z,z^\prime)$ is given explicitly by (\ref{d2m}) -- (\ref{hj})
of Section \ref{Sec3} where the statement (\ref{St-3}) is proven.

\textsc{\newline\newline (V)~Correlation Functions of Complex
Eigenvalues in The Spectra Free of Real Eigenvalues.} Let
${\boldsymbol {\cal H}_0}$ be an $n\times n$ random real matrix with
no real eigenvalues such that its entries are statistically
independent random variables picked from a normal distribution
$\textsf{N}(0,1)$. Then, the $p$-point correlation function ($1\le
p\le \ell$) of its complex eigenvalues, defined by (\ref{rpb1}),
equals \numparts
\begin{eqnarray}\fl
R^{({\boldsymbol {\cal H}}_0)}_{0,p}(z_1,\cdots, z_p;n) = p_{n,n}\, \frac{\prod_{j=0}^{\ell-1} r_j}
    {\prod_{j=1}^n \Gamma(j/2)}
    \prod_{j=1}^p {\rm erfc}\left( \frac{z_j-{\bar
    z}_j}{i\sqrt{2}}\right)\, \exp\left(
    -\frac{z_j^2+{\bar z}_j^2}{2}
    \right) \nonumber \\
    \times\, {\rm pf\,}
    \left[
    \begin{array}{cc}
      \kappa_{\ell}(z_i, z_j) & \kappa_{\ell}(z_i, {\bar z}_j) \\
      \kappa_{\ell}({\bar z}_i, z_j) & \kappa_{\ell}({\bar z}_i, {\bar z}_j) \\
    \end{array}
    \right]_{2p \times 2p}.
\end{eqnarray}
Here, $n=2\ell$ and the `pre-kernel' $\kappa_\ell$ equals
\begin{eqnarray}
    \kappa_\ell (z,z^\prime) = i \sum_{j=0}^{\ell-1} \frac{1}{r_j} \bigg[
        p_{2j}(z) p_{2j+1}(z^\prime) - p_{2j}(z^\prime)
    p_{2j+1}(z)
    \bigg].
\end{eqnarray}
The polynomials $p_j(z)$ are skew orthogonal in the complex
half-plane ${\rm Im\,}z >0$,
\begin{eqnarray}
    \langle p_{2j+1}, p_{2k} \rangle_{\rm c} = - \langle p_{2k}, p_{2j+1}
    \rangle_{\rm
    c} = i\,r_j \, \delta_{jk},  \\
    \label{sp2}
    \langle p_{2j+1}, p_{2k+1} \rangle_{\rm c} = \langle p_{2j}, p_{2k}
    \rangle_{\rm
    c}=0,
\end{eqnarray}
with respect to the skew product
\begin{eqnarray} \fl
    \langle f, g \rangle_{\rm c} = \int_{{\rm Im\,}z>0} d^2 z\,
    {\rm erfc} \left(
        \frac{z-{\bar z}}{i\sqrt{2}}
    \right)\,
    \exp\left(-\frac{z^2+{\bar z}^2}{2}\right)\,
    \left[
        f(z)g({\bar z}) - f({\bar z})g(z)
    \right].
\end{eqnarray}
\endnumparts
For detailed derivation, a reader is referred to Section \ref{Sec8}
which also addresses the problem of calculating the probability
$p_{n,0}$ to find no real eigenvalues in the spectrum of
GinOE.\newline

\subsubsection{How to integrate a Pfaffian?\newline}
\textsf{\newline} All the results announced so far would have not
been derived without a Pfaffian integration theorem that we consider
to be a major technical achievement of our study. Conceptually, it
is based on a new, topological, interpretation of the ordered
Pfaffian expansion as introduced in Section \ref{Sec5}.
\textsc{\newline\newline (VI)~The Pfaffian Integration Theorem.} Let
$d\pi(z)$ be any benign measure on $z \in {\mathbb C}$, and the
function $Q_n(x,y)$ be an antisymmetric function of the form
\numparts
\begin{eqnarray}
    Q_n(x,y) = \frac{1}{2} \sum_{j,k=0}^{n-1} q_j(x) \, {\hat \mu}_{jk} \,
    q_k(y)
\end{eqnarray}
where $q_j$'s are arbitrary polynomials of $j$-th order, and
${\boldsymbol {\hat \mu}}$ is an antisymmetric matrix. Then the
integration formula
\begin{eqnarray} \fl
    \label{theorem-a}
    \left(\frac{2}{i}\right)^\ell \prod_{j=1}^\ell \, \int_{z_j\in {\mathbb
    C}} d\pi(z_j)\,\,
    &{\rm pf\,} \left[
    \begin{array}{cc}
      Q_{n}(z_i, z_j) & Q_{n}(z_i, {\bar z}_j) \\
      Q_{n}({\bar z}_i, z_j) & Q_{n}({\bar z}_i, {\bar z}_j) \\
    \end{array}
    \right]_{2\ell \times 2\ell} \nonumber \\
    &=
    Z_{(1^\ell)}\left(
    \frac{1}{2} {\rm tr}_{(0,n-1)}
    {\boldsymbol {\hat \upsilon}}^1,\cdots,\frac{1}{2} {\rm tr}_{(0,n-1)}
    {\boldsymbol {\hat \upsilon}}^\ell
    \right)
\end{eqnarray}
holds, provided the integrals in its~l.h.s.~exist. Here,
$Z_{(1^\ell)}$ are zonal polynomials whose $\ell$ arguments are
determined by a matrix ${\boldsymbol {\hat \upsilon}}$ with the
entries
\begin{eqnarray}
    \label{sigma=def}
    {\hat \upsilon}_{\alpha,\beta} = i \sum_{k=0}^{n-1} {\hat \mu}_{\alpha,k}
    \int_{z\in {\mathbb C}} d\pi(z) \left[
        q_k(z)\, q_\beta({\bar z}) - q_\beta(z) \, q_k({\bar z})
    \right].
\end{eqnarray}
\endnumparts
This theorem that can be viewed as a generalisation of the Dyson
integration theorem (Dyson 1970, Mahoux and Mehta 1991) will be
proven in Section \ref{Sec5}.

Interestingly, the Pfaffian integration theorem is not listed in the
classic reference book (Mehta 2004) on the Random Matrix Theory
(RMT). Also, we are not aware of any other RMT literature reporting
this result which may have implications far beyond the scope of the
present paper.

\subsection{A guide through the paper}

Having announced the main results of our study, we defer plunging
into formal mathematical proofs of the above six statements until
Section 3. Instead, in Section \ref{Sec2}, we deliberately draw the
reader's attention to a comparative analysis of GinOE and two other
representatives of non-Hermitean random matrix models known as
Ginibre's Unitary (GinUE) and Ginibre's Symplectic (GinSE) Ensemble.
Starting with the definitions of the three ensembles, we briefly
discuss their diverse physical applications, pinpoint {\it
qualitative} differences between their spectra, and present a
detailed {\it comparative analysis} of major {\it structural}
results obtained for GinUE, GinSE and GinOE since 1965. We took
great pains to write a review-style Section \ref{Sec2} in order (i)
to help the reader better appreciate a profound difference between
GinOE and the two other non-Hermitean random matrix ensembles on
both qualitative and structural levels as well as (ii) place our own
work in a more general context.

A formal analysis starts with Section \ref{Sec3} devoted to a
general consideration of statistics of real eigenvalues. Its first
part, Section \ref{SubSec31}, summarises previously known analytic
results (Edelman 1997) for the probability function $p_{n,k}$ of the
fluctuating number of real eigenvalues in the spectrum of GinOE.
Section \ref{SubSec32} deals with the joint probability density
function of complex eigenvalues of GinOE random matrices that have a
given number of real eigenvalues. The Pfaffian representation
(\ref{St-3}) is the main outcome of Section \ref{SubSec32}. This
result is further utilised in Section \ref{SubSec33} where the
probability function $p_{n,k}$ is put into the form of a `Pfaffian
integral' (\ref{pnk-comp}). The analysis of the latter expression
culminates in concluding that the Dyson integration theorem, a
standard tool of Random Matrix Theory, is inapplicable for treating
the Pfaffian integral obtained. The latter task will be accomplished
in Section \ref{Sec5}.

Section \ref{Sec4} attacks the probability function $p_{n,k}$ for a
few particular values of $k$. The probabilities $p_{n,n-2}$,
$p_{n,n-4}$ and $p_{n,n-6}$ of one, two, and three pairs of complex
conjugated eigenvalues occurring are treated in Section
\ref{SubSec41}, Section \ref{SubSec42} and \ref{SubSec43},
respectively. This is done by explicit calculation of the Pfaffian in (\ref{pnk-comp})
followed by a term-by-term integration of the resulting Pfaffian expansion.
In Section \ref{SubSec44}, we briefly discuss a faster-than-exponential growth of the
number of terms in this expansion caused by further decrease of $k$.

Section {\ref{Sec5}}, devoted to the Pfaffian integration theorem,
is central to the paper. Its main objective is to introduce a
topological interpretation of the terms arising in a permutational
expansion of the Pfaffian in the l.h.s. of (\ref{theorem-a}). Such a
topological interpretation turns out to be the proper language in
the subsequent proof of the Pfaffian integration theorem. In Section
\ref{SubSec51}, the Pfaffian integration theorem is formulated and
discussed in the light of the Dyson integration theorem. In Section
\ref{SubSec52}, an ordered permutational Pfaffian expansion is
defined and interpreted in {\it topological terms}. The notions of
{\it strings}, {\it substrings}, {\it loop-like strings} and {\it
loop-like substrings} for certain subsets of terms arising in an
ordered Pfaffian expansion are introduced and illustrated on simple
examples in Sections \ref{SubSec521} and \ref{SubSec522}. Further,
{\it equivalent strings} and {\it equivalent classes of strings} are
defined and counted. The issue of decomposition of strings into a
set of loop-like substrings is also considered in detail (Lemma
5.2). Section {\ref{SubSec53}} is devoted to the counting of
loop-like strings. In Section \ref{SubSec54}, the notion of {\it
adjacent} strings is introduced and illustrated. Adjacent strings
are counted in Lemma 5.4. Their relation to loop-like strings is
discussed in Lemma 5.5. Section {\ref{SubSec55}} deals with a
characterisation of adjacent strings by their {\it handedness};
adjacent strings of a given handedness are counted there, too. In
Section {\ref{SubSec56}}, {\it equivalent classes of adjacent
strings} are defined, counted, and explicitly built. Section
{\ref{SubSec57}} and Section {\ref{SubSec58}} are preparatory for
Section {\ref{SubSec59}}, where the {\it Pfaffian integration
theorem} is eventually proven. For the readers' benefit, a
vocabulary of the topological terms we use is summarised in Table
\ref{strings-table}.

Section \ref{Sec6} utilises the Pfaffian integration theorem to
obtain a general solution for the sought probability function
$p_{n,k}$ (Section \ref{SubSec61}), derive a determinantal
expression for the entire generating function of $p_{n,k}$'s
(Section \ref{SubSec62}), and address the issue of integer moments
of the fluctuating number of real eigenvalues in GinOE spectra
(Section \ref{SubSec63}).

Section \ref{Sec7} is devoted to the probability $p_{n,n-2}$ for two
complex conjugated pairs of eigenvalues to occur; a large-$n$
analysis of $p_{n,n-2}$ is also presented there.

Section \ref{Sec8} discusses the Pfaffian structure of the $p$-point
correlation functions of complex eigenvalues belonging to spectra of
a subclass of GinOE matrices without real eigenvalues. The same
section addresses the problem of calculating the probability
$p_{n,0}$ to find no real eigenvalues in spectra of GinOE.

Section 9 contains conclusions with the emphasis placed on open
problems. The most involved technical calculations are collected in
four appendices.

\section{Comparative analysis of GinOE, GinUE, and GinSE}\label{Sec2}

\subsection{Definition and consequences of violated Hermiticity}

Ginibre's three random matrix models -- GinOE, GinUE, and GinSE --
were derived from the celebrated Gaussian Orthogonal (GOE), Gaussian
Unitary (GUE), and Gaussian Symplectic (GSE) random matrix ensembles
in a purely formal way by dropping the Hermiticity constraint.
Consequently, the non-Hermitean descendants of GOE, GUE and GSE share
the {\it same} Gaussian probability
density function
\begin{eqnarray} \label{G-def} \fl
    P_{\beta}[{\boldsymbol {\cal H}}] = (\pi a_\beta)^{-\beta
    n^2/2} \exp
    \left[ - a_\beta^{-1} \, {\rm tr\,} \left( {\boldsymbol {\cal H}}
    {\boldsymbol {\cal
    H}}^\dagger \right)\right], \;\;\;
    {\boldsymbol {\cal H}}^\dagger\neq
{\boldsymbol {\cal H}}, \;\;\; \beta=1,2,4,
\end{eqnarray}
for a matrix ``Hamiltonian'' ${\boldsymbol {\cal H}} \in {\mathbb
T}_\beta(n)$ to occur; the constant $a_\beta$ is chosen to be
$a_\beta = 2-\delta_{\beta, 2}$ (this slightly differs from the
convention used in the original paper by Ginibre). However, the
spaces ${\mathbb T}_\beta$ on which the matrices ${\boldsymbol {\cal
H}}$ vary are {\it different}: ${\mathbb T}_1(n)$, ${\mathbb
T}_2(n)$, and ${\mathbb T}_4(n)$ span all $n\times n$ matrices with
real (GinOE, $\beta=1$), complex (GinUE, $\beta=2$), and real
quaternion (GinSE, $\beta=4$) entries, respectively.

The violated Hermiticity, ${\boldsymbol {\cal H}}^\dagger\neq
{\boldsymbol {\cal H}}$, brings about the two major phenomena: (i)
complex-valuedness of the random matrix {\it spectrum} and (ii)
splitting the random matrix {\it eigenvectors} into a bi-orthogonal
set of {\it left} and {\it right} eigenvectors. Statistics of
complex eigenvalues, including their joint probability density
function, and statistics of left and right eigenvectors of a random
matrix ${\boldsymbol {\cal H}}\in {\mathbb T}_\beta$ drawn from
(\ref{G-def}) are of primary interest.

Only the spectral statistics will be addressed in the present paper.
(For studies of the eigenvector statistics in GinUE, the reader is
referred to the papers by Chalker and Mehlig (1998), Mehlig and
Chalker (2000), and Janik {\it et al} (1999). We are not aware of
any results for eigenvector statistics in GinSE and GinOE.)

\subsection{Physical applications}
While the ensemble definition (\ref{G-def}) was born out of pure
mathematical curiosity~\footnote[4]{J. Ginibre, private
communication.}, non-Hermitean random matrices have surfaced in
various fields of knowledge by E. Wigner's ``miracle of the
appropriateness'' (Wigner 1960). From the physical point of view,
non-Hermitean random matrices have proven to be as important as
their Hermitean counterparts. (For a detailed exposition of physical
applications of Hermitean RMT we refer to the review by Guhr {\it et
al} (1998)).

Random matrices drawn from GinUE appear in the description of
dissipative quantum maps (Grobe {\it et al} 1988, Grobe and Haake
1989) and in the characterisation of two-dimensional random
space-filling cellular structures (Le C\"aer and Ho 1990, Le C\"aer
and Delannay 1993).

Ginibre's Orthogonal Ensemble of random matrices arises in the
studies of dynamics (Sommers {\it et al} 1988, Sompolinsky {\it et
al} 1988) and of synchronisation effect (Timme {\it et al} 2002,
Timme {\it et al} 2004) in random networks; GinOE is also helpful in
the statistical analysis of cross-hemisphere correlation matrix of
the cortical electric activity (Kwapie\'{n} {\it et al} 2000) as
well as in the understanding of inter-market financial correlations
(Kwapie\'{n} {\it et al} 2006).

All three Ginibre ensembles (GinOE, GinUE, GinSE) arise in the
context of directed ``quantum chaos'' (Efetov 1997a, Efetov 1997b,
Kolesnikov and Efetov 1999, Fyodorov {\it et al} 1997, Markum {\it
et al} 1999). Their chiral counterparts (Stephanov 1996, Halasz {\it
et al} 1997, Osborn 2004, Akemann 2005) help elucidate universal
aspects of the phenomenon of spontaneous chiral symmetry breaking in
quantum chromodynamics (QCD) with chemical potential: the presence
or absence of real eigenvalues in the complex spectrum singles out
different chiral symmetry breaking patterns. For a review of QCD
applications of non-Hermitean random matrices with built-in
chirality, the reader is referred to Akemann (2007).

Other recent findings (Zabrodin 2003) associate statistical models
of non-Hermitean normal random matrices with integrable structures
of conformal maps and interface dynamics at both classical
(Mineev-Weinstein {\it et al} 2000) and quantum (Agam {\it et al}
2002) scales. For a comprehensive review of these and other physical
applications, the reader is referred to the survey paper by Fyodorov
and Sommers (2003).

\subsection{Spectral statistical properties of Ginibre's random matrices}
\label{SubSec23}

\begin{figure}[t]
\centering
\includegraphics[scale=.38]{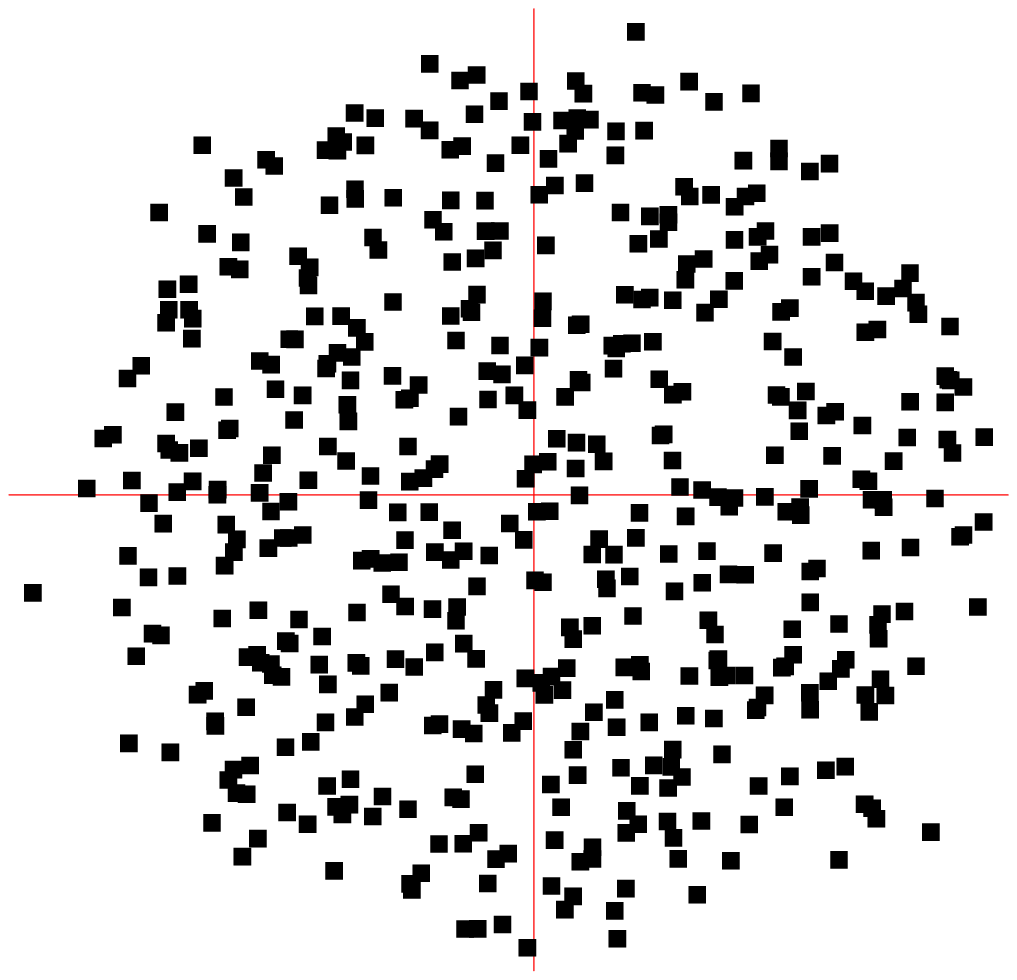} \;
\includegraphics[scale=.365]{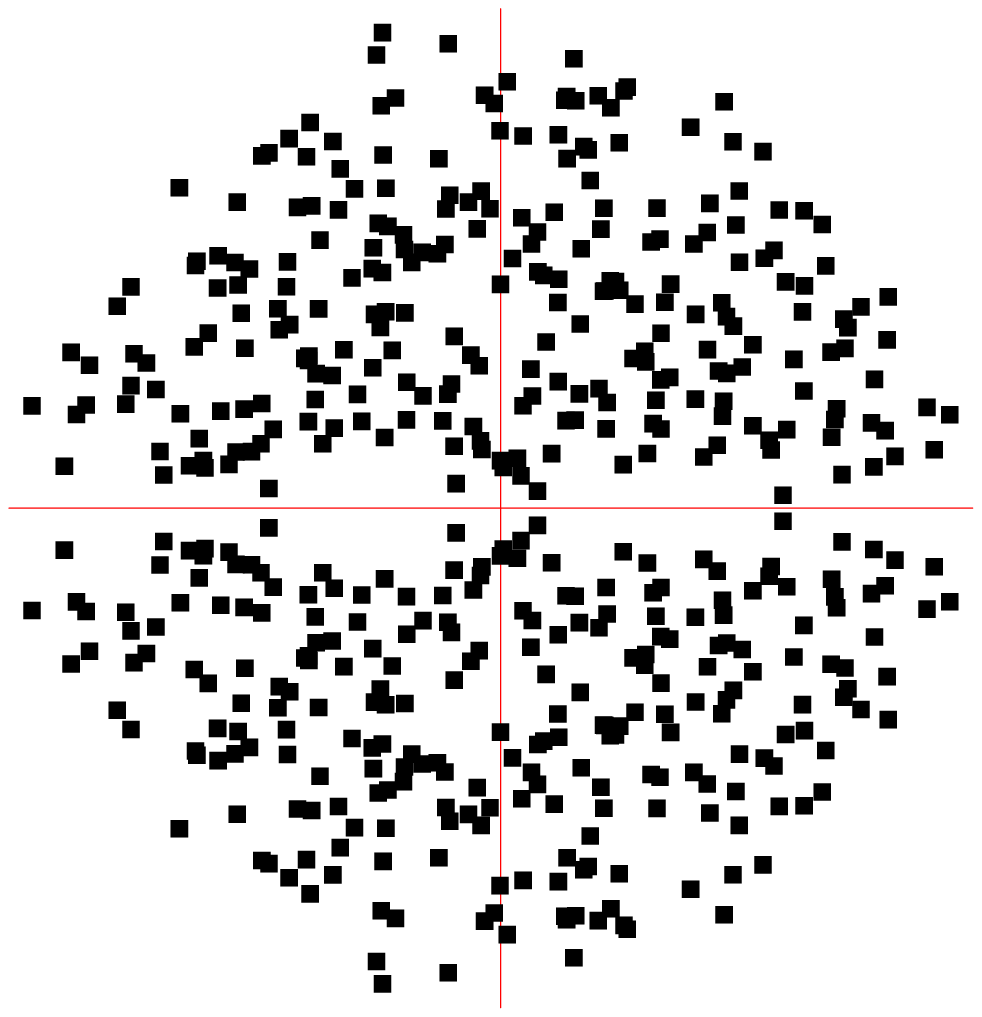}\;
\includegraphics[scale=.4]{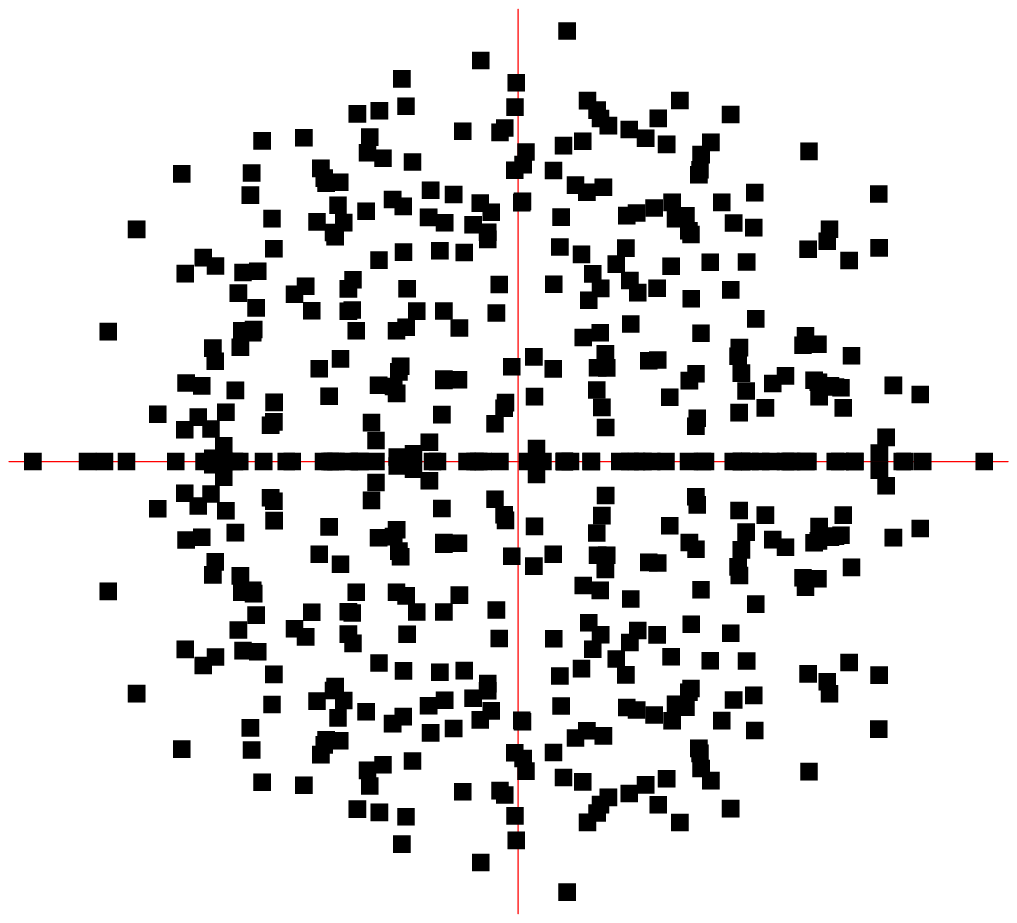} \;
 \caption{Numerically simulated distributions of complex eigenvalues in
            GinUE (left panel), GinSE (middle panel), and GinOE (right panel).
            Three different eigenvalue patterns are
            clearly
            observed. Eigenvalues are scattered almost uniformly in GinUE,
            depleted from the real axis in GinSE, and accumulated
            along the real axis in GinOE.}\label{dos-simulation}
\end{figure}

A profound difference between spectral patterns of the three
non-Hermitean random matrix models has been realised long ago.
Anticipated in the early papers by Ginibre (1965) and Mehta and
Srivastava (1966), it was further confirmed analytically by using
varied techniques \footnote[5]{The difference in spectral patterns
of non-Hermitean {\it chiral} random matrix models arising in the
QCD context was first studied numerically by Halasz {\it et al}
(1997). A review of recent theoretical developments can be found in
Akemann (2007).} (Edelman 1997, Efetov 1997a, Efetov 1997b,
Kolesnikov and Efetov 1999, Kanzieper 2002a, Kanzieper 2002b,
Nishigaki and Kamenev 2002, Splittorff and Verbaarschot 2004,
Kanzieper 2005, Akemann and Basile 2007).

{\it Qualitatively}, there is a general consensus that (i) the
spectrum of GinUE is approximately characterised by a {\it uniform}
density of complex eigenvalues. This is not the case for the two
other ensembles. (ii) In GinSE, the density of complex eigenvalues
is smooth but the probability density of real eigenvalues tends to
zero. This corresponds to a {\it depletion} of the eigenvalues along
the real axis. (iii) On the contrary, the density of eigenvalues in
GinOE exhibits an {\it accumulation} of the eigenvalues along the
real axis. {\it It is the latter phenomenon that will be quantified
in our paper.}

Our immediate goal here is to highlight the inter-relation between
these {\it qualitative features} of the complex spectra and the {\it
formal structures} underlying Ginibre's random matrix ensembles. To
this end, we present a brief comparative review of the major {\it
structural} results obtained for all three Ginibre's ensembles
(GinUE, GinSE, and GinOE) since 1965, in the order of increasing
difficulty of their treatment.
\newline
\vspace{0.1cm}\newline\textsc{Joint Probability Density Function of
All $n$ Eigenvalues.} In this subsection, we collect explicit
results for the joint probability density functions of all $n$
complex eigenvalues of a random matrix ${\boldsymbol {\cal H}} \in
{\mathbb T}_\beta(n)$ drawn from any of the three Ginibre random
matrix ensembles.
\newline

\begin{itemize}

\item {\bf GinUE:} The spectrum of a random matrix ${\boldsymbol {\cal H}} \in
  {\mathbb T}_2(n)$ drawn from GinUE consists of $n$ complex
eigenvalues $(z_1,\cdots z_n)$ whose joint probability density
function mirrors (Ginibre 1965) that of GUE (see, e.g., Mehta
2004),
        \begin{eqnarray}
        \label{jpdf=2} \fl
            P_n^{(2)}(z_1,\cdots,z_n) = \left( \pi^n \prod_{\ell=1}^n \ell!
            \right)^{-1}
            \prod_{\ell_1>\ell_2=1}^n |z_{\ell_1} -
            z_{\ell_2}|^2 \prod_{\ell=1}^n e^{-z_\ell{\bar z}_\ell}.
        \end{eqnarray}
 Similarly to the electrostatic model introduced by E. Wigner
    (1957), J. Ginibre pointed out that the j.p.d.f.
    (\ref{jpdf=2}) can be thought of as that describing the
    distribution of the positions of charges of a
    two-dimensional Coulomb gas in an harmonic oscillator
    potential $U(z)= |z|^2/2$, at the inverse temperature
    $1/T=2$.
    \newline

\item{\bf GinSE:}
The spectrum of a random matrix ${\boldsymbol {\cal H}} \in
  {\mathbb T}_4(n)$ drawn from GinSE consists of $n$ pairs of
  complex conjugated eigenvalues $(z_1,{\bar z}_1,\cdots,
  z_n,{\bar z}_n)$. The corresponding joint probability density
  function was derived by Ginibre (1965),
        \begin{eqnarray}
        \label{jpdf=4}
        \fl
        P_n^{(4)}(z_1,\cdots,z_n) &= \left(  (2\pi)^n n! \prod_{\ell=1}^n
(2\ell-1)!\right)^{-1} \,
        \prod_{\ell_1>\ell_2=1}^n |z_{\ell_1} -
        z_{\ell_2}|^2 |z_{\ell_1} - {\bar z}_{\ell_2}|^2 \nonumber
        \\
        &\times
        \prod_{\ell=1}^n |z_\ell - {\bar z}_\ell|^2\, \exp(-z_\ell{\bar
        z}_\ell).
        \end{eqnarray}
    Notice that the factor
        $\prod_{\ell=1}^n |z_\ell - {\bar z}_\ell|^2\,$ in
    (\ref{jpdf=4}) is directly responsible for the {\it
    depletion} of the eigenvalues along the real axis. For
    GinSE, a physical analogy with a two-dimensional Coulomb gas
    is much less transparent; it has been discussed by
    P.~Forrester (2005). \newline

\item{\bf GinOE:} Contrary to GinUE and GinSE, the complex spectrum
  $(w_1,\cdots, w_n)$ of a random matrix ${\boldsymbol {\cal
H}}\in {\mathbb T}_1(n)$ drawn from GinOE generically contains a
{\it finite fraction} of real eigenvalues; the remaining complex
eigenvalues always form complex conjugated {\it pairs}. Indeed,
no other option is allowed by the {\it real} secular equation
${\rm det\,}(w - {\boldsymbol {\cal H}})=0$ determining the
eigenvalues of ${\boldsymbol {\cal H}}$.
\newline\newline
This very peculiar feature of GinOE, that we call {\it
accumulation} of the eigenvalues along the real axis, can
conveniently be accommodated by dividing the entire space
${\mathbb T}_1(n)$ spanned by all real $n\times n$ matrices
${\boldsymbol {\cal H}}\in {\mathbb T}_1(n)$ into $(n+1)$
mutually exclusive sectors ${\mathbb T}_1(n/k)$ associated with
the matrices ${\boldsymbol{\cal H}_k} \subset {\boldsymbol {\cal
H}}$ having exactly $k$ real eigenvalues, such that ${\mathbb
T}_1(n)=\bigcup_{k=0}^n {\mathbb T}_1(n/k)$. The sectors
${\mathbb T}_1(n/k)$, characterised by the partial j.p.d.f.'s
$P_{{\boldsymbol {\cal H}}\in {\mathbb T}_1(n/k)} (w_1,\cdots,
w_n)$, can be explored separately because they contribute
additively to the j.p.d.f. of all $n$ eigenvalues of
${\boldsymbol {\cal H}}$ from ${\mathbb T}_1(n)$:
\begin{eqnarray}
    P_n^{(1)}(w_1,\cdots, w_n) =
    \sum_{k=0}^n P_{{\boldsymbol {\cal H}}\in {\mathbb T}(n/k)}
    (w_1,\cdots, w_n).
\label{jpdf=1}
\end{eqnarray}
In entire generality, the partial j.p.d.f.'s have been
determined by Lehmann and Sommers (1991) who proved, a quarter
of a century after Ginibre's work, that the $k$-th partial
j.p.d.f. ($0\le k \le n$) equals
\begin{eqnarray} \label{b=1}
    \fl
    P_{{\boldsymbol {\cal H}}\in {\mathbb T}(n/k)}
     (\lambda_1,\cdots,\lambda_k;
    z_1,\cdots, z_\ell) \nonumber \\
    =
    \frac{2^{\ell - n(n+1)/4}}{i^\ell \,k! \,\ell!\, \prod_{j=1}^n \Gamma(j/2)}
    \prod_{i>j=1}^k |\lambda_i - \lambda_j|
    \, \prod_{j=1}^k \, \exp(-\lambda_j^2/2) \nonumber \\
    \times
    \prod_{j=1}^k \prod_{i=1}^\ell (\lambda_j-z_i)(\lambda_j-{\bar
    z}_i) \, \prod_{i>j=1}^{\ell} |z_i - z_j|^2 |z_i - {\bar
    z}_j|^2\,
     \nonumber \\
    \times  \prod_{j=1}^\ell \, (z_j-{\bar z}_j)\,
     {\rm erfc} \left(
        \frac{z_j-{\bar z}_j}{i\sqrt{2}}
    \right)\,
    \exp\left(-\frac{z_j^2+{\bar z}_j^2}{2}\right).
\end{eqnarray}
Here, the parameterisation
$(w_1,\cdots,w_n)=(\lambda_1,\cdots,\lambda_k; z_1,{\bar
z}_1,\cdots, z_\ell,{\bar z}_\ell)$ was used to indicate that
the spectrum is composed of $k$ real and $2\ell$ complex
eigenvalues so that $k+2\ell=n$. The above j.p.d.f. is
supported for $(\lambda_1,\cdots,\lambda_k)\in {\mathbb R}^k$,
$({\rm Re}\, z_1,\cdots,{\rm Re}\, z_\ell)\in {\mathbb R}^\ell$,
and $({\rm Im}\, z_1,\cdots, {\rm Im}\, z_\ell)\in ({\mathbb
R}^+)^\ell$. Notice that since the complex eigenvalues come in
$\ell$ conjugated pairs, the identity $n=2\ell+k$ implies that
half of the sets  ${\mathbb T}_1(n/k)$ are empty: this happens
whenever $n$ and $k$ are of different parity.
\newline
\newline
In writing (\ref{b=1}), we have used a representation due to
Edelman (1997) who rediscovered the result by Lehmann and
Sommers (1991) a few years later. The particular case $k=n$ of
(\ref{b=1}), corresponding to the matrices ${\boldsymbol {\cal
H}} \in {\mathbb T}(n/n)$ with all eigenvalues real, was first
derived by Ginibre (1965). No physical interpretation of the
distribution (\ref{b=1}) in terms of a two-dimensional Coulomb
gas is known as yet.
\end{itemize}\vspace{0.3cm}
\textsc{Eigenvalue Correlation Functions and Inapplicability of the
Dyson Integration Theorem to the Description of GinOE.} Spectral
statistical properties of random matrices can be retrieved from a
set of spectral correlation functions defined as
\begin{eqnarray}
\label{rpb=2} \fl
    R_p^{(\beta)}(z_1,\cdots,z_p;n) = \frac{n!}{(n-p)!} \,
    \prod_{j=p+1}^n \int_{\mathbb C} d^2z_j \, P_n^{(\beta)}(z_1, \cdots, z_n), \;\;\; \beta=2,\; 4
\end{eqnarray}
for GinUE ($\beta=2$) and GinSE ($\beta=4$), and
\begin{eqnarray}
\label{rpb1} \fl
    R_{p,q}^{({\boldsymbol {\cal H}}_k)}(\lambda_1,\cdots,\lambda_p;z_1,\cdots,z_q ;n)
=\frac{k!}{(k-p)!} \frac{\ell!}{(\ell-q)!} \nonumber\\
\hspace{-1cm}
\times \,
    \prod_{j=p+1}^k \int_{\mathbb R} d\lambda_j  \prod_{m=q+1}^\ell \int_{\mathbb C} d^2 z_m\,
    \, P_{{\boldsymbol {\cal H}}\in {\mathbb T}(n/k)}
(\lambda_1,\cdots,\lambda_k;z_1,\cdots, z_\ell),
\end{eqnarray}
for GinOE ($\beta=1$). The GinOE correlation function refers to the
spectrum of matrices ${\boldsymbol{\cal H}_k} \subset {\boldsymbol
{\cal H}}$ having exactly $k$ real eigenvalues.

The analytic calculation of the above correlation functions is one
of the major operational tasks of the non-Hermitean RMT. Whenever
feasible, such a calculation either explicitly rests on or can
eventually be traced back to the {\it three concepts}: (i) a
determinant (or Pfaffian) representation of the j.p.d.f.'s of all
eigenvalues, (ii) a projection property of the kernel function
associated with the aforementioned determinant representation, and
(iii) the Dyson integration theorem (Dyson 1970, Mahoux and Mehta
1991) that makes use of both (i) and (ii). Highly successful in the
Hermitean RMT, the above three concepts are not always at work in
the {\it non-Hermitean} RMT. Particularly, the Dyson integration
theorem, being effective for GinUE and GinSE (see below), fails to
work for GinOE. \newline\newline{\it It will be argued that a mixed
character of the GinOE spectrum consisting of both complex and
purely real eigenvalues is the direct cause of the failure.} For the
readers convenience as well as for the future reference, we cite the
Dyson integration theorem below \footnote[6]{While the formulation
in Mehta (2004) refers to the flat measure $d\pi(x) = dx$, the
Theorem 2.1 stays valid for any benign measure $d\pi(x)$.} (Mehta
(1976); see also Theorem 5.1.4 in Mehta's book
(2004)).\vspace{0.5cm}
\newline{\bf Theorem 2.1 (Dyson integration theorem).} {\it Let $f(x,y)$ be a
function with real, complex or quaternion values, such that
\numparts
\begin{eqnarray}\label{dt-1}
{\bar f}(x,y) = f(y,x),
\end{eqnarray}
where ${\bar f} = f$ if $f$ is real, ${\bar f}$ is the complex
conjugate if it is complex, and ${\bar f}$ is the dual of $f$ if it
is quaternion. Assume that
\begin{eqnarray}\label{dt-2}
    \int \,d\pi(y)\, f(x,y) \, f(y,z)\,  = f(x,z) + \lambda\, f(x,z) -
    f(x,z) \,\lambda,
\end{eqnarray}
where $\lambda$ is a constant quaternion and $d\pi$ is a suitable
measure. Let $[f(x_i,x_j)]_{n\times n}$ denote the $n\times n$
matrix with its $(i,j)$ element equal to $f(x_i,x_j)$. Then,
\begin{eqnarray}\label{dt-3}\fl
    \int\,  d\pi(x_n)\, {\rm det} [f(x_i,x_j)]_{n \times n} \, = (c-n+1)\,
    {\rm det}
    [f(x_i,x_j)]_{(n-1)\times (n-1)}
\end{eqnarray}
with
\begin{eqnarray} \label{dt-4}
    c= \int\, d\pi(x)\, f(x,x).
\end{eqnarray}
\endnumparts
When $f(x,y)$ is real or complex, the quaternion constant $\lambda$
vanishes. For $f(x,y)$ taking quaternion values, $\det$ should be
replaced by ${\rm qdet}$, the quaternion determinant (Dyson 1972).
}\vspace{0.5cm}\newline This theorem prompts the following
definition.\vspace{0.5cm}\newline{\bf Definition 2.1.} {\it A
function $f(x,y)$ satisfying the first and the second equation in
the Dyson integration theorem is said to obey the projection
property.}\vspace{0.5cm}\newline

Being equipped with the above
reminder, we are ready to present, and discuss, a collection of
formulae available for the $p$-point correlation functions in
Ginibre's ensembles. \newline

\begin{itemize}
  \item {\bf GinUE:} The joint probability density function (\ref{jpdf=2}) of
    all $n$ eigenvalues
    is reducible to a {\it
  determinant} form (Ginibre 1965)
    \begin{eqnarray}
    \label{jpdf=2=det}
        P_n^{(2)}(z_1,\cdots,z_n) = \frac{1}{n!}\,{\rm det} \left[
        K_n^{(2)}(z_k, z_\ell)
    \right]_{n\times n}  \prod_{j=1}^n e^{-z_j{\bar z}_j}
    \end{eqnarray}
with $K_n^{(2)}(z,z^\prime)$ being a two-point {\it scalar
kernel}
\begin{eqnarray}
\label{sk=2}
    K_n^{(2)}(z,z^\prime) = \frac{1}{\pi}
    \sum_{\ell=0}^{n-1} \frac{(z{\bar
    z}^\prime)^\ell}{\Gamma(\ell+1)}.
\end{eqnarray}
Since it obeys the projection property, the Dyson Integration
Theorem brings a {\it determinant} expression for the $p$-point
correlation function:
\begin{eqnarray}
\label{pp=2}
    R_p^{(2)}(z_1,\cdots, z_p; n) = {\rm det}
    \left[ K_n^{(2)}(z_k,z_\ell) \right]_{p\times p}  \prod_{j=1}^p e^{-z_j{\bar z}_j}.
\end{eqnarray}
These results, first derived by J. Ginibre (1965), provide a
comprehensive description of spectral fluctuations in GinUE.
\newline

  \item {\bf GinSE:} The joint probability density function
  (\ref{jpdf=4}) of all $n$ eigenvalues is reducible to a {\it quaternion
  determinant} form (Mehta and Srivastava 1966)
    \begin{eqnarray} \fl
    \label{jpdf=4=det}
    P_n^{(4)}(z_1,\cdots,z_n) =\frac{1}{n!}\,{\rm qdet} \left[
        K_n^{(4)}(z_k, z_\ell)
    \right]_{n \times n} \prod_{j=1}^n ({\bar z}_j-z_j)\,
    e^{-z_j{\bar z}_j},
    \end{eqnarray}
    where $K_n^{(4)}(z,z^\prime)$ is a quaternion whose $2\times
2$ matrix
    representation reads (Kanzieper 2002a)
\begin{eqnarray}
\label{mk=4}
    \Theta [K_n^{(4)} (z, z^\prime)] =
    \left(
      \begin{array}{cc}
        -\kappa_n^{(4)}({\bar z}, z^\prime) & -\kappa_n^{(4)}({\bar z}, {\bar
        z}^\prime) \\
        \kappa_n^{(4)}(z, z^\prime) & \kappa_n^{(4)}(z, {\bar z}^\prime)
     \end{array}
    \right).
\end{eqnarray}
Here,
\begin{eqnarray}
\label{kappa-n=4}
    \kappa_n^{(4)}(z, z^\prime) = \frac{1}{2\pi} \sum_{k=0}^{n-1}
    \sum_{\ell=0}^{k}
    \frac{z^{2k+1} (z^\prime)^{2\ell} - (z^\prime)^{2k+1} z^{2\ell} }{(2k+1)!!
    \, (2\ell)!!}.
\end{eqnarray}
Alternatively, but equivalently, (\ref{jpdf=4=det}) can be
    reduced to the Pfaffian form (Akemann and Basile 2007)
    \begin{eqnarray}\fl
    \label{4=pfaff}
    P_n^{(4)}(z_1,\cdots,z_n) &=\frac{1}{n!}\;{\rm pf\,}
    \left[
      \begin{array}{cc}
        \kappa_n^{(4)}(z_i, z_j) & \kappa_n^{(4)}(z_i,{\bar z}_j) \\
        \kappa_n^{(4)}({\bar z}_i, z_j) & \kappa_n^{(4)}({\bar z}_i, {\bar
        z}_j)
     \end{array}
    \right]_{2n \times 2n} \nonumber \\
    &\times \prod_{j=1}^n
    \, ({\bar z}_j-z_j)\, e^{-z_j {\bar z}_j}
    \end{eqnarray}
which is instructive to compare with (\ref{St-3}). \newline
\newline As soon as the quaternion kernel $K_n^{(4)}(z,z^\prime)$ satisfies the
projection property, the $p$-point correlation functions take a {\it
quaternion determinant/Pfaffian} form:
\begin{eqnarray}
\label{pp=4} \fl
    R_p^{(4)}(z_1,\cdots, z_p; n) = {\rm qdet} \left[
        K_n^{(4)}(z_k, z_\ell)
    \right]_{p \times p} \prod_{j=1}^p ({\bar z}_j - z_j)\, e^{-z_j{\bar z}_j}.
\end{eqnarray}
This result is due to Mehta and Srivastava (1966). \newline

  \item {\bf GinOE:} To the best of our knowledge, structural aspects of
  correlation functions in GinOE have never been addressed (see,
  however, a recent paper by Sinclair (2006)); consequently, no
  analogues of the above GinUE and GinSE formulae [(\ref{pp=2})
  and (\ref{pp=4})] are available. This gap will partially be
  filled in the present paper where we derive a {\it quaternion
  determinant/Pfaffian} expression (\ref{St-3}) for the j.p.d.f.
  of {\it all complex eigenvalues} of a random matrix
  ${\boldsymbol {\cal H}}\in {\mathbb T}_1(n/k)$. Importantly,
  the $2 \times 2$ kernel therein does {\it not} possess the
  projection property, hereby making the Dyson integration
  theorem inapplicable for the calculation of associated
  correlation functions. We reiterate that a mixed character of
  the GinOE spectrum, composed of both complex and purely real
  eigenvalues is behind the statement made
  \footnote[7]{Intriguingly, it will be shown in Section
  \ref{Sec8} that the matrices ${\cal H}\in {\mathbb T}_1(n/0)$
  exhibit GinSE-like correlations; this contrasts the well known
  correlations of the GOE type (Ginibre 1965) for the matrices
  ${\cal H}\in {\mathbb T}_1(n/n)$.}.

\end{itemize}
\vspace{0.5cm}\textsc{Quantifying the Qualitative Differences
Between Spectra of GinUE, GinSE, and GinOE.} The mean density of
eigenvalues $R_1^{(\beta)}(z=x+iy;n)$ is the simplest spectral
statistics exemplifying differences in spectral patterns of GinUE,
GinSE, and\linebreak GinOE. Below we collect, and comment on, the
exact and the large-$n$ results for the mean density of eigenvalues
of ${\boldsymbol {\cal H}} \in {\mathbb T}_\beta(n)$.
\newline

\begin{itemize}
    \item{\bf GinUE:} In accordance with (\ref{pp=2}) and (\ref{sk=2}), the
    exact result for the mean spectral
    density reads (Ginibre 1965)
    \begin{eqnarray}
    \label{dos1ex}
    R_1^{(2)}(z;n)=\frac{\Gamma(n,x^2+y^2)}{\pi\, \Gamma(n)}.
    \end{eqnarray}
    Here, $\Gamma(a,\phi)$ is the upper incomplete gamma function
    \begin{eqnarray}
        \Gamma(a,\phi) = \int_\phi^\infty dt\, t^{a-1}e^{-t}. \nonumber
    \end{eqnarray}
    For a large-$n$ GinUE matrix and $z=x+iy$ fixed
      \footnote[8]{That is, $z$ does not scale with $n\gg
1$.}, the mean spectral density approaches the constant
\begin{eqnarray}
    R_1^{(2)}(z; n \gg 1) \simeq \frac{1}{\pi}.
\end{eqnarray}
This suggests that that the eigenvalues of a
    large-$n$ GinUE matrix
    are distributed almost {\it uniformly} within the two-dimensional
    disk
    of  the radius $\sqrt{n}$. More rigorously, this statement follows from
    the {\it macroscopic}
    limit of (\ref{dos1ex})
    \begin{eqnarray}
        \label{disk-2}
        \lim_{n\rightarrow \infty} R_1^{(2)}(
z={\hat z}\sqrt{n};n) =
        \frac{1}{\pi}
            \left\{
                      \begin{array}{cc}
                        1 & {\rm if\;} |{\hat z}|<1 \\
                        0 & {\rm if\;} |{\hat z}|>1 \\
                      \end{array}
            \right.
    \end{eqnarray}
    known as the Girko circular law (Girko 1984, Girko 1986, Bai 1997). The
    density of eigenvalues away from
    the disk is exponentially suppressed (Kanzieper 2005).
    \newline

\begin{figure}[t]
\includegraphics[scale=.14, bb=0 0 800 600]{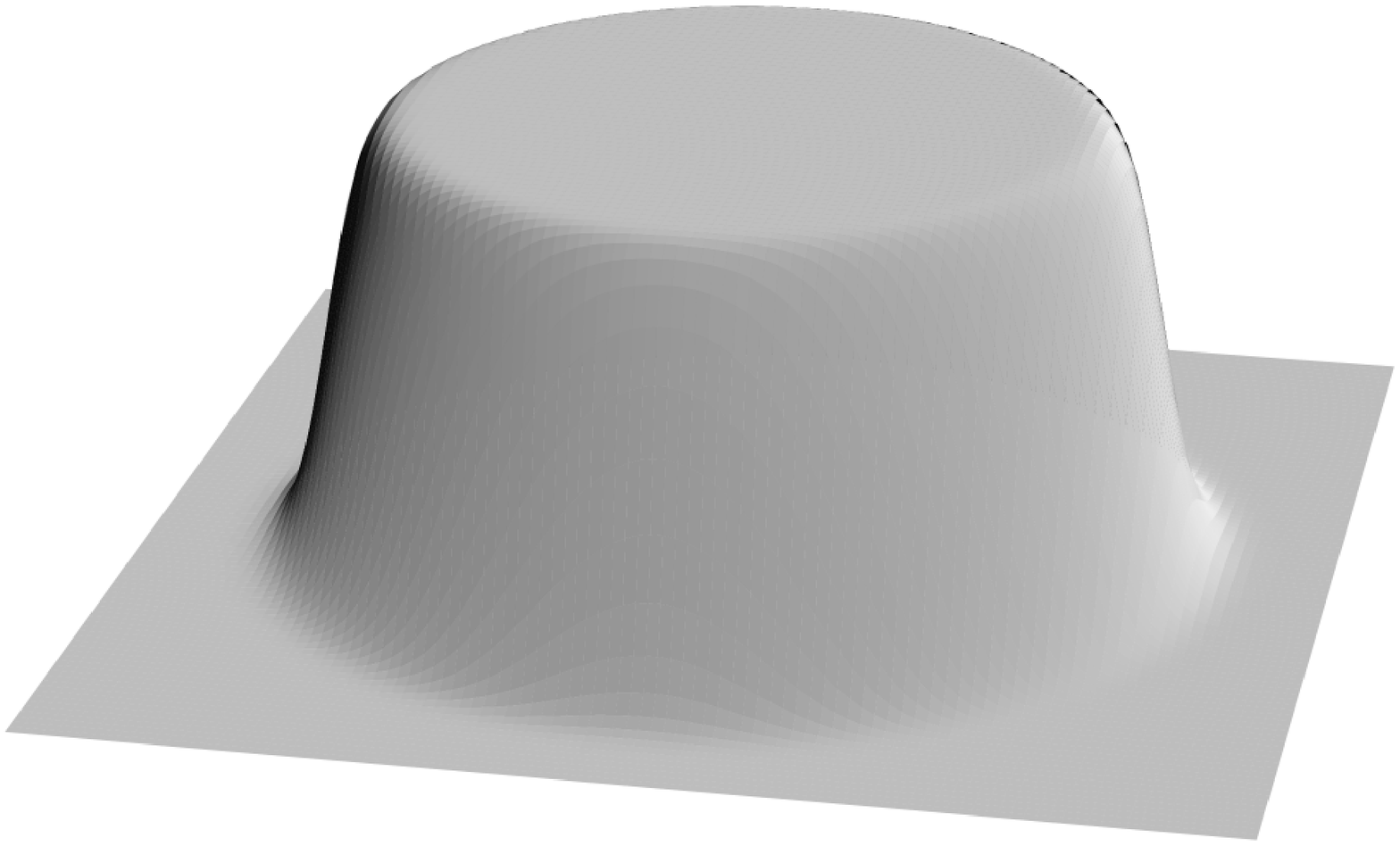}\;\;\;
\includegraphics[scale=.14, bb=0 0 800 600]{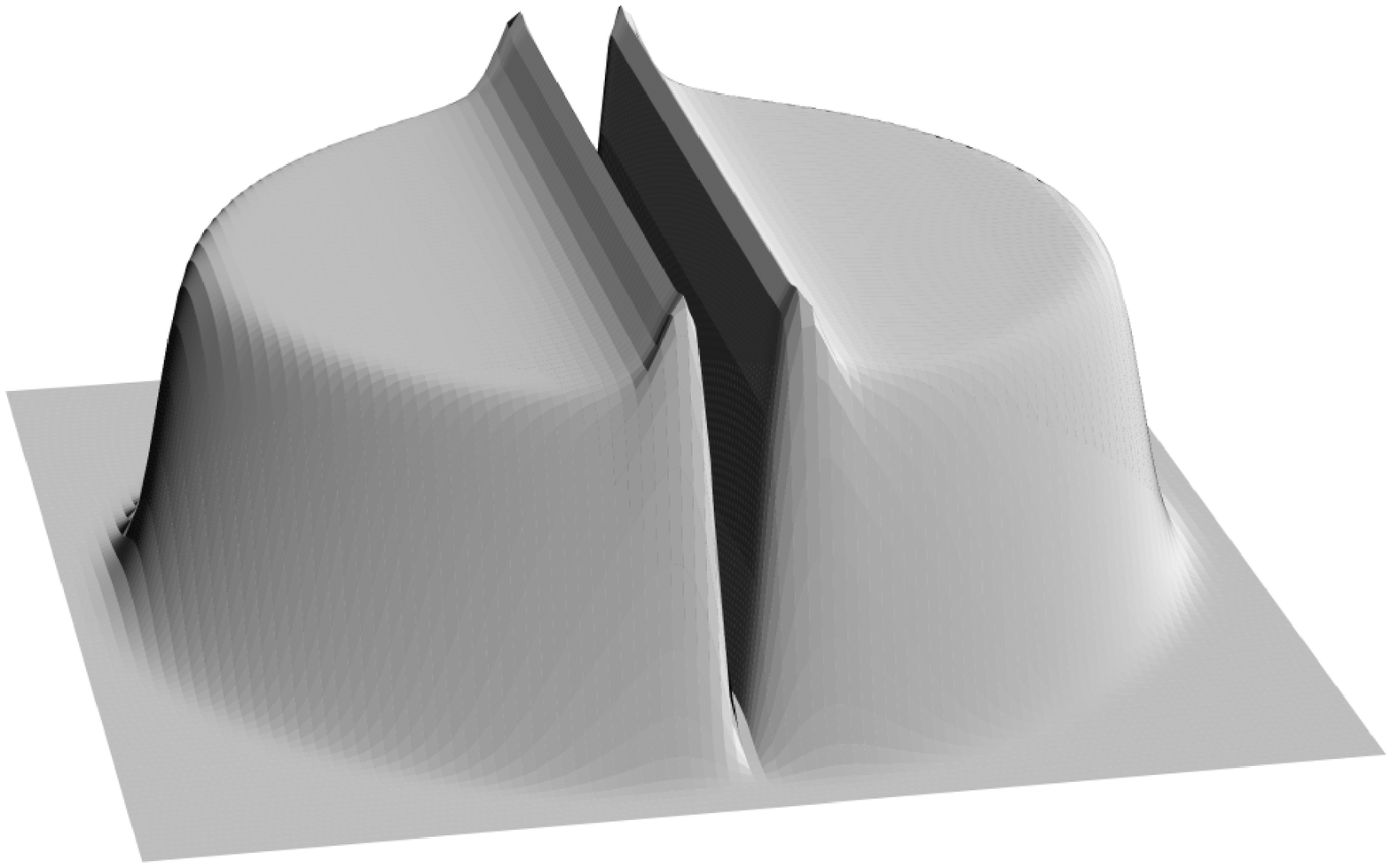}\;
\includegraphics[scale=.17, bb=0 0 800 600]{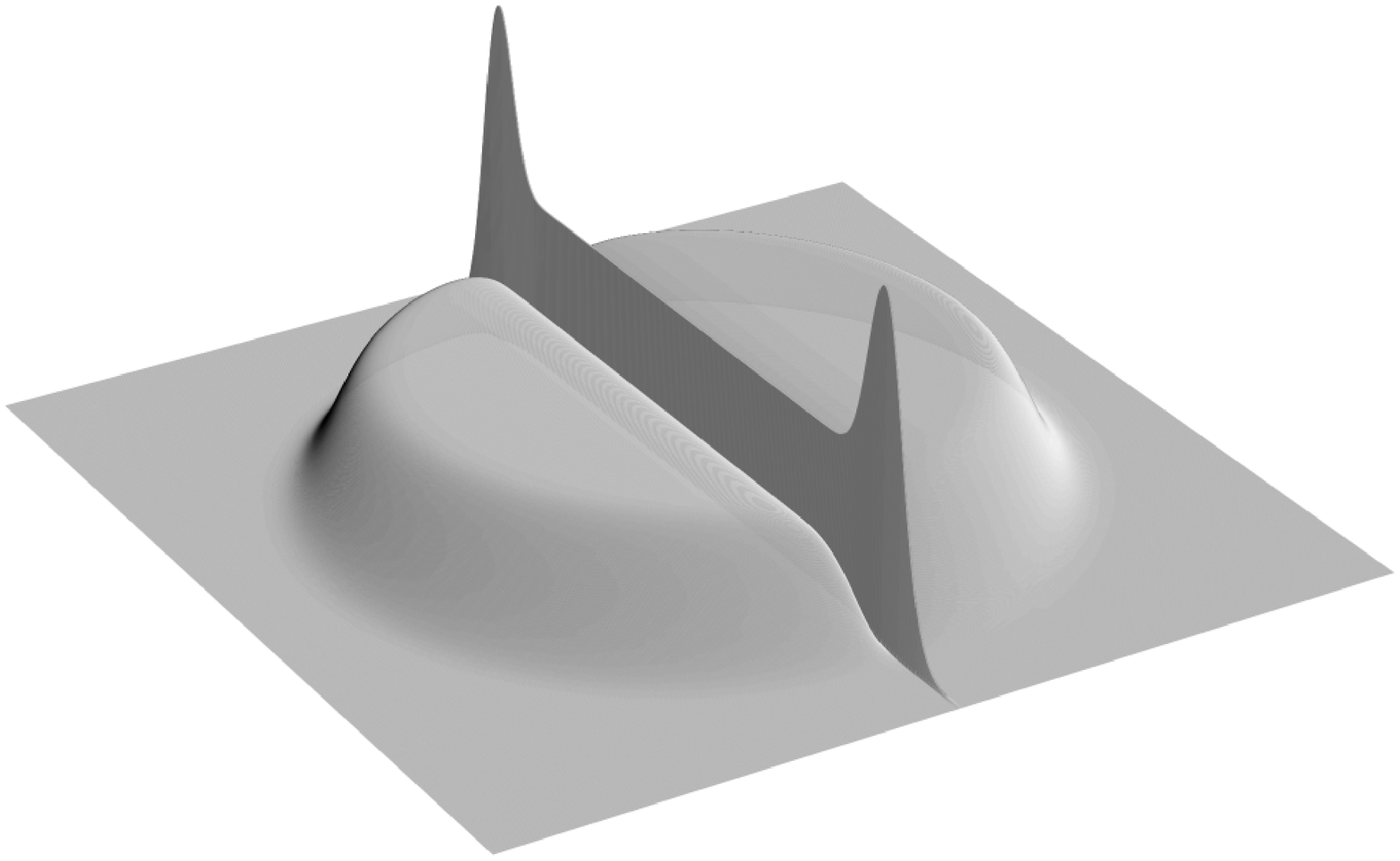}
    \caption{Profiles of eigenlevel densities $R_1^{(\beta)}(z;n)$, plotted as functions of
            the complex energy $z=x+iy$ for $n$ fixed, show a nearly
            uniform eigenlevel distribution in GinUE (left panel), a depletion of eigenvalues
            along the real axis in GinSE as exemplified by the density drop at $y=0$ (middle panel),
            and accumulation of real eigenvalues in GinOE displayed as a wall
            at $y=0$ (right panel); the wall height imitates the density of real eigenvalues.}\label{dos-profiles}
\end{figure}

\item{\bf GinSE:} In accordance with (\ref{mk=4}) -- (\ref{pp=4}), the exact result for the mean spectral
    density reads (Mehta and Srivastava 1966, Kanzieper 2002a)
    \begin{eqnarray}
    \label{r14}
    R_1^{(4)}(z;n)= 2y \, e^{-(x^2+y^2)}\,{\rm Im\,}\kappa_n^{(4)}(x+iy,x-iy)
    \end{eqnarray}
    with $\kappa_n^{(4)}$ given by (\ref{kappa-n=4}). For a
    large-$n$ GinSE matrix and $z=x+iy$ fixed, it reduces to
    (Kanzieper, 2002a)
    \begin{eqnarray}
    \label{dos=4} R_1^{(4)}(z;n \gg 1) \simeq \frac{y}{\sqrt{2\pi}}
    \, e^{-2y^2} {\rm\, erfi} \big(y\sqrt{2}
    \big).
    \end{eqnarray}
    Here, ${\rm erfi\,}(\phi)$ is the imaginary error function
    \begin{eqnarray}
    {\rm erfi}(\phi) = \frac{2}{\sqrt{\pi}} \int_0^\phi dt \,
    e^{t^2}. \nonumber
    \end{eqnarray}
    Both results [(\ref{r14}) and (\ref{dos=4})] suggest that the mean eigenvalue density is no longer
    uniform but exhibits a {\it depletion} of eigenvalues along the real
    axis. Similarly to the GinUE, the mean spectral
    density in GinSE is suppressed away from a disk of the radius $\sqrt{2n}$
as suggested by the circular law
\begin{eqnarray}
        \label{disk-4}
        \lim_{n\rightarrow \infty} R_1^{(4)}({\hat z}\sqrt{2n};n) =
        \frac{1}{2\pi}
            \left\{
                      \begin{array}{cc}
                        1 & {\rm if\;} |{\hat z}|<1 \\
                        0 & {\rm if\;} |{\hat z}|>1 \\
                      \end{array}
            \right.
    \end{eqnarray}
due to Khoruzhenko and Mezzadri (2005).
\newline

\item{\bf GinOE:} A mixed character of the GinOE spectrum consisting of
both complex and real eigenvalues makes the RMT techniques based
on the Dyson integration theorem inapplicable to the description
of spectral statistical properties of GinOE.

To evaluate the mean eigenvalue density in the finite-$n$ GinOE,
a totally different approach has been invented by A. Edelman and
co-workers (Edelman {\it et al} 1994, Edelman 1997). Starting
directly with the definition (\ref{G-def}) taken at $\beta=1$
and applying the methods of multivariate statistical analysis
(Muirhead 1982), these authors have separately determined the
exact mean densities \footnote[1]{The two are related to the
$(p,q)$ correlation functions $R_{p,q}^{({\boldsymbol {\cal
H}}_k)}$ for the matrices ${\boldsymbol {\cal H}}_k$ restricted
to have exactly $k$ real eigenvalues [see the definition
(\ref{rpb1})] as follows:
\begin{eqnarray} \fl
    R^{(1)}_{1,{\rm real}}(z;n) = \delta(y)\sum_{k=1}^n R_{1,0}^{({\boldsymbol {\cal H}}_k)}
    (x;n),\;\;\;
    R^{(1)}_{1,{\rm complex}}(z;n) = \sum_{\ell=1}^{\lfloor n/2 \rfloor} R_{0,1}^{({\boldsymbol {\cal H}}_{n-2\ell})}
    (z;n). \nonumber
\end{eqnarray}
} of purely {\it real} eigenvalues (Edelman {\it et al} 1994)
\pagebreak
\begin{eqnarray}
\label{dre} \fl
    R_{1,{\rm real}
}^{(1)}(z;n) = \frac{
\delta(y)}{\sqrt{2\pi}}
    \Bigg[\frac{\Gamma(n-1,x^2)}{\Gamma(n-1)}
         + \frac{2^{n/2-3/2}}{\Gamma(n-1)} \,
        |x|^{n-1} \, e^{-x^2/2}\,
\gamma \left( \frac{n-1}{2}, \frac{x^2}{2}
        \right) \Bigg]\nonumber \\{}
\end{eqnarray}
and of strictly {\it nonreal} \footnote[2]{In the formulae, we
use the subscript ``complex'' to identify eigenvalues with zero
real part.} eigenvalues (Edelman 1997):
\begin{eqnarray} \fl
    \label{dce}
    R^{(1)}_{1,{\rm complex}
} (z;n) =
     \sqrt{\frac{2}{\pi}}\,\frac{\Gamma(n-1,x^2+y^2)}{\Gamma(n-1)}\,
     \, y\,  e^{2y^2}\,{\rm erfc\,}
    \big( y \sqrt{2} \big).
\end{eqnarray}
Here $x$ and $y$ are real and imaginary parts of $z=x+iy$. The
function $\gamma(a,\phi)$ in (\ref{dre}) is the lower incomplete
gamma function
\begin{eqnarray}
\label{gf=def}
    \gamma(a,\phi) = \int_0^\phi dt\, t^{a-1}e^{-t}.
    \nonumber
\end{eqnarray}
Importantly, {\it no}
reference was made to the j.p.d.f. (\ref{jpdf=1}) and (\ref{b=1}) in
deriving (\ref{dre}) and (\ref{dce}).
\newline\newline
For a large-$n$ GinOE matrix and $z=x+iy$ fixed, the above two
formulae yield the mean density of eigenvalues in the form
\begin{eqnarray}
\label{de-large}
    R_{1}^{(1)}(z;n\gg 1) \simeq \frac{1}{\sqrt{2\pi}}
     \delta(y) +
    \sqrt{\frac{2}{\pi}}\, y\, e^{2y^2}\,{\rm erfc\,}
    \big( y \sqrt{2} \big).
\end{eqnarray}
Similarly to GinUE and GinSE, the circular law (Girko 1984, Sommers
{\it et al} 1988, Bai 1997, Edelman 1997)
\begin{eqnarray}
    \label{disk-1}
      \lim_{n\rightarrow \infty} \,
R_{1}^{(1)} ({\hat z}\sqrt{n};n)= \frac{1}{\pi}
            \left\{
                      \begin{array}{cc}
                        1 & {\rm if\;} |{\hat z}|<1 \\
                        0 & {\rm if\;} |{\hat z}|>1 \\
                      \end{array}
            \right.
\end{eqnarray}
holds.
\end{itemize}

\subsection{Statistical description of the eigenvalue accumulation in GinOE}

In essence, the approach culminating in the explicit formula
(\ref{dre}) for the mean density of real eigenvalues represents the
simplest possible quantitative description of the phenomenon of
eigenvalue {\it accumulation} along the real axis. Complementarily,
the one might be interested in the full statistics of the number
${\cal N}_r$ (${\cal N}_c$) of real (complex) eigenvalues occurring
in the GinOE spectrum. In the latter context, the result (\ref{dre})
can only supply the first moment of ${\cal N}_r$ -- the expected
number $E_n={\mathbb E}[{\cal N}_r]$ of real eigenvalues. Indeed,
integrating out the eigenlevel densities (\ref{dre}) and/or
(\ref{dce}) over the entire complex plane,
\begin{eqnarray}
    \label{En}
    E_n = \int_{{\mathbb C}}d^2z \, R_{1,{\rm\,
    real}}^{(1)}(z;n) = n - \int_{{\mathbb C}\setminus {\mathbb R}}d^2 z\, R_{1,{\rm\,
    complex}}^{(1)}(z;n), \nonumber
\end{eqnarray}
Edelman {\it et al} (1994) have obtained the remarkable result
\begin{eqnarray}
    \label{En-explicit}
    E_n = \frac{1}{2} + \sqrt{2}\, \frac{{}_2 F_1\,
(1,-1/2;n;1/2)}{B(n,1/2)}
\end{eqnarray}
expressed in terms of the Gauss hypergeometric function and the
Euler Beta function. As $n\rightarrow \infty$, it furnishes the
asymptotic series
\begin{eqnarray} \fl
    \label{EnAs}
    E_n =\sqrt{\frac{2n}{\pi}} \, \left(
    1 - \frac{3}{8n} -\frac{3}{128 n^2} + \frac{27}{1024n^3} +
    \frac{499}{32768 n^4} + {\cal O}(n^{-5})
    \right) + \frac{1}{2}.
\end{eqnarray}
The nonperturbative $\sqrt{n}$-dependence of leading term in
(\ref{EnAs}) had been earlier detected numerically by Sommers {\it
et al} (1988).

What is about a more detailed statistical description of the number
${\cal N}_r$ (${\cal N}_c$) of real (complex) eigenvalues? Can all
the moments and the entire probability function of the discrete
random variable ${\cal N}_r$ (${\cal N}_c$) be determined? The
preceding discussion, particularly highlighting a patchy knowledge
of the spectral {\it correlations} in GinOE, suggests that both
spectral characteristics, which can be expressed as \footnote[3]{The
coefficient $S(p,k)$ is the Stirling number of the second kind.}
\begin{eqnarray}
    \label{enp}
    {\mathbb E}[{\cal N}_r^p] = \sum_{k=1}^p S(p,k) \,
    \int_{{\mathbb C}} d^2 z_1 \cdots  \int_{{\mathbb C}} d^2 z_k \,
    R_{k, {\rm \, real}}^{(1)} (z_1,\cdots,z_k;n)
\end{eqnarray}
and
\begin{eqnarray} \fl
    \label{enpc}
    {\mathbb E}[{\cal N}_c^p] = \sum_{\ell=1}^p S(p,\ell) \,
    \int_{{\mathbb C}\setminus {\mathbb R}} d^2 z_1 \cdots  \int_{{\mathbb C}\setminus {\mathbb R}} d^2 z_\ell \,
    R_{\ell, {\rm \, complex}}^{(1)} (z_1, \cdots,z_\ell; n),
\end{eqnarray}
are not within immediate reach. First, only one correlation function
out of $p$ involved in either (\ref{enp}) or (\ref{enpc}) is
explicitly known. Second, even if all the multipoint correlation
functions were readily available, the integrations in (\ref{enp})
and (\ref{enpc}) would not be trivial either because of the
anticipated inapplicability of the Dyson integration theorem as
discussed in Section \ref{SubSec23}.

Some of the difficulties outlined here will be overcome in the
present paper. For a list of our major results the reader is
referred back to the Section \ref{SubSec12}. Proofs and derivations
are given in the Sections to follow.

\section[Statistics of real eigenvalues in GinOE
spectra]{Statistics of real eigenvalues in GinOE spectra: A Pfaffian
integral representation for the probability $p_{n,k}$} \label{Sec3}

\subsection{Generalities and known results}
\label{SubSec31} Instead of targeting the moments ${\mathbb E}[{\cal
N}_r^p]$ of a fluctuating number of real eigenvalues as discussed in
the previous section, we are going to directly determine the entire
probability function $p_{n,k} = {\rm Prob}({\cal N}_r=k)$. The
definition of the $p_{n,k}$, describing the probability of finding
exactly $k$ real eigenvalues in the spectrum of an $n\times n$ real
Gaussian random matrix, can be deduced from (\ref{jpdf=1}) and
(\ref{b=1}),
\begin{eqnarray} \fl
    \label{pnk}
    p_{n,k} = {\rm Prob}({\boldsymbol {\cal H}}\in {\mathbb T}(n/k))
    = \prod_{i=1}^k \int_{{\mathbb R}} d\lambda_i
    \prod_{j=1}^\ell \int_{{\rm Im\,}z_j > 0} d^2 z_j\,
    P_{{\boldsymbol {\cal H}}\in {\mathbb T}(n/k)}.
\end{eqnarray}
Here, $\ell$ is the number of pairs of complex conjugated
eigenvalues in the spectrum of an $n \times n$ matrix ${\boldsymbol
{\cal H}}\in {\mathbb T}(n/k)$ having exactly $k$ real eigenvalues,
and $P_{{\boldsymbol {\cal H}}\in {\mathbb T}(n/k)}$ is the j.p.d.f.
of all $n$ eigenvalues of such a matrix. Obviously, the identity
$n=k+2\ell$ holds.

Previous attempts to determine the probability function $p_{n,k}$
based on (\ref{pnk}) brought no explicit formula for $p_{n,k}$ for
generic $k$. The only analytic results available are due to Edelman
(1997) who proved the following properties of the above probability
function \footnote[4]{Table \ref{prob-table} provides a useful
illustration of both properties.}:
\newline\newline
\textsc{Property 1.}
The probability of having all $n$ eigenvalues real
equals
\begin{eqnarray}
\label{ed-res}
 p_{n,n}=2^{-n(n-1)/4}.
\end{eqnarray}
\newline
\textsc{Property 2.} For all $0\le k\le n$, the $p_{n,k}$ are of the
form
\begin{eqnarray}
p_{n,k}=r_{n,k}+ s_{n,k}\sqrt{2},
\end{eqnarray}
where $r_{n,k}$ and $s_{n,k}$ are rational.
\newline\newline
The first result is a simple consequence of (\ref{pnk}) which, at
$k=n$, reduces to a known Selberg integral. All known examples
suggest that this is the smallest probability out of all
$p_{n,k}$'s. The second result is based on more involved
considerations of (\ref{pnk}) and (\ref{jpdf=1}).

\subsection{Joint probability density function of complex eigenvalues of
${\boldsymbol {\cal H}} \in {\mathbb T}(n/k)$} \label{SubSec32}

To evaluate the sought probability function $p_{n,k}$, we start with
the definition (\ref{pnk}). In a first step, we carry out the
$\lambda$-integrations therein to assess the j.p.d.f. of all complex
eigenvalues of a random matrix ${\boldsymbol {\cal H}}\in {\mathbb
T}(n/k)$ having exactly $k$ real eigenvalues:
\begin{eqnarray}
    \label{phk-comp} \fl
    P_{{\boldsymbol {\cal H}_k}}
(z_1,\cdots, z_\ell)
    = \prod_{i=1}^k \int_{{\mathbb R}} d\lambda_i \, P_{{\boldsymbol {\cal
    H}}\in {\mathbb T}(n/k)}
 (\lambda_1,\cdots,\lambda_k; z_1,\cdots, z_\ell).
\end{eqnarray}
To proceed, we consult (\ref{b=1}) to spot that a part of it,
\begin{eqnarray} \fl
\label{piece}
    {\cal I}_{k,\ell}(\{z,{\bar z}\}) \equiv
    \prod_{i=1}^k \int_{{\mathbb R}} d\lambda_i  \,
    \prod_{j=1}^k \, e^{-\lambda_j^2/2}
    \prod_{i>j}^k | \lambda_i - \lambda_j | \,
    \prod_{j=1}^k \prod_{i=1}^\ell (\lambda_j - z_i) (\lambda_j - {\bar
    z}_i),
\end{eqnarray}
coincides up to a prefactor to be specified below with the average
characteristic polynomial \footnote[5]{Equations (\ref{phk-comp})
and (\ref{piece}) suggest that ${\cal P}_{0,n}=1$.}
\begin{eqnarray} \fl
    {\cal P}_{k,\ell}(\{z,{\bar z}\})=\left(
                \int
     D{{\boldsymbol {\cal O}}}
        \, e^{-{\rm tr\,} {\boldsymbol {\cal O}}^2/2}
    \right)^{-1} \nonumber \\
    \times
    \int
     D{{\boldsymbol {\cal O}}}
        \prod_{j=1}^\ell
        {\rm det\,} (z_j - {\boldsymbol {\cal O}})\,
        {\rm det\,} ({\bar z}_j - {\boldsymbol {\cal O}})
        \, e^{-{\rm tr\,} {\boldsymbol {\cal O}}^2/2}
\end{eqnarray}
of a $k\times k$ real symmetric matrix ${\boldsymbol {\cal
O}}={\boldsymbol {\cal O}}^T$ drawn from the GOE. More precisely,
\begin{eqnarray}
    {\cal I}_{k,\ell}(\{z,{\bar z}\}) = {\mathfrak{s}}_k \, k!\, {\cal
    P}_{k,\ell}(\{z,{\bar
    z}\}), \nonumber
\end{eqnarray}
where ${\mathfrak{s}}_k$ is given by the Selberg integral (Mehta
2004)
\begin{eqnarray}
     {\mathfrak{s}}_k =  \frac{1}{k!}\prod_{i=1}^k \int_{{\mathbb R}}
    d\lambda_i  \,
    \prod_{j=1}^k \, e^{-\lambda_j^2/2}
    \prod_{i>j}^k | \lambda_i - \lambda_j | =
    2^{k/2} \,  \prod_{j=1}^k \Gamma(j/2).
\end{eqnarray}
Consequently, the j.p.d.f. of all complex eigenvalues of
${\boldsymbol {\cal H}} \in {\mathbb T}(n/k)$ is expressed in terms
of the average GOE characteristic polynomial ${\cal
P}_{k,\ell}(\{z,{\bar z}\})$ as
\begin{eqnarray}
        \label{phk-pol} \fl
P_{{\boldsymbol {\cal H}_k}} (z_1,\cdots, z_\ell)
    &= \frac{2^\ell}{i^\ell \ell!}\,
    \frac{{\mathfrak{s}}_k}{{\mathfrak{s}}_n}
    \, p_{n,n} \;
    {\cal P}_{k,\ell}(\{z,{\bar
    z}\}) \prod_{i>j=1}^{\ell} |z_i - z_j|^2 |z_i - {\bar
    z}_j|^2\,
     \nonumber \\
     &\times  \prod_{j=1}^\ell \, (z_j-{\bar z}_j)\,
     {\rm erfc} \left(
        \frac{z_j-{\bar z}_j}{i\sqrt{2}}
    \right)\,
    \exp\left(-\frac{z_j^2+{\bar z}_j^2}{2}\right).
\end{eqnarray}
A little bit more spadework is needed to appreciate the beauty
hidden in this representation. Borrowing the result due to Borodin
and Strahov (2005), who discovered a Pfaffian formula for a general
averaged GOE characteristic polynomial (see also an earlier paper by
Nagao and Nishigaki (2001)), we may write ${\cal P}_{k,\ell}$ in the
form
\begin{eqnarray}
    \label{chp} \fl
    {\cal P}_{k,\ell} (\{z,{\bar z}\}) =
    \frac{{\mathfrak{s}}_n/{\mathfrak{s}}_k}
    {\Delta_{2\ell}(\{z,{\bar z}\})}\, (-1)^\ell  \nonumber \\
    \times \prod_{j=1}^\ell \,
    \exp\left(
        \frac{z_j^2 + {\bar z}_j^2}{2}\right)
    \,
     {\rm pf\,}
    \left[
    \begin{array}{cc}
      {\cal D}_{n}(z_i, z_j) & {\cal D}_{n}(z_i, {\bar z}_j) \\
      {\cal D}_{n}({\bar z}_i, z_j) & {\cal D}_{n}({\bar z}_i, {\bar z}_j) \\
    \end{array}
    \right]_{2\ell \times 2\ell}.
\end{eqnarray}
Here, ${\cal D}_n(z_i,z_j)$ is the so-called $D$-part of the
$2\times 2$ GOE matrix kernel (Tracy and Widom 1998) to be defined
in (\ref{d2m}) and (\ref{d2m+1}) below; $\Delta_{2\ell}(\{z,{\bar
z}\})$ is the Vandermonde determinant \footnote[6]{Throughout the
paper, we adopt the definition $\Delta_k(x) = \prod_{i>j}^k
(x_i-x_j)$ which differs from that of Borodin and Strahov (2005) who
use the same notation $\Delta_k(x)$ for the double product with
$i<j$.}
\begin{eqnarray} \fl
    \label{vnd}
    \Delta_{2\ell}(\{z,{\bar z}\}) = \Delta_{2\ell}(z_1,{\bar z}_1,\cdots,
    z_\ell,{\bar z}_\ell)
    =
    \prod_{i>j}^\ell |z_i - z_j|^2 \, \prod_{i>j}^\ell |z_i - {\bar
    z}_j|^2 \, \prod_{i=1}^\ell ({\bar z}_i - z_i).
\end{eqnarray}
Combining (\ref{phk-pol}), (\ref{chp}) and (\ref{vnd}), we obtain:
\begin{eqnarray}
    \label{St-3-der}
    \fl
    P_{{\boldsymbol {\cal H}_k}}
(z_1\cdots, z_\ell)
&=   \frac{p_{n,n}}{\ell!}\,\left(\frac{2}{i}\right)^\ell
    \, \nonumber \\
    &\times \prod_{j=1}^\ell \, {\rm erfc}
    \left( \frac{z_j - {\bar z}_j}{i\sqrt{2} } \right)
    \,
    {\rm pf\,} \left[
    \begin{array}{cc}
      {\cal D}_{n}(z_i, z_j) & {\cal D}_{n}(z_i, {\bar z}_j) \\
      {\cal D}_{n}({\bar z}_i, z_j) & {\cal D}_{n}({\bar z}_i, {\bar z}_j) \\
    \end{array}
    \right]_{2\ell \times 2\ell}
\end{eqnarray}
where $({\rm Re}\, z_1,\cdots,{\rm Re}\, z_\ell)\in {\mathbb
R}^\ell$ and $({\rm Im}\, z_1,\cdots, {\rm Im}\, z_\ell)\in
({\mathbb R}^+)^\ell$ by derivation.

Equation (\ref{St-3-der}) is the central result of this section.
Announced in (\ref{St-3}), it describes the j.p.d.f. of $\ell$ pairs
of complex conjugated eigenvalues $\{z_j,{\bar z}_j\}$ of an
$n\times n$ random matrix ${\boldsymbol {\cal H}}_k$ whose $k$
remaining eigenvalues are real, $k+2\ell=n$.

To make the expression for the j.p.d.f. $P_{{\boldsymbol {\cal
H}}_k}$ explicit, we have to specify the kernel function ${\cal
D}_n(z_i,z_j)$. The latter turns out to be sensitive to the parity
of $n$ (see, e.g., Adler {\it et al} (2000)). For $n=2m$ even, the
kernel function is given by
\begin{eqnarray} \fl
\label{d2m}
    {\cal D}_{2m}(x,y) =
    \frac{1}{2}
    e^{-(x^2+y^2)/2}
    \sum_{j=0}^{m-1} \frac{q_{2j+1}(x)\, q_{2j}(y) - q_{2j}(x)\,
    q_{2j+1}(y)}{h_j}
\end{eqnarray}
while for $n=2m+1$ odd, it equals
\begin{eqnarray} \fl
\label{d2m+1}
    {\cal D}_{2m+1}(x,y) =  \frac{1}{2}
        e^{-(x^2+y^2)/2}
    \sum_{j=0}^{m-1} \frac{{\tilde q}_{2j+1}(x)\, {\tilde q}_{2j}(y)
    -
    {\tilde q}_{2j}(x)\, {\tilde q}_{2j+1}(y)}{h_j}.
\end{eqnarray}
Both representations (\ref{d2m}) and (\ref{d2m+1}) involve the
polynomials $q_j(x)$ skew orthogonal on ${\mathbb R}$ with respect
to the GOE skew product (Mehta 2004)
\begin{eqnarray}
    \left<f,\, g\right> = \frac{1}{2} \int_{{\mathbb R}}dx\, e^{-x^2/2}
        \int_{{\mathbb R}}dy\, e^{-y^2/2}
        {\rm sgn}(y-x)\, f(x)\, g(y)
\end{eqnarray}
such that
\begin{eqnarray} \fl
\label{sops}
    \left< q_{2k}, q_{2\ell+1} \right> =
    -
    \left< q_{2k+1}, q_{2\ell} \right> = h_k \delta_{k,\ell},\;\;\;
    \left< q_{2k}, q_{2\ell} \right> = \left< q_{2k+1}, q_{2\ell+1} \right>
    =
    0.
\end{eqnarray}
The skew orthogonal polynomials $q_j(x)$ can be expressed
\footnote[7]{The representation (\ref{skew-hermite}) is not unique;
see, e.g., Eynard (2001).} in terms of Hermite polynomials as
\footnote[8]{Equation (\ref{skew-hermite}) assumes that
$H_{-1}(x)\equiv 0$.}
\begin{eqnarray}
    \label{skew-hermite}
    q_{2j}(x) = \frac{1}{2^{2j}}\, H_{2j}(x), \nonumber \\
    q_{2j+1}(x) = \frac{1}{2^{2j+1}}\, \big[
    H_{2j+1}(x) - 4j\, H_{2j-1}(x)\big]
\end{eqnarray}
while ``tilded'' polynomials ${\tilde q}_j(x)$ \footnote[1]{Note
that the ${\tilde q}_{2j}(x)$ is no longer a polynomial of the
degree $2j$.} entering (\ref{d2m+1}) are related to $q_j(x)$ via
\begin{eqnarray}
\label{sh-odd}
    {\tilde q}_{2j}(x) = q_{2j}(x) - \frac{(2j)!}{2^{2j} j!}\,
    \frac{2^{2m} m!}{(2m)!}\,q_{2m}(x), \nonumber \\
    {\tilde q}_{2j+1}(x) = q_{2j+1}(x).
\end{eqnarray}
Specifying the normalisation
\begin{eqnarray}
\label{hj}
    h_j = \left< q_{2j},\, q_{2j+1}\right>=\frac{\sqrt{\pi}\,
(2j)!}{2^{2j}}
\end{eqnarray}
completes our derivation of (\ref{St-3-der}).

\subsection{Probability function $p_{n,k}$ as a Pfaffian integral and
inapplicability of the Dyson integration theorem for its
calculation} \label{SubSec33}

The results obtained in Section \ref{SubSec32} allow us to express
the probability function $p_{n,k}$ in the form of an $\ell$-fold
integral
\begin{eqnarray} \fl
    \label{pnk-comp}
    p_{n,k} = \frac{p_{n,n}}{\ell!} \, \left(\frac{2}{i}\right)^\ell
    \,
    \prod_{j=1}^\ell \int_{{\rm Im\,}z_j >0} d^2z_j\, \nonumber \\
    \times\; {\rm erfc}
    \left( \frac{z_j - {\bar z}_j}{i\sqrt{2} } \right)
    \,
    {\rm pf\,} \left[
    \begin{array}{cc}
      {\cal D}_{n}(z_i, z_j) & {\cal D}_{n}(z_i, {\bar z}_j) \\
      {\cal D}_{n}({\bar z}_i, z_j) & {\cal D}_{n}({\bar z}_i, {\bar z}_j) \\
    \end{array}
    \right]_{2\ell \times 2\ell}
\end{eqnarray}
involving a Pfaffian. It can also be rewritten as a quaternion
determinant
\begin{eqnarray}
    {\rm pf\,} \left[
    \begin{array}{cc}
      {\cal D}_{n}(z_i, z_j) & {\cal D}_{n}(z_i, {\bar z}_j) \\
      {\cal D}_{n}({\bar z}_i, z_j) & {\cal D}_{n}({\bar z}_i, {\bar z}_j) \\
    \end{array}
    \right]_{2\ell \times 2\ell} =
    {\rm qdet} [{\hat {\cal D}}_n(z_i, z_j)]_{\ell \times \ell} \nonumber
\end{eqnarray}
where the self-dual quaternion ${\hat {\cal D}}_n(z_i, z_j)$ has a
$2\times 2$ matrix representation
\begin{eqnarray}
    \Theta [{\hat {\cal D}}_n(z_i, z_j)] = \left(
    \begin{array}{rr}
      -{\cal D}_{n}(\bar z_i, z_j) & -{\cal D}_{n}({\bar z}_i, {\bar z}_j) \\
      {\cal D}_{n}(z_i, z_j) & {\cal D}_{n}(z_i, {\bar z}_j) \\
    \end{array}
    \right). \nonumber
\end{eqnarray}
The Pfaffian/quaternion determinant form of the integrand in
(\ref{pnk-comp}), closely resembling the structure of both the
j.p.d.f. (\ref{jpdf=4=det}) of all complex eigenvalues and the
$p$-point correlation function (\ref{pp=4}) in GinSE, makes it
tempting to attack the $\ell$-fold integral (\ref{pnk-comp}) with
the help of the Dyson integration theorem (see Section
\ref{SubSec23}). Unfortunately, the key condition of this theorem --
the projection property -- is not fulfilled.

To see this point, we represent the kernel function ${\cal
D}_n(x,y)$ in the form
\begin{eqnarray}
\label{DnTW}
    {\cal D}_{n}(x,y) =
    \frac{1}{2} \, e^{-(x^2+y^2)/2}
    \sum_{j,k=0}^{n-1} q_{j}(x)\, {\hat \mu}_{jk} \,q_{k}(y),
\end{eqnarray}
see the primary definition (\ref{d2m}) and (\ref{d2m+1}). In
(\ref{DnTW}), the real antisymmetric matrix ${\boldsymbol {\hat
\mu}}$ of the size $n \times n$ depends on the parity of $n$. It
equals
\begin{eqnarray}
\label{mu-even}
    {\hat \mu}^{\rm even}_{jk} =
    \left(
      \begin{array}{ccc}
        -{\boldsymbol {\hat e}}_2 h_0^{-1} & {} & {}  \\
        {}  &  \ddots & {} \\
        {} & {} & -{\boldsymbol {\hat e}}_2 h_{m-1}^{-1}  \\
        \end{array}
    \right)
\end{eqnarray}
and
\begin{eqnarray}
\label{mu-odd}
    {\hat \mu}^{\rm odd}_{jk} =
    \left(
      \begin{array}{cccc}
        -{\boldsymbol {\hat e}}_2 h_0^{-1} & {} & {} & -{\boldsymbol {\hat \varsigma}}_0^{\rm T}\\
        {}  &  \ddots & {} & \vdots \\
        {} & {} & -{\boldsymbol {\hat e}}_2 h_{m-1}^{-1} &-{\boldsymbol {\hat \varsigma}}_{m-1}^{\rm T} \\
        {\boldsymbol {\hat \varsigma}}_0 & \cdots & {\boldsymbol {\hat \varsigma}}_{m-1} &0 \\
        \end{array}
    \right)
\end{eqnarray}
for $n=2m$ and $n=2m+1$, respectively. We remind that $h_j$ is
defined by (\ref{hj}); also we have used the notation
\begin{eqnarray}
\label{cm}
    {\boldsymbol {\hat e}}_2 =
    \left(
      \begin{array}{cc}
        0 & 1\\
        -1 &  0\\
        \end{array}
    \right), \;\;\;  {\boldsymbol {\hat \varsigma}}_j = c_m\, \big(
    0,\,\frac{1}{j!}
    \big), \;\;\; c_m = \frac{m!}{h_m}.
\end{eqnarray}
Actually, the representation (\ref{DnTW}) can be put into a more
general form due to Tracy and Widom (1998) that would contain
arbitrary, not necessarily skew orthogonal, polynomials upon a
proper redefinition of the matrix ${\boldsymbol {\hat \mu}}$.

For the Dyson integration theorem to be applicable, the projection
property for the self-dual quaternion ${\hat {\cal D}}_n(z_i,z_j)$
must hold. For this to be the case, the integral identity
\begin{eqnarray} \fl
    \label{pp-dyson}
    \int d\alpha(w) \,
    \left[
        {\cal D}_n(z_1, w) \, {\cal D}_n({\bar w},z_2) -
        {\cal D}_n(z_2, w) \, {\cal D}_n({\bar w},z_1)
    \right] \stackrel{?}{=} - {\cal D}_n (z_1, z_2)
\end{eqnarray}
should be satisfied for the measure
\begin{eqnarray}
\label{alpha-measure}
    d\alpha(w) = {\rm erfc}\,\left( \frac{w-{\bar w}}{i\sqrt{2}}
     \right) \theta({\rm Im\,} w) \, d^2w.
\end{eqnarray}
Here, $\theta(\phi)$ is the Heaviside step function,
\begin{eqnarray}
    \theta(\phi) = \left\{
                      \begin{array}{cc}
                        1 & {\rm if\;} \phi > 0 \\
                        0 & {\rm if\;} \phi <0 \\
                      \end{array}
            \right.. \nonumber
\end{eqnarray}
Straightforward calculations based on (\ref{DnTW}) show that the
integral on the l.h.s. of (\ref{pp-dyson}) equals
\begin{eqnarray}
\label{DnTW-2}
    \frac{1}{2} \, e^{-(z_1^2+z_2^2)/2}
    \sum_{j,k=0}^{n-1} q_{j}(z_1)\, (\boldsymbol{ {\hat \mu}\, {\hat \chi}\,
    {\hat \mu}})_{jk}
    \,q_{k}(z_2),
\end{eqnarray}
where $n\times n$ matrix $\boldsymbol{{\hat \chi}}$ has the entries
\begin{eqnarray}
    \label{chi-matrix}
    {\hat \chi}_{jk} = \frac{1}{2} \int d\alpha(w) \, e^{-(w^2+{\bar w}^2)/2}
    \left[
        q_j(w)\, q_k({\bar w}) - q_j({\bar w})\, q_k(w)
    \right].
\end{eqnarray}
Since \footnote[2]{That $-({\boldsymbol{{\hat \mu}{\hat \chi}}})$
cannot be a unit matrix ${\boldsymbol{\hat 1}}$, follows from the
fact that ${\hat \chi}_{jk}$ is purely imaginary
[(\ref{chi-matrix})] while ${\hat \mu}_{jk}$ are real valued
[(\ref{mu-even}) and (\ref{mu-odd})]. See Appendix B for an explicit
calculation.} $\boldsymbol{({\hat \mu}\, {\hat \chi})} \neq -
{\boldsymbol {\hat 1}}_n$, the l.h.s. of (\ref{pp-dyson}) given by
(\ref{DnTW-2}) differs from the r.h.s. (\ref{pp-dyson}). As a
result, the kernel function ${\cal D}_n(z_1, z_2)$ does not satisfy
the projection property \footnote[3]{As soon as the integral on the
l.h.s. of (\ref{pp-dyson}) combines a $D$-part of the GOE $2\times
2$ matrix kernel originally introduced for the GOE's {\it real}
spectrum with the GinOE-induced measure $d\alpha(w)$ supported in
the {\it complex} half-plane ${\rm Im\,} w
> 0$, a violation of the projection property is not unexpected.}. Consequently,
the Dyson integration theorem is inapplicable for the calculation of
$p_{n,k}$ in the form of the Pfaffian integral (\ref{pnk-comp}).

\section{Probability function $p_{n,k}$: Sensing a structure through
  particular cases}
\label{Sec4} Before turning to the derivation of the general formula
for $p_{n,k}$ (see the results announced in Section \ref{SubSec12}),
it is instructive to consider a few particular cases corresponding
to low values of $\ell$, the number of pairs of complex conjugated
eigenvalues. Below, the cases of $\ell=1,\, 2$ and $3$ are treated
explicitly.

\subsection{What is the probability to find exactly one pair of
complex conjugated eigenvalues?} \label{SubSec41}

As a first nontrivial application of the Pfaffian integral
representation (\ref{pnk-comp}) for the probability function
$p_{n,k}$, let us consider the next-to-the-simplest \footnote[4]{In
the simplest case of $\ell=0$, our representation (\ref{pnk-comp})
reproduces the result (\ref{ed-res}) first derived by Edelman
(1997).} case of $\ell=1$ corresponding to the occurrence of exactly
one pair of complex conjugated eigenvalues. Since the $D$-kernel
(\ref{DnTW}) is antisymmetric under exchange of its arguments,
\begin{eqnarray}
{\cal D}_n(x,y)=-{\cal D}_n(y,x),
\end{eqnarray}
the Pfaffian in (\ref{pnk-comp}) reduces to
\begin{eqnarray}
{\rm pf\,} \left[
    \begin{array}{cc}
      0 & {\cal D}_{n}(z, {\bar z}) \\
      -{\cal D}_{n}(z,{\bar z}) & 0 \\
    \end{array}
    \right] = {\cal D}_n(z,{\bar z}) \nonumber
\end{eqnarray}
resulting in
\begin{eqnarray}
    \label{pnk-L=1}
    p_{n,n-2} =     p_{n,n} \, \frac{2}{i} \,
    \,
    \int_{{\rm Im\,}z >0} d^2z\,\, {\rm erfc}
    \left( \frac{z - {\bar z}}{i\sqrt{2} } \right)
    \, {\cal D}_n(z,{\bar z}).
\end{eqnarray}
To calculate the integral
\begin{eqnarray}
{\cal I}_1= \int_{{\rm Im\,}z >0} d^2z\,\, {\rm erfc}
    \left( \frac{z - {\bar z}}{i\sqrt{2} } \right)
    \, {\cal D}_n(z,{\bar z})=\int d\alpha(z)
    \, {\cal D}_n(z,{\bar z})    \nonumber
\end{eqnarray}
(see (\ref{alpha-measure})), we rewrite it in a more symmetric
manner
\begin{eqnarray}
    \frac{1}{2}\,
    \int d\alpha(z)\,\Big[ {\cal D}_n(z,{\bar z}) - {\cal D}_n({\bar
    z},z)\Big],
    \nonumber
\end{eqnarray}
and make use of (\ref{chi-matrix}), (\ref{alpha-measure}) and
(\ref{DnTW}) to deduce that it equals
\begin{eqnarray}
    - \frac{1}{2}\,\sum_{j,k=0}^{n-1} \mu_{jk}\chi_{kj}  = -\frac{1}{2} {\rm
    tr}_{(0,
    n-1)}({\boldsymbol{{\hat \mu}{\hat \chi}}}),
    \nonumber
\end{eqnarray}
or, equivalently,
\begin{eqnarray}
\label{int-1} {\cal I}_1 =
 \int d\alpha(z)\, {\cal D}_n(z,{\bar z}) =
\frac{i}{4}\, {\rm tr}_{(0,n-1)} {\boldsymbol {\hat \sigma}}.
\end{eqnarray}
Here, ${\boldsymbol {\hat \sigma}}$ is given by ${\boldsymbol {\hat
\sigma}}=2i{\boldsymbol{{\hat \mu}{\hat \chi}}}$ (see also Appendix
B). We therefore conclude that the probability sought equals
\begin{eqnarray}
    \label{pnn-2tr}
    p_{n,n-2} = \frac{1}{2}\, p_{n,n}
    \,{\rm tr}_{(0,n-1)} {\boldsymbol {\hat \sigma}}.
\end{eqnarray}
Due to the trace identity (\ref{trid}) proven in Appendix C, we
eventually derive:
\begin{eqnarray}
    \label{pnn-2fin}
    p_{n,n-2} =  p_{n,n}
    \,{\rm tr}_{(0,\lfloor n/2 \rfloor -1)} {\boldsymbol {\hat
    \varrho}},
\end{eqnarray}
reducing the size of the matrix by two. Here, the smaller matrix
${\boldsymbol {\hat \varrho}}$ depends on the parity of $n$, as
defined in the Section \ref{SubSec12} (see also (\ref{rho-even-app})
and (\ref{rho-odd-app})).

Remarkably, the trace in (\ref{pnn-2fin}) can explicitly be
calculated (see Appendix D) to yield a closed expression for the
probability to find exactly one pair of complex conjugated
eigenvalues:
\begin{eqnarray}
    \label{pnn-2fin-2}
    p_{n,n-2} = 2\, p_{n,n}\,
    \int_0^\infty dy\, y \, e^{y^2}\, {\rm
    erfc}(y \sqrt{2}) \,
     L_{n-2}^2(-2y^2).
\end{eqnarray}
Yet another, though equivalent, representation for the probability
$p_{n,n-2}$ is given in Section \ref{Sec7} that addresses the
large-$n$ behaviour of $p_{n,n-2}$.

\subsection{Two pairs of complex conjugated eigenvalues ($\ell=2$)}
\label{SubSec42}
For $\ell=2$, the Pfaffian in (\ref{pnk-comp})
\begin{eqnarray}
\label{L2PFAFF}
    {\rm pf\,} \left[
    \begin{array}{cccc}
      0 & {\cal D}_{n}(z_1, {\bar z}_1) & {\cal D}_{n}(z_1, z_2) & {\cal D}_{n}(z_1, {\bar z}_2)\\
      -{\cal D}_{n}(z_1,{\bar z}_1) & 0 & {\cal D}_{n}({\bar z}_1, z_2) & {\cal D}_{n}({\bar z}_1, {\bar z}_2)\\
      -{\cal D}_{n}(z_1, z_2) & -{\cal D}_{n}({\bar z}_1,z_2) & 0 & {\cal D}_{n}(z_2, {\bar z}_2)\\
      -{\cal D}_{n}(z_1,{\bar z}_2) & -{\cal D}_{n}({\bar z}_1, {\bar z}_2) & -{\cal D}_{n}(z_2,{\bar z}_2) & 0\\
    \end{array}
    \right]
\end{eqnarray}
reduces to
\begin{eqnarray} \label{L=2}\fl
    {\cal D}_n(z_1,{\bar z}_1)\, {\cal D}_n(z_2,{\bar
              z}_2)
    +  {\cal D}_n(z_1,{\bar z}_2)\, {\cal D}_n({\bar z}_1,z_2)
              - {\cal D}_n(z_1,z_2)\, {\cal D}_n({\bar z}_1, {\bar
              z}_2),
\end{eqnarray}
so that
\begin{eqnarray}
    \label{pnk-L=2} \fl
    p_{n,n-4} &=    - 2 \,p_{n,n}
    \,
    \int d\alpha(z_1)
    \,
    \int d\alpha(z_2)\, \nonumber \\
    \fl &\times
    \Big[
        {\cal D}_n(z_1,{\bar z}_1)\, {\cal D}_n(z_2,{\bar
              z}_2)
    +  {\cal D}_n(z_1,{\bar z}_2)\, {\cal D}_n({\bar z}_1,z_2)
              - {\cal D}_n(z_1,z_2)\, {\cal D}_n({\bar z}_1, {\bar
              z}_2)
    \Big].
\end{eqnarray}
Apart from a known integral taking the form of (\ref{int-1}), a new
integral
\begin{eqnarray} \label{int-22} \fl
{\cal I}_2 =
    \int d\alpha(z_1)
    \,
    \int d\alpha(z_2)\,
    \Big[ {\cal D}_n(z_1,{\bar z}_2)\,
    {\cal D}_n({\bar z}_1,z_2)
              - {\cal D}_n(z_1,z_2)\, {\cal D}_n({\bar z}_1, {\bar
              z}_2)
    \Big]
\end{eqnarray}
appears in (\ref{pnk-L=2}). Somewhat lengthy but straightforward
calculations based on (\ref{chi-matrix}), (\ref{alpha-measure}) and
(\ref{DnTW}) result in
\begin{eqnarray} \label{int-2}
 {\cal I}_2 = \frac{1}{8} \, {\rm tr}_{(0, n-1)} ({\boldsymbol {\hat
\sigma}}^2).
\end{eqnarray}
Combining (\ref{pnk-L=2}), (\ref{int-1}), (\ref{int-22}) and
(\ref{int-2}), we obtain:
\begin{eqnarray}
    p_{n,n-4} = p_{n,n}\, \left[
        \frac{1}{8}\, \left( {\rm tr}_{(0,n-1)} {\boldsymbol {\hat
        \sigma}}\right)^2 - \frac{1}{4}\, {\rm tr}_{(0,n-1)} ({\boldsymbol
    {\hat \sigma}}^2)
    \right].
\end{eqnarray}
Due to the trace identity (\ref{trid}) proven in Appendix C, the
latter reduces to
\begin{eqnarray}
\label{ppnn4}
    p_{n,n-4} = \frac{1}{2}\, p_{n,n}\, \left[
         \left( {\rm tr}_{(0,\lfloor n/2 \rfloor -1)} {\boldsymbol {\hat
        \varrho}}\right)^2 - {\rm tr}_{(0,\lfloor n/2 \rfloor-1)} ({\boldsymbol
    {\hat \varrho}}^2)
    \right].
\end{eqnarray}
Interestingly, the expression in the parenthesis of (\ref{ppnn4})
coincides with $Z_{(1^2)}(p_1, p_2)$ after the substitution ${\rm
tr\,}{\boldsymbol {\hat \varrho}}^j=p_j$ (see Table
\ref{zonal-table}).

\subsection{Three pairs of complex conjugated eigenvalues ($\ell=3$)}
\label{SubSec43} The complexity of the integrand in (\ref{pnk-comp})
grows rapidly with increasing $\ell$. For $\ell=3$, that is three
pairs of complex conjugated eigenvalues in the matrix spectrum, the
Pfaffian of the $6\times 6$ antisymmetric matrix
\begin{eqnarray} \fl
    \left(
      \begin{array}{cccccc}
      0 & {\cal D}_{n}(z_1, {\bar z}_1) & {\cal D}_{n}(z_1, z_2) & {\cal
        D}_{n}(z_1, {\bar z}_2)
        & {\cal D}_{n}(z_1, z_3) & {\cal D}_{n}(z_1, {\bar z}_3) \\
      -{\cal D}_{n}(z_1,{\bar z}_1) & 0 & {\cal D}_{n}({\bar z}_1, z_2) &
        {\cal D}_{n}({\bar z}_1, {\bar z}_2)
        & {\cal D}_{n}({\bar z}_1, z_3) & {\cal D}_{n}({\bar z}_1, {\bar
        z}_3)\\
      -{\cal D}_{n}(z_1, z_2) & -{\cal D}_{n}({\bar z}_1,z_2) & 0 & {\cal
        D}_{n}(z_2, {\bar z}_2)
        & {\cal D}_{n}(z_2,z_3) & {\cal D}_{n}(z_2,{\bar z}_3)\\
      -{\cal D}_{n}(z_1,{\bar z}_2) & -{\cal D}_{n}({\bar z}_1, {\bar z}_2) &
        -{\cal D}_{n}(z_2,{\bar z}_2) & 0
        & {\cal D}_{n}({\bar z}_2,z_3) & {\cal D}_{n}({\bar z}_2,{\bar z}_3)\\
        -{\cal D}_{n}(z_1, z_3) & -{\cal D}_{n}({\bar z}_1, z_3) & -{\cal
        D}_{n}(z_2,z_3) & -{\cal D}_{n}({\bar z}_2,z_3) & 0 & {\cal
        D}_{n}(z_3,{\bar z}_3) \\
        -{\cal D}_{n}(z_1, {\bar z}_3) & -{\cal D}_{n}({\bar z}_1, {\bar z}_3)
        & -{\cal D}_{n}(z_2,{\bar z}_3) & -{\cal D}_{n}({\bar z}_2,{\bar z}_3)
        & -{\cal D}_{n}(z_3, {\bar z}_3) & 0 \\
    \end{array}
    \right) \nonumber
\end{eqnarray}
is getting involved. It can be calculated with some effort to give
$15$ terms which can be attributed to three different groups. The
first group consists of the single term
\begin{eqnarray}
    \label{gr-1}
    G_1 = {\cal D}_{n}(z_1,{\bar z}_1)\, {\cal D}_{n}(z_2,{\bar z}_2)\,{\cal
                       D}_{n}(z_3,{\bar
                       z}_3).
\end{eqnarray}
The second group contains $6$ terms,
\begin{eqnarray}
    \label{gr-2} \fl
    G_2 &= {\cal D}_n(z_1,{\bar z}_1)\,\Big[ {\cal D}_n(z_2,{\bar z}_3)\,{\cal D}_n({\bar
                           z}_2,z_3)\,
                           -  {\cal D}_n(z_2,z_3)\,{\cal D}_n({\bar
                           z}_2,{\bar z}_3)\Big]\, \nonumber \\
                         \fl &+ {\cal D}_n(z_2,{\bar z}_2)\, \Big[ {\cal D}_n(z_1,{\bar
                           z}_3)\,{\cal D}_n({\bar
                           z}_1,z_3)\,
                           -  {\cal D}_n(z_1,z_3)\,{\cal D}_n({\bar
                           z}_1,{\bar z}_3)\Big]\, \nonumber \\
                           \fl  &+ {\cal D}_n(z_3,{\bar z}_3)\,\Big[ {\cal D}_n(z_1,{\bar
                           z}_2)\,{\cal D}_n({\bar
                           z}_1,z_2)\,
                           -  {\cal D}_n(z_1,z_2)\,{\cal D}_n({\bar
                           z}_1,{\bar z}_2)\Big],
\end{eqnarray}
while the third group counts $8$ terms:
\begin{eqnarray}
    \label{gr-3} \fl
    G_3 &= {\cal D}_{n}(z_1,{\bar z}_3)\, {\cal D}_n({\bar
                               z}_2, z_3)\,{\cal D}_n({\bar z}_1, z_2)\,
                               -
                               {\cal D}_n({\bar z}_1,z_3)\, {\cal D}_n(z_2,{\bar z}_3)\,{\cal D}_n(z_1,{\bar z}_2)\, \nonumber
                                \\
                              \fl
            &+ {\cal D}_n(z_1,{\bar z}_2)\, {\cal D}_n(z_2,z_3)\,{\cal D}_n({\bar z}_1,{\bar
                           z}_3)
            - {\cal D}_n({\bar z}_1,z_2)\, {\cal D}_n({\bar z}_2,{\bar z}_3)\,{\cal D}_n(z_1,z_3)\,
                             \nonumber \\
                              \fl &+
                     {\cal D}_n({\bar z}_1,{\bar z}_2)\, {\cal D}_n(z_2,{\bar
                               z}_3)\,{\cal D}_n(z_1,z_3)\,
                               -{\cal D}_n(z_1,z_2)\, {\cal D}_n({\bar z}_2,z_3)\,{\cal D}_n({\bar z}_1,{\bar z}_3)\,
                                \nonumber \\
                              \fl &+
                {\cal D}_n(z_1,z_2)\, {\cal D}_n({\bar z}_2,{\bar z}_3)\,{\cal D}_n({\bar
                           z}_1, z_3) -
                           {\cal D}_n({\bar z}_1,{\bar z}_2)\, {\cal D}_n(z_2,z_3)\,{\cal D}_n(z_1, {\bar z}_3).
\end{eqnarray}
In the above notation, the probability function $p_{n,n-6}$ takes
the form
\begin{eqnarray} {\label{pnn-6}} \fl
    p_{n, n-6} = \frac{4i}{3}\,p_{n,n}\, \int d\alpha(z_1) \, \int
                d\alpha(z_2)\, \int d\alpha(z_3)\,
                \Big[ G_1 + G_2 + G_3 \Big].
\end{eqnarray}
The integrals containing $G_1$ and $G_2$ can easily be performed
with the help of (\ref{int-1}), (\ref{int-22}) and (\ref{int-2}) to
bring
\begin{eqnarray}
\label{6f1}
        \int d\alpha(z_1) \, \int
                d\alpha(z_2)\, \int d\alpha(z_3)\, G_1 = {\cal
I}_1^3 = \left(
    \frac{i}{4}\, {\rm tr}_{(0,n-1)} {\boldsymbol {\hat \sigma}}
\right)^3
\end{eqnarray}
and
\begin{eqnarray}\label{6f2} \fl
        \int d\alpha(z_1) \, \int
                d\alpha(z_2)\, \int d\alpha(z_3)\, G_2 &= 3\, {\cal
                I}_1 \, {\cal I}_2 \nonumber \\ \fl
     &= 3\, \left(
    \frac{i}{4}\, {\rm tr}_{(0,n-1)} {\boldsymbol {\hat \sigma}}
\right) \, \left( \frac{1}{8}\, {\rm tr}_{(0,n-1)} ({\boldsymbol
{\hat \sigma}}^2) \right).
\end{eqnarray}
The remaining integral involving $G_3$ can be evaluated similarly to
${\cal I}_1$ and ${\cal I}_2$, the result being
\begin{eqnarray}
\label{6f3}
    {\cal I}_3 = \int d\alpha(z_1) \, \int
                d\alpha(z_2)\, \int d\alpha(z_3)\, G_3 =
        \frac{i^3}{8} \, {\rm tr}_{(0,n-1)} ({\boldsymbol {\hat
\sigma}}^3).
\end{eqnarray}
Combining (\ref{pnn-6}), (\ref{6f1}), (\ref{6f2}) and (\ref{6f3}),
we derive:
\begin{eqnarray}
    \label{pnn6-1} \fl
    p_{n,n-6} = p_{n,n}\, \left[
        \frac{1}{48}\, \left( {\rm tr}_{(0,n-1)} {\boldsymbol {\hat \sigma}}
        \right)^3
        -
        \frac{1}{8} \, {\rm tr}_{(0,n-1)} {\boldsymbol {\hat
        \sigma}}\, {\rm tr}_{(0,n-1)} ({\boldsymbol {\hat \sigma}}^2)
        + \frac{1}{6}\, {\rm tr}_{(0,n-1)} ({\boldsymbol {\hat
        \sigma}}^3)
    \right]. \nonumber \\ {}
\end{eqnarray}
Finally, we apply the trace identity (\ref{trid}) to end up with the
formula
\begin{eqnarray} \fl
    \label{pnn6-2}
    p_{n,n-6} &= \frac{1}{6}\, p_{n,n}\,
    \Big[
         \left( {\rm tr}_{(0,\lfloor n/2 \rfloor -1)} {\boldsymbol {\hat
        \varrho}}\right)^3 \nonumber \\
    \fl &- 3\,
    {\rm tr}_{(0,\lfloor n/2 \rfloor-1)} {\boldsymbol
    {\hat \varrho}}\,
    {\rm tr}_{(0,\lfloor n/2 \rfloor-1)} ({\boldsymbol
    {\hat \varrho}}^2) + 2\,
    {\rm tr}_{(0,\lfloor n/2 \rfloor-1)} ({\boldsymbol
    {\hat \varrho}}^3)
    \Big].
\end{eqnarray}
The expression in the parenthesis of (\ref{pnn6-2}) is seen to
coincide with $Z_{(1^3)}(p_1, p_2, p_3)$ after the substitution
${\rm tr\,}{\boldsymbol {\hat \varrho}}^j=p_j$ (see Table
\ref{zonal-table}).

\subsection{Higher $\ell$}
\label{SubSec44} The three examples considered clearly demonstrate
that the calculational complexity grows enormously with increasing
$\ell$, the number of complex conjugated eigenvalues in the random
matrix spectrum. Indeed, the number of terms ${\cal N}_{\ell}$ in
the expansion of the Pfaffian in (\ref{pnk-comp}) equals ${\cal
N}_{\ell}=(2\ell-1)!!$ and exhibits a faster-than-exponential
growth, ${\cal N}_{\ell} \approx 2^{\ell+1/2}\, e^{\ell\, (\ln \ell
-1)}$, for $\ell \gg 1$. For this reason, one has to invent a
classification of the terms arising in the Pfaffian expansion that
facilitates their effective computation. This will be done in
Section \ref{Sec5}, where we introduce a topological interpretation
of the Pfaffian expansion, and prove the Pfaffian integration
theorem which can be viewed as a generalisation of the Dyson
integration theorem (see Section \ref{SubSec23}).

\section[Pfaffian integration theorem]{Topological interpretation of the
  Pfaffian expansion, and the Pfaffian integration theorem}
\label{Sec5}

\subsection{Statement of the main result and its discussion}
\label{SubSec51} Before stating the main result of this section, the
Pfaffian integration theorem, we wish to start with presenting a
simple Corollary to the Dyson integration theorem.
\vspace{0.3cm}\\{\bf Corollary 5.1.} {\it Let $f(x,y)$ be a function
with real, complex or quaternion values satisfying the conditions
(\ref{dt-1}) and (\ref{dt-2}) of the Theorem 2.1, and $d\pi$ be a
suitable measure. Then \numparts
\begin{eqnarray}
\label{col-1}
    \int \prod_{j=1}^\ell d\pi(x_j)\, {\rm det\,} [f(x_i,x_j)]_{\ell\times
    \ell}
    = \frac{\Gamma(c+1)}{\Gamma(c+1-\ell)},
\end{eqnarray}
where
\begin{eqnarray}
\label{col-2}
    c = \int d\pi(x)\, f(x,x).
\end{eqnarray}
\endnumparts
For $f$ taking quaternion values, the ${\rm det}$ should be
interpreted as ${\rm qdet}$, the quaternion determinant (Dyson
1972).}
\newline\newline
{\bf Proof.} Repeatedly apply the Dyson integration theorem to the
l.h.s. of (\ref{col-1}) to arrive at its r.h.s. $\;\;\;\blacksquare$
\newline\newline
Importantly, the above Corollary exclusively applies to functions
$f(x,y)$ satisfying the projection property as defined in Section
\ref{SubSec23} (see Definition 2.1 therein). However, guided by our
study of the integrable structure of GinOE, we are going to ask if
the integrals of the kind (\ref{col-1}) can explicitly be calculated
if the projection property is relaxed. In general, the answer is
positive. In particular, for $f(x,y)$ being a self-dual quaternion,
the following integration theorem will be proven.
\vspace{0.3cm}\\{\bf Theorem 5.1 (Pfaffian integration theorem).}
{\it Let $d\pi(z)$ be any benign measure on $z \in {\mathbb C}$, and
the function $Q_n(x,y)$ be an antisymmetric function of the form
\numparts
\begin{eqnarray} \label{th-1}
    Q_n(x,y) = \frac{1}{2} \sum_{j,k=0}^{n-1} q_j(x) \, {\hat \mu}_{jk} \,
    q_k(y)
\end{eqnarray}
where the $q_j(x)$ are arbitrary polynomials of $j$-th order, and
${\boldsymbol {\hat \mu}}$ is an antisymmetric matrix. Then the
integration formula
\begin{eqnarray} \fl
    \label{th-2}
    \int_{{\mathbb C}}\,  \prod_{j=1}^\ell \,  d\pi(z_j)\,\,
    {\rm pf\,} \left[
    \begin{array}{cc}
      Q_{n}(z_i, z_j) & Q_{n}(z_i, {\bar z}_j) \\
      Q_{n}({\bar z}_i, z_j) & Q_{n}({\bar z}_i, {\bar z}_j) \\
    \end{array}
    \right]_{2\ell \times 2\ell} \nonumber \\
    =
    \left(\frac{i}{2}\right)^\ell
    Z_{(1^\ell)}\left(
    \frac{1}{2} {\rm tr}_{(0,n-1)}
    {\boldsymbol {\hat \upsilon}}^1,\cdots,\frac{1}{2} {\rm tr}_{(0,n-1)}
    {\boldsymbol {\hat \upsilon}}^\ell
    \right)
\end{eqnarray}
holds, provided the integrals in its~l.h.s.~exist. Here,
$Z_{(1^\ell)}$ are zonal polynomials whose $\ell$ arguments are
determined by a matrix ${\boldsymbol {\hat \upsilon}}$ with the
entries
\begin{eqnarray}
    \label{th-3}
    {\hat \upsilon}_{\alpha,\beta} = i \sum_{k=0}^{n-1} {\hat \mu}_{\alpha,k}
    \int_{z\in {\mathbb C}} d\pi(z) \left[
        q_k(z)\, q_\beta({\bar z}) - q_\beta(z) \, q_k({\bar z})
    \right].
\end{eqnarray}
}\endnumparts
\newline\newline
As we integrate over all variables, the Pfaffian integration theorem
can be viewed as a generalisation of the Corollary 5.1 proven a few
lines above, for the case of a kernel not satisfying the projection
property. This follows from the identity
\begin{eqnarray}
    {\rm pf\,} \left[
    \begin{array}{cc}
      Q_{n}(z_i, z_j) & Q_{n}(z_i, {\bar z}_j) \\
      Q_{n}({\bar z}_i, z_j) & Q_{n}({\bar z}_i, {\bar z}_j) \\
    \end{array}
    \right]_{2\ell \times 2\ell} =
    {\rm qdet\,} \left[
    {\hat Q}_n(z_i,z_j)
    \right]_{\ell \times \ell}, \nonumber
\end{eqnarray}
where the quaternion ${\hat Q}_n(z_i,z_j)$ has the $2\times 2$
matrix representation:
\begin{eqnarray}\label{th-4}
    f(z_i, z_j) =
    \Theta [{\hat Q}_n(z_i, z_j)] = \left(
    \begin{array}{rr}
      -Q_{n}(\bar z_i, z_j) & -Q_{n}({\bar z}_i, {\bar z}_j) \\
      Q_{n}(z_i, z_j) & Q_{n}(z_i, {\bar z}_j) \\
    \end{array}
    \right).
\end{eqnarray}
To see that the Pfaffian integration theorem reduces to the
Corollary for the particular case of a kernel with restored
projection property, we spot that the latter is equivalent to the
statement
\begin{eqnarray}
    \label{pp-v}
    {\boldsymbol {\hat \upsilon}} = - 2 i\, {\boldsymbol {\hat 1}}_n
\end{eqnarray}
as can be deduced from the discussion below (\ref{pp-dyson}),
Section \ref{SubSec33}. As a result,
\begin{eqnarray}
    {\rm tr\,}_{(0,n-1)} \, {\boldsymbol {\hat \upsilon}\,}^j =
    n\, \left(\frac{2}{i}\right)^{\,j},
\end{eqnarray}
and the r.h.s. of (\ref{th-2}) reduces to (Macdonald 1998)
\begin{eqnarray} \label{pfth-comp}\fl
    \left(\frac{i}{2}\right)^{\ell}
    Z_{(1^\ell)}\left(
    \frac{n}{2}\, \left(\frac{2}{i}\right)^1,\cdots,\frac{n}{2} \,
\left(\frac{2}{i}\right)^\ell
    \right) = \left( \frac{i}{2}\frac{\partial}{\partial z}\right)^\ell \,
    \exp \Bigg(
        \frac{n}{2}\sum_{r\ge 1} (-1)^{r-1} \frac{(-2i \, z)^r}{r}
    \Bigg)\Bigg|_{z=0}\nonumber \\
    =\left(\frac{\partial}{\partial z}\right)^\ell (1+z)^{n/2}\Bigg|_{z=0}=
    \frac{\Gamma(n/2+1)}{\Gamma(n/2+1-\ell)}.
\end{eqnarray}
Finally, noticing from (\ref{th-4}) that
\begin{eqnarray} \fl
    \int\, d\pi(z) f(z,z) = \int\, d\pi(z) \, \left(
    \begin{array}{rr}
      Q_{n}(z,\bar z) & 0 \\
      0 & Q_{n}(z, {\bar z}) \\
    \end{array}
    \right) \nonumber \\= \frac{i}{4}\,  {\boldsymbol {\hat e}_0}\,{\rm
    tr\,}_{(0,n-1)}{\boldsymbol {\hat
    \upsilon}}
    = (n/2)\,  {\boldsymbol {\hat e}_0}
\end{eqnarray}
with $ {\boldsymbol {\hat e}_0} = {\rm diag}(1,1)$, we conclude that
the constant $c$ in the Corollary [(\ref{col-2})] equals $c=n/2$ so
that the result (\ref{pfth-comp}) brought by the Pfaffian
integration theorem is identically equivalent to the one
[(\ref{col-1})] following from the Dyson integration theorem. We
stress that this is only true for the kernel $Q_n(z_i,z_j)$
satisfying the projection property.

To prove the Theorem 5.1, we will invent a formalism based on a
topological interpretation of the ordered Pfaffian expansion. For
the readers' benefit, a vocabulary of the topological terms to be
defined and used in the following sections is summarised in Table
\ref{strings-table}.

\subsection{Topological interpretation of the ordered Pfaffian expansion}
\label{SubSec52}

\begin{table}
\caption{\label{strings-table} The vocabulary of topological terms
defined to interpret the ordered Pfaffian expansion. The notation
used is: D -- Definition, E -- Example, L -- Lemma, T -- Theorem, C
-- Corollary, F -- Figure.}

\begin{indented}
\lineup
\item[]
\begin{tabular}{@{}*{3}{l}} \br
Term & \centre{1}{Notation} & Appearance \cr \mr

String & \centre{1}{${\cal S}_i$} & D5.1, E5.1, F3, F4\vspace{0.1cm} \cr

Length of a string & \centre{1}{$\|{\cal S}_i\|$} & D5.2, E5.2
\vspace{0.1cm}\cr

Equivalent strings & \centre{1}{${\cal S}_i \sim {\cal S}_j$} &
D5.3, E5.3, L5.1, F3 \vspace{0.1cm} \cr

Equivalence class of strings & \centre{1}{${\cal C}_j$} & D5.4,
E5.4, L5.1, F3, F5 \vspace{0.1cm}\cr

Size of equivalence class & \centre{1}{$\|{\cal C}_j\|$} & D5.4,
L5.1, F3 \vspace{0.1cm}\cr

Substring & \centre{1}{${\cal S}_i^{(p)}$} & D5.5, E5.5 \vspace{0.1cm}\cr

Length of a substring & \centre{1}{$p=\|{\cal S}_i^{(p)}\|$} & D5.5,
E5.5 \vspace{0.1cm}\cr

Loop-like substring& \centre{1}{${\cal S}_i^{(p)}$} & D5.6, E5.6,
L5.2, L5.5,\vspace{0.0cm}\cr {} & {} & F3, F4 \vspace{0.1cm}\cr

Longest loop-like string& \centre{1}{${\cal S}_i^{(\ell)}$} & F3,
E5.7, L5.4\vspace{0.1cm}\cr

Adjacent loop-like string& \centre{1}{${\cal S}_i^{(p)}$} & D5.7, E5.7, E5.9,
E5.11, \vspace{0.0cm}\cr
{}& {} & L5.4--L5.6, C5.2, F4, F5 \vspace{0.1cm}\cr

Handedness of adjacent (sub)string& \centre{1}{${\mathbb H}(\alpha_L,
  \alpha_R)$} & D5.8, E5.8, E5.9, E5.11, \vspace{0.0cm}\cr
{}& {} & L5.6, C5.2, F5 \vspace{0.1cm}\cr

Equivalent adjacent (sub)strings& \centre{1}{${\cal S}_i \sim {\cal S}_j$} &
D5.9, E5.10, E5.11, \vspace{0.0cm}\cr
{}& {} & L5.7, F5 \vspace{0.1cm}\cr

Equivalence class of adjacent (sub)strings& \centre{1}{${\cal
AC}_j$} & D5.10 \vspace{0.1cm}\cr

Compound string& \centre{1}{${\cal S}_i$} & D5.11, T5.3, F3, F6
\vspace{0.1cm}\cr

Topology class& \centre{1}{${\boldsymbol \lambda}$} & F6
\vspace{0.1cm}\cr

\br
\end{tabular}
\end{indented}
\end{table}

To integrate the Pfaffian in (\ref{th-2}), we start with its {\it
ordered} expansion
\begin{eqnarray}
\label{pf-ord} \fl
    {\rm pf\,} \left[
    \begin{array}{cc}
      Q_{n}(z_i, z_j) & Q_{n}(z_i, {\bar z}_j) \\
      Q_{n}({\bar z}_i, z_j) & Q_{n}({\bar z}_i, {\bar z}_j) \\
    \end{array}
    \right]_{2\ell \times 2\ell} =
    \frac{1}{2^\ell \ell!} \sum_{\sigma\in S_{2\ell}}
    {\rm sgn} (\sigma) \, \prod_{j=0}^{\ell-1}
    Q_n(w_{\sigma(2j+1)},w_{\sigma(2j+2)}).\nonumber \\
    {}
\end{eqnarray}
Here, the summation extends over all permutations $\sigma\in
S_{2\ell}$ of $2\ell$ objects
\begin{eqnarray}
\label{perm}
    \{ w_1=z_1,w_2={\bar z}_1,\cdots, w_{2\ell-1} = z_\ell, w_{2\ell}= {\bar
    z}_{\ell} \}
\end{eqnarray}
so that the total number of terms in (\ref{pf-ord}) is $(2\ell)!$.
\pagebreak

\subsubsection{Strings and their equivalence
classes\newline\newline}\label{SubSec521}
\begin{figure}[t]
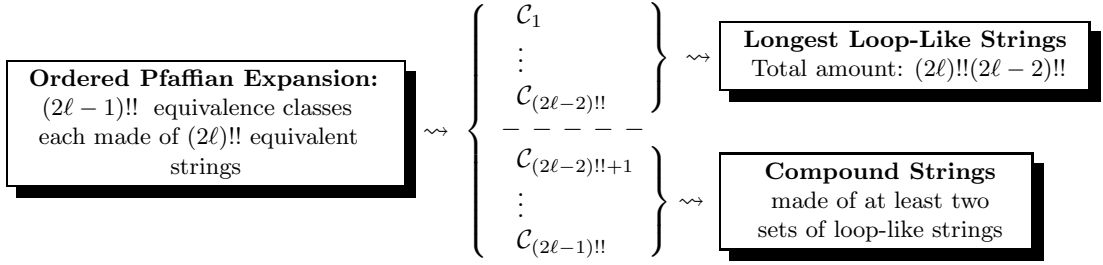

\vspace{0.5cm}
\begin{minipage}{4.4in}
$\raisebox{-0.8cm}{\shadowbox{
\begin{minipage}{1.9in}
\begin{center}
\small{ {\bf Ordered Pfaffian Expansion:}\newline $(2\ell-1)!!\;$
{\rm equivalence classes}
\newline{\rm each made of $(2\ell)!!$ equivalent}
\newline{\rm strings}}
\end{center}
\end{minipage}
}} \rightsquigarrow \; \left\{
\begin{array}{ll}
\left.
   \begin{array}{ll}
  {\cal C}_1 & {} \\
  \vdots & {} \\
  {\cal C}_{(2\ell-2)!!} & {}
  \end{array}  \right\}\;\rightsquigarrow
  \raisebox{-0.5cm}
  {\shadowbox{
    \begin{minipage}{1.8in}
    \begin{center}
    \small{
   {\bf Longest Loop-Like Strings}
   \newline{Total amount: $(2\ell)!!(2\ell-2)!!$}}
   \end{center}
   \end{minipage}}}
  & {} \\
  -\,-\,-\,-\,- & {} \\
 \left. \begin{array}{l}
  {\cal C}_{(2\ell-2)!!+1}  \\
  \vdots  \\
  {\cal C}_{(2\ell-1)!!}
  \end{array}\right\} \rightsquigarrow \; \raisebox{-0.7cm}{\shadowbox{
    \begin{minipage}{1.5in}
    \begin{center}
    \small{
   {\bf{Compound Strings}} \\ {\rm made of at least two \\ sets of loop-like
   strings}}
   \end{center}
   \end{minipage}}}
\end{array}
\right. $
\newline\newline
\end{minipage}
 \caption{The $(2\ell)!$ terms, or strings, of an ordered Pfaffian expansion can be assigned to $(2\ell-1)!!$ equivalence classes
 $\{{\cal C}_1,\cdots,{\cal C}_{(2\ell-1)!!}\}$, each one containing $(2\ell)!!$ equivalent strings (Lemma 5.1). Any given string
 of the length $\ell$ can be decomposed into a set of loop-like substrings of respective lengths $\{\ell_j\}$ such that $\sum_{j}\ell_j
 =\ell$ (Lemma 5.2). The sets consisting of only one loop-like string are called {\it longest loop-like strings}; the sets made of more than one
 loop-like substring are called {\it compound strings}. An amount of longest loop-like strings is counted in Lemma 5.3.}
 \label{Topology-1}
\end{figure}

\hspace{-0.3cm}{\bf
Definition 5.1.} {\it Each term of the ordered Pfaffian expansion is
called a {\bf string}. The $i$-th string ${\cal S}_i$ equals
\begin{eqnarray}
\label{str-i}
 {\cal S}_i =
 {\rm sgn} (\sigma_i) \, \prod_{j=0}^{\ell-1}
    Q_n(w_{\sigma_i (2j+1)},w_{\sigma_i(2j+2)}).
\end{eqnarray}
where $\sigma_i$ is the $i$-th permutation out of $(2\ell)!$
possible permutations $\sigma\in S_{2\ell}$. Notice that a sign is
attached to each string.}\newline\newline {\bf Example 5.1.} The
ordered expansion of the Pfaffian for $\ell=2$ [see (\ref{L2PFAFF})]
contains $4!=24$ strings that we assign to three different groups
(their meaning will become clear below):
\begin{eqnarray} \label{def-1-ex}
            \begin{array}{ccc}
            {} & {} & {} \\
            \shadowbox{Group 1} & \shadowbox{Group 2} & \shadowbox{Group 3}\\
            {} & {} & {} \\
            ~~{\bf +(1{\bar 1})(2{\bar 2})}~~ & ~~{\bf +(1{\bar 2})({\bar 1}2)}~~ & ~~{\bf -(12)({\bar 1}{\bar 2})}~~ \\
            -(1{\bar 1})({\bar 2}2) & -(1{\bar 2})(2{\bar 1}) & +(12)({\bar 2}{\bar 1}) \\
            +({\bar 1}1)({\bar 2}2) & +({\bar 2}1)(2{\bar 1}) & -(21)({\bar 2}{\bar 1}) \\
            -({\bar 1}1)(2{\bar 2}) & -({\bar 2}1)({\bar 1}2) & +(21)({\bar 1}{\bar 2}) \\
            +(2{\bar 2})(1{\bar 1}) & +({\bar 1}2)(1{\bar 2}) & -({\bar 1}{\bar 2})(12) \\
            -(2{\bar 2})({\bar 1}1) & -({\bar 1}2)({\bar 2}1) & +({\bar 1}{\bar 2})(21) \\
            +({\bar 2}2)({\bar 1}1) & +(2{\bar 1})({\bar 2}1) & -({\bar 2}{\bar 1})(21) \\
            -({\bar 2}2)(1{\bar 1}) & -(2{\bar 1})(1{\bar 2}) & +({\bar 2}{\bar 1})(12) \\
            {} & {} & {} \\
          \end{array}
\end{eqnarray}
For brevity, the obvious notation $\pm (pq)({\bar p}{\bar q})$ was
used to denote the string $\pm Q_n(z_p,z_q)Q_n({\bar z}_p, {\bar
z}_q)$. The three strings shown in bold are those that previously
appeared in (\ref{L=2}) when treating the probability $p_{n,n-4}$.
\newline\newline
{\bf Definition 5.2.} {\it The {\bf length} $\|{\cal S}_i\|$ of a
string ${\cal S}_i$ equals the number of kernels it is composed
of.}\newline\newline {\bf Example 5.2.} The string ${\cal S}_i$ in
(\ref{str-i}) is of the length $\ell$: $\|{\cal S}_i\|=\ell$. All
strings in (\ref{def-1-ex}) are of the length $2$.
\newline\newline
{\bf Definition 5.3.} {\it Two strings ${\cal S}_i$ and ${\cal S}_j$
of the ordered Pfaffian expansion are said to be {\bf equivalent
strings}, ${\cal S}_i \sim {\cal S}_j$, if they can be obtained from
each other by (i) permutation of kernels and/or (ii) permutation of
arguments inside kernels (these will also be called intra-kernel
permutations). }
\newline\newline
{\bf Example 5.3.} For $\ell=2$, three different groups of
equivalent strings can be identified as suggested by
(\ref{def-1-ex}). The first group, exemplified by the string $ {\bf
+(1{\bar 1})(2{\bar 2})} $ involves $8$ equivalent strings; two
other groups, each consisting of $8$ strings as well, are
represented by the strings ${\bf +(1{\bar 2})({\bar 1}2)}$ and ${\bf
-(12)({\bar 1}{\bar 2})}$, respectively.
\newline\newline
{\bf Definition 5.4.} {\it A group of equivalent strings is called
the {\bf equivalence class of strings}. The $j$-th equivalence class
to be denoted as ${\cal C}_j$ consists of $\|{\cal C}_j \|$
equivalent strings.} The number $\|{\cal C}_j\|$ is called the {\bf
size of equivalence class}.
\newline\newline
{\bf Example 5.4.} For $\ell=2$, exactly three different equivalence
classes of strings (\ref{def-1-ex}) can be identified in the ordered
Pfaffian expansion (\ref{pf-ord}).
\newline\newline
In the context of the initiated classification of strings arising in
the ordered Pfaffian expansion (\ref{pf-ord}), a natural question to
ask is this: Can the total number of equivalence classes and the
number of of equivalent strings in each class be determined? The
answer is provided by Lemma 5.1.
\newline\newline
{\bf Lemma 5.1.} {\it All terms of the ordered Pfaffian expansion
can be assigned to $(2\ell-1)!!$  equivalence classes $\{{\cal
C}_1,\cdots,{\cal C}_{(2\ell-1)!!}\}$, each containing $(2\ell)!!$
equivalent strings: $\|{\cal C}_j\| = (2\ell)!!$ for all $j\in
1,\cdots, (2\ell-1)!!$.}
\newline\newline
{\bf Proof}. Consider a string ${\cal S}_i$ belonging to the
equivalence class ${\cal C}_j$ and composed of $\ell$ specific
kernels, $\|{\cal S}_i\|=\ell$. There exist $\ell !$ possible
permutations of kernels and $2^\ell$ intra-kernel permutations of
arguments. As a result, the total number of strings generated by
these two operations from a given string ${\cal S}_i$ equals $\|
{\cal C}_j\|=2^\ell \ell! = (2\ell)!!$. Since the total number of
strings in the ordered expansion is $(2\ell)!$, one concludes that
there exist $(2\ell)!/(2\ell)!! = (2\ell-1)!!$ different equivalence
classes.
 $\;\;\;\blacksquare$
\newline\newline
The results of this and the two subsequent Sections are summarised
in Fig. \ref{Topology-1}.

\subsubsection{Decomposing strings into loop-like
substrings\newline\newline}\label{SubSec522} 
\begin{figure}[t]
\hspace{3cm}\includegraphics[scale=.80]{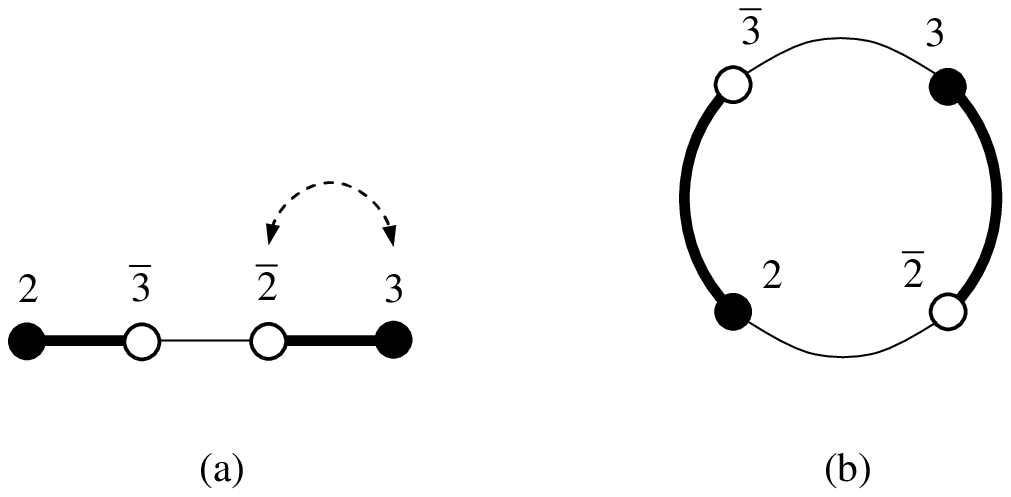}
 \caption{Graphic explanation of the notion ``loop-like substring''.
 (a) The substring $(2{\bar 3})({\bar 2}3)$ can be transformed into the form of an adjacent substring
    (in the sense of Definition 5.7) by flipping arguments in the
    second kernel, $({\bar 2}3) \mapsto (3{\bar 2})$. (b) The
    emerging adjacent substring $(2{\bar 3})(3{\bar 2})$ can clearly be
    depicted in the form of a loop by gluing the arguments $2$ and $\bar{2}$ together.}\label{loop-like-string}
\end{figure}

\hspace{-0.3cm}Having
assigned all $(2\ell)!$ strings of the ordered Pfaffian expansion to
$(2\ell-1)!!$ equivalence classes each containing $(2\ell)!!$
strings, we wish to concentrate on the structure of the strings
themselves. Below, we shall prove that any string ${\cal S}_i$ (of
the length $\|{\cal S}_i\|=\ell$) can be decomposed into a certain
number (between $1$ and $\ell$) of {\it loop-like substrings}, see
the Lemma 5.2. To prepare the reader to the definition of a {\it
loop-like substring}, we first define the notion of a {\it
substring} itself.
\newline\newline
{\bf Definition 5.5.} {\it A product ${\cal S}_i^{(p)}$ of $p$
kernels $Q_n$ is called a {\bf substring} of the string
\begin{eqnarray}
\label{si-string} {\cal S}_i =
 {\rm sgn} (\sigma_i) \, \prod_{j=0}^{\ell-1}
    Q_n(w_{\sigma_i (2j+1)},w_{\sigma_i(2j+2)}), \;\;\; \| {\cal
    S}_i\| =\ell,
\end{eqnarray}
specified in Definition 5.1, if it takes the form
\begin{eqnarray}
{\cal S}_i^{(p)} = \prod_{k=1}^p Q_n(w_{\sigma_i
(2j_k+1)},w_{\sigma_i(2j_k+2)})
\end{eqnarray}
where $j_1 \neq j_2 \neq \cdots \neq j_p$. The length $\| {\cal
S}_i^{(p)}\|$ of the substring is $\| {\cal S}_i^{(p)}\|=p$ with
$1\le p\le \ell$. Notice that no sign is assigned to a substring.}
\newline\newline
{\bf Example 5.5.} The string
\begin{eqnarray}
+(1{\bar 1})(2{\bar 3})({\bar 2}3) \nonumber
\end{eqnarray}
arising in the $\ell=3$ Pfaffian expansion (see the first term in
(\ref{gr-2})) can exhaustively be decomposed into seven substrings
\footnote[5]{It is easy to see that the number of substrings of the
length $p$ equals $
    {\ell \choose p}\nonumber \nonumber
$ so that the total amount of all possible substrings of a string of
the length $\ell$ is
\begin{eqnarray}
    \sum_{p=1}^\ell {\ell \choose p} = 2^\ell-1. \nonumber
\end{eqnarray}
}
\begin{eqnarray} \fl
\underbrace{(1{\bar 1}),\;\; (2{\bar 3}),\;\; ({\bar
2}3)}_{p=1},\;\; \underbrace{(1{\bar 1})(2{\bar 3}),\;\; (1{\bar
1})({\bar 2}3),\;\; (2{\bar 3})({\bar 2}3)}_{p=2},\;\; \underbrace{
(1{\bar 1})(2{\bar 3})({\bar 2}3) }_{p=3} \nonumber
\end{eqnarray}
of lengths $p=1,2$ and $3$, respectively.
\newline\newline
{\bf Definition 5.6.} {\it A substring
\begin{eqnarray}
{\cal S}_i^{(p)} = \prod_{k=1}^p Q_n(w_{\sigma_i
(2j_k+1)},w_{\sigma_i(2j_k+2)}) \nonumber
\end{eqnarray}
is said to be a {\bf loop-like substring} of the length $p= \| {\cal
S}_i^{(p)}\|$, if the two conditions are satisfied:}
\begin{enumerate}
  \item {\it The set of all $2p$ arguments
\begin{eqnarray}
    {\cal W}_{2p} = \{
    w_{\sigma_i (2j_1+1)},w_{\sigma_i(2j_1+2)},\cdots,
    w_{\sigma_i (2j_p+1)},w_{\sigma_i(2j_p+2)} \} \nonumber
\end{eqnarray}
collected from the substring ${\cal S}_i^{(p)}$ remains unchanged
under the operation of complex conjugation
\begin{eqnarray}
    {\bar {\cal W}}_{2p} = \{
    {\bar w}_{\sigma_i (2j_1+1)},{\bar w}_{\sigma_i(2j_1+2)},\cdots,
    {\bar w}_{\sigma_i (2j_p+1)},{\bar w}_{\sigma_i(2j_p+2)} \}. \nonumber
\end{eqnarray}
That means the two sets ${\bar {\cal W}}_{2p}$ and ${\cal W}_{2p}$
of arguments are identical, up to their order. (This property will
be referred to as invariance under complex conjugation.)}
  \item {\it For all subsets $\delta {\cal S}_i^{(q)}$ consisting of $q$ kernels
  $Q_n$ with $1\le q \le p-1$, the substring ${\cal S}_i^{(p)}\backslash \delta {\cal S}_i^{(q)}$
  of the length $p-q$ obtained by removal of $\delta {\cal
  S}_i^{(q)}$ from ${\cal S}_i^{(p)}$ is not invariant under the
  operation of complex conjugation of its arguments.}\newline
\end{enumerate}
{\bf Example 5.6.} Out of seven substrings of the string $+(1{\bar
1})(2{\bar 3})({\bar 2}3)$ detailed in the Example 5.5, the two
substrings
\begin{eqnarray}
    (1{\bar 1}), \;\; (2{\bar 3})({\bar 2}3) \nonumber
\end{eqnarray}
are loop-like (of the lengths $p=1$ and $2$, respectively). The
remaining five substrings are not loop-like. Four of them,
\begin{eqnarray}
    (2{\bar 3}), \;\; ({\bar 2}3), \;\; (1{\bar 1})(2{\bar 3}),\;\;
    (1{\bar 1})({\bar 2}3), \nonumber
\end{eqnarray}
are not loop-like substrings because the property (i) of Definition
5.6 is not satisfied. The fifth substring (of the length $p=3$)
\begin{eqnarray}
   (1{\bar 1})(2{\bar 3}) ({\bar 2}3) \nonumber
\end{eqnarray}
is not loop-like because the property (ii) of Definition 5.6 is
violated. Indeed,
 there {\it does } exist a subset $\delta {\cal S}_i^{(1)}$ consisting of one kernel, represented by the pair
 of arguments $(1{\bar 1})$, whose removal
 would {\it not} destroy the property (i) for the reduced substring
 $(2{\bar 3}) ({\bar 2}3)$.
\newline\newline
The example presented shows that a particular string of the ordered
Pfaffian expansion could be decomposed into a set of loop-like
substrings. Is such a decomposition possible in general? The answer
is given by the following Lemma.
\newline\newline
{\bf Lemma 5.2.} {\it Any given string ${\cal S}_i$ of the length
$\|{\cal S}_i\|=\ell$ from the ordered Pfaffian expansion can be
decomposed into a set of loop-like substrings ${\cal
S}_i^{(\ell_j)}$ of respective lengths $\| {\cal S}_i^{(\ell_j)}\|
=\ell_j$,
\begin{eqnarray}
    {\cal S}_i = \bigcup_{j} {\cal S}_i^{(\ell_j)}
    \nonumber
\end{eqnarray}
such that $\sum_{j} \ell_j=\ell$.}
\newline\newline
{\bf Proof.} We use induction to prove the above statement.
\begin{enumerate}
 \item
{\it Induction Basis.} For $\ell=1$, the Lemma obviously holds
since the strings $(1,{\bar 1})$ and $({\bar 1},1)$ are
loop-like by Definition 5.6.

 \item {\it Induction Hypothesis.} The Lemma is supposed to hold for any
   string ${\cal S}_i$ of the length $\|{\cal
S}_i\|=\ell$:
\begin{eqnarray} \label{si-dec}
    {\cal S}_i = \bigcup_{j} {\cal S}_i^{(\ell_j)}, \;\;\; {\rm
    with}\;\;\;
    \sum_{j}\ell_j=\ell.
\end{eqnarray}

\item {\it Induction Step.} Consider a given string $\tilde{\cal S}_i$ of the
length $\|\tilde{\cal S}_i\|=\ell+1$. Given the induction
hypothesis, we are going to prove that such a string $\tilde{\cal
S}_i$ can be decomposed into a set of loop-like substrings.
\newline\newline
To proceed, we note that any given string $\tilde{\cal S}_i$ of the
length $\|\tilde{\cal S}_i\|=\ell+1$ can be generated from some
string ${\cal S}_i$ of length $\ell$ (see \ref{si-string}) by adding
to it an additional pair of arguments $(z_{\ell+1},{\bar
z}_{\ell+1})$:
\begin{eqnarray}
    {\tilde {\cal S}}_i = (z_{\ell+1},{\bar z}_{\ell+1})
    \;\;\; \otimes \;\;\underbrace{\prod_{k=1}^\ell (w_{\sigma_i
    (2j_k+1)},w_{\sigma_i(2j_k+2)})}_{{\rm the\; string\; } {\cal
    S}_i\; {\rm with \;  sgn} (\sigma_i)\; {\rm dropped} }
\end{eqnarray}
with (or without) further exchange of either $z_{\ell+1}$ or ${\bar
z}_{\ell+1}$ with one of the arguments belonging to the string
${\cal S}_i$ of length $\ell$.
\begin{enumerate}
    \item If no exchange is made, the given string $\tilde{\cal
    S}_i$ is a unit of a single loop-like string ${\Pi}^{(1)}=(z_{\ell+1},{\bar
    z}_{\ell+1})$ and of a string ${\cal S}_i$ admitting the
    decomposition (\ref{si-dec}). As a result, the string ${\tilde {\cal S}}_i$
    of the length $(\ell+1)$ is decomposed into a set of loop-like substrings
\begin{eqnarray}
    \tilde{{\cal S}}_i =
    \Pi^{(1)} \bigcup \left(\bigcup_{j} {\cal S}_i^{(\ell_j)}\right).
\end{eqnarray}

\item If either $z_{\ell+1}$ or ${\bar z}_{\ell+1}$ was swapped with one of
the arguments belonging to a loop-like substring $S_i^{(\ell_{j_0})}
\subset {\cal S}_i$ of the string ${\cal S}_i$, such an exchange
will give rise to a new loop-like substring
$S_i^{(\ell_{j_0}^\prime)} \subset {\tilde S}_i$ of the length
\begin{eqnarray}
\|S_i^{(\ell_{j_0}^\prime)}\| = \|S_i^{(\ell_{j_0})}\|+1=
\ell_{j_0}+1. \nonumber
\end{eqnarray}
Consequently, the given string ${\tilde {\cal S}}_i$ of the length
$(\ell+1)$ is then decomposed into a set of loop-like substrings
\begin{eqnarray}
\label{si-dec-1}
    {\tilde {\cal S}}_i =
    {\cal S}_i^{(\ell_{j_0}^\prime)} \bigcup \left(\bigcup_{j \neq j_0} {\cal S}_i^{(\ell_j)}\right).
\end{eqnarray}
To prove that the substring ${\cal S}_i^{(\ell_{j_0}^\prime)}$ is
indeed loop-like, two properties have to be checked in accordance
with the Definition 5.6. \newline\newline First, one has to show
that the set of $2(\ell_{j_0}+1)$ arguments collected from the
substring ${\cal S}_i^{(\ell_{j_0}^\prime)}$ is invariant under the
operation of complex conjugation; this is obviously true because
${\cal S}_i^{(\ell_{j_0})}$ is loop-like. Second, one has to
demonstrate that the removal of any subset $\delta {\cal S}$ from
${\cal S}_i^{(\ell_{j_0}^\prime)}$ will destroy the {\it invariance
property of the remaining substring} ${\cal
S}_i^{(\ell^\prime_{j_0})} \backslash \delta{\cal S}$. Three
different cases are to be considered here:
\newline\newline
(b1) If the subset $\delta {\cal S}$ does not contain the
    fragments $(z_{\ell+1},\cdots)$ and $(\cdots,\bar{z}_{\ell+1})$,
    it is also a subset of $S_i^{(\ell_{j_0})}$. Since the latter is
    loop-like, the invariance property is destroyed.
\newline\newline
(b2) If the subset $\delta {\cal S}$ contains only one of the
fragments $(z_{\ell+1},\cdots)$ or $(\cdots,\bar{z}_{\ell+1})$, the
invariance property is obviously destroyed.
\newline\newline
(b3) If the subset $\delta {\cal S}$ contains both fragments
$(z_{\ell+1},u)$ and $(v,\bar{z}_{\ell+1})$, the invariance property
is also destroyed. To prove it, we use the {\it reductio ad
absurdum}. Indeed, let us assume that there exists a subset $\delta
{\cal S} \subset{\cal S}_i^{(\ell_{j_0}^\prime)}\subset {\tilde
S}_i$, containing both fragments $(z_{\ell+1},u)$ and
$(v,\bar{z}_{\ell+1})$, whose removal does not destroy the
invariance property of ${\cal S}_i^{(\ell^\prime_{j_0})} \backslash
\delta{\cal S}$. The existence of such a subset $\delta {\cal S}$
implies the existence of yet another subset $\delta {\cal S}^\prime
\subset {\cal S}_i^{(\ell_{j_0})} \subset S_i$,
\begin{eqnarray}
    \delta {\cal S}^\prime = \Big\{ \delta {\cal S} \backslash \left\{
        (z_{\ell+1},u) \cup(v, \bar{z}_{\ell+1})
    \right\}\Big\} \cup \{(v,u)\},
\end{eqnarray}
whose removal {\it does} destroy the invariance of ${\cal
S}_i^{(\ell_{j_0})}\backslash \delta {\cal S}^\prime$ under the
operation of complex conjugation (this claim is obviously true
because the substring ${\cal S}_i^{(\ell_{j_0})}$ is loop-like).
Then, the identity
\begin{eqnarray}
    {\cal
S}_i^{(\ell_{j_0}^\prime)}\backslash \delta {\cal S} = {\cal
S}_i^{(\ell_{j_0})}\backslash \delta {\cal S}^\prime
\end{eqnarray}
suggests that the removal of $\delta {\cal S}$ from ${\cal
S}_i^{(\ell^\prime_{j_0})}$ {\it must} destroy the invariance of
${\cal S}_i^{(\ell^\prime_{j_0})} \backslash \delta{\cal S}$ as
well. Since this contradicts to the assumption made, one concludes
that the invariance property is indeed destroyed.
\end{enumerate}
\end{enumerate}
End of the proof. $\;\;\;\blacksquare$
\newline\newline
We close this Section by providing a brief explanation of the origin
of the term ``loop-like substring'' introduced in Definition 5.6. In
Section \ref{SubSec54}, it will be proven that any loop-like
substring can be brought to the form of an adjacent substring
(Definition 5.7) by means of proper (i) permutation of kernels
and/or (ii) permutation of intra-kernel arguments (Lemma 5.5). The
latter can graphically be represented as a {\it loop}. Indeed, the
loop-like substring $(2{\bar 3})({\bar 2}3)$ considered in Example
5.6 can be transformed into the form of an adjacent substring (in
the sense of Definition 5.7) by flipping arguments in the second
kernel, $({\bar 2}3) \mapsto (3{\bar 2})$:
\begin{eqnarray}
    \underbrace{(2{\bar 3})({\bar 2}3)}_{{\rm loop-like\; substring}}
    \mapsto \underbrace{(2{\bar 3})(3{\bar 2})}_{{\rm adjacent\; substring}}. \nonumber
\end{eqnarray}
The resulting adjacent string $(2{\bar 3})(3{\bar 2})$ can be drawn
in the form of a loop, see Fig. \ref{loop-like-string}. This is
precisely the reason why the substring $(2{\bar 3})({\bar 2}3)$ is
called loop-like.

\subsubsection{Counting longest loop-like substrings ${\cal S}_i^{(\ell)}$ of the length $\ell$\newline\newline}
\label{SubSec53}\hspace{-0.3cm}Although in this subsection, we are
going to concentrate on the longest loop-like substrings
\footnote[6]{As soon as substrings of the longest possible length
$\ell$ are considered, they are strings themselves.} of the length
$\ell$, our main counting result given by Lemma 5.3 stays valid for
loop-like substrings of a smaller length $1\le p < \ell$.
\newline\newline
{\bf Lemma 5.3.} {\it The ordered Pfaffian expansion contains
$(2\ell)!!\,(2\ell-2)!!$ longest loop-like strings
 of the length
$\ell$.}
\newline\newline
{\bf Proof.} Let ${\cal N}_L(\ell)$ be the total number of loop-like
strings of the length $\ell$ in the ordered Pfaffian expansion and
let $n_L(\ell)$ denote the number of equivalence classes all longest
loop-like strings can be assigned to. Following Lemma 5.1, the two
are related to each other as
\begin{eqnarray}
\label{NLnL}
    {\cal N}_L(\ell) = (2\ell)!!\, n_L(\ell),
\end{eqnarray}
because each equivalence class ${\cal C}_j$ contains precisely
$(2\ell)!!$ equivalent strings (see Definition 5.3). It thus remains
to determine $n_L(\ell)$ that can equally be interpreted as a number
of {\it inequivalent} \footnote[7]{In view of Definition 5.3, the
two strings are inequivalent if they cannot be reduced to each other
by means of (i) permutation of kernels and/or (ii) intra-kernel
permutation of arguments.} longest loop-like strings of the length
$\ell$.

The latter can be evaluated by counting the number of ways,
$n_L(\ell+1)$, all {\it inequivalent} longest loop-like strings of
the length $\ell+1$ can be generated from those longest of the
length $\ell$. Since both numbers, $n_L(\ell)$ and $n_L(\ell+1)$,
refer to the {\it longest} loop-like strings, the two correspond to
the Pfaffians of matrices of the size $\ell\times \ell$ and
$(\ell+1)\times (\ell+1)$, correspondingly. In the language of
strings, an increase of the matrix size by one leads to the
appearance of an additional pair of arguments $(z_{\ell+1}, {\bar
z}_{\ell+1})$ in a string of the length $\ell+1$.

We claim that
\begin{eqnarray} \label{nL}
    n_L(\ell+1) = 2\ell \, n_L(\ell).
\end{eqnarray}
To prove this, we concentrate on a given longest loop-like string of
the length $\ell$ and add to it an additional pair of arguments
$(z_{\ell+1},{\bar z}_{\ell+1})$:
\begin{eqnarray}
\label{pr-str-1}
    (z_{\ell+1},{\bar z}_{\ell+1})\;\;\; \otimes \underbrace{\prod_{k=1}^\ell (w_{\sigma_i
    (2j_k+1)},w_{\sigma_i(2j_k+2)})}_{{\rm longest\; loop-like\; string\; of \; the\; length\;} \ell}.
\end{eqnarray}
Here, $\sigma_i$ is a particular permutation of $2\ell$ arguments
(\ref{perm}) corresponding to a longest loop-like string of the
length $\ell$. The resulting string (\ref{pr-str-1}) is {\it not} a
longest loop-like string of the length $\ell+1$ (rather, it is
composed of two loop-like strings of the lengths $1$ and $\ell$,
respectively). Since a loop-like string necessarily assumes the
presence of a fragment $(z_{\ell+1}, \cdots)\, (\cdots, {\bar
z}_{\ell+1})$ somewhere in the string, one has to exchange either
$z_{\ell+1}$ or ${\bar z}_{\ell+1}$ with one of the arguments
belonging to the original longest loop-like string of the length
$\ell$. Clearly, there exist $2\ell$ exchange options for each
argument, $z_{\ell+1}$ (or ${\bar z}_{\ell+1}$). As a result, we
arrive at the relation (\ref{nL}). Given $n_L(1)=1$, we derive the
desired result by induction:
\begin{eqnarray}
    n_L(\ell) = (2\ell-2)!!
\end{eqnarray}
Combining it with (\ref{NLnL}) completes the proof. \footnote[8]{It
is instructive to turn to Example 5.1 that discusses the ordered
Pfaffian expansion for $\ell=2$. The Lemma 5.3 predicts existence of
${\cal N}_L(2)=4!!\, 2!! = 16$ longest loop-like strings that can be
assigned to $n_L(2)=2!!=2$ equivalence classes, each composed of
$(2\ell)!! = 4!! = 8$ equivalent strings. This is in line with
direct counting (\ref{def-1-ex}): the two equivalence classes are
represented by the longest loop-like strings $+(1{\bar 2})({\bar
1}2)$ and $-(12)({\bar 1}{\bar 2})$ (see the second and third
column); each equivalence class contains $8$ equivalent
strings.}$\;\;\;\blacksquare$

\subsubsection{Adjacent vs non-adjacent
loop-like substrings\newline\newline}
\label{SubSec54}\hspace{-0.3cm}Further classification of loop-like
substrings is needed in order to prepare ourselves to the proof of
the Pfaffian integration theorem.
\newline\newline
{\bf Definition 5.7.} {\it A loop-like substring ${\cal S}_i^{(p)}$
of the length $p= \| {\cal S}_i^{(p)}\|$ is called} {\bf adjacent
loop-like substring},{\it\; or simply a} {\bf loop}, {\it if it is
represented by a product
\begin{eqnarray}
{\cal S}_i^{(p)} = \prod_{k=1}^p Q_n(w_{\sigma_i
(2j_k+1)},w_{\sigma_i(2j_k+2)}) \nonumber
\end{eqnarray}
of $p$ kernels such that:}
\begin{enumerate}
  \item {\it The first argument of the first kernel and the second argument of
the last, $p$-th kernel, are complex conjugate of each other,
\begin{eqnarray}
    w_{\sigma_i (2j_1+1)} = {\bar w}_{\sigma_i(2j_p+2)}. \nonumber
\end{eqnarray}
}
\item {\it For each pair of neighbouring kernels in the string, the second
argument of the left kernel in the pair and the first argument of
the right kernel in the pair are complex conjugate of each
other,
\begin{eqnarray}
    w_{\sigma_i (2j_k+2)} = {\bar w}_{\sigma_i(2j_{k+1}+1)}, \;\;\;\; k=1,\cdots, p-1. \nonumber
\end{eqnarray}
}\newline
\end{enumerate}
{\bf Example 5.7.} Out of $16$ longest loop-like strings arising in
the Pfaffian expansion for $\ell=2$ \footnote[1]{For an example,
please refer to the second and third column in (\ref{def-1-ex}).
Also, see Lemma 5.3 for the explanation of the number $16=4!!\, 2!!$
and Definition 5.6 for the notion of a loop-like string.}, the
following eight are adjacent:
\begin{eqnarray}
   -(1{\bar 2})(2{\bar 1}),\;\; -({\bar 2}1)({\bar 1}2),\;\; -({\bar 1}2)({\bar
   2}1),\;\; -(2{\bar 1})(1{\bar 2}), \nonumber \\
   +(12)({\bar 2}{\bar 1}),\;\; +(21)({\bar 1}{\bar 2}),\;\; +({\bar 1}{\bar
   2})(21),\;\; +({\bar 2}{\bar 1})(12). \nonumber
\end{eqnarray}
Notice that although {\it longest} loop-like strings have been
considered in the above example, the notion of an adjacent string is
equally relevant for a loop-like string of {\it smaller} length.
\newline\newline
{\bf Lemma 5.4.} {\it Out of $(2\ell)!!(2\ell-2)!!$ longest
loop-like strings of the length $\ell$ associated with an ordered
Pfaffian expansion, exactly $(2\ell)!!$ are adjacent.}
\newline\newline
{\bf Proof.} To count the total number ${\cal N}_A(\ell)$ of all
adjacent loop-like strings of the length $\ell$, we consider a
specific pair of adjacent loop-like strings of the length $\ell$
represented by the sequences of arguments
\begin{eqnarray}
   (z_{j_1},\underbrace{{\bar z}_{j_2})(z_{j_2}}_{{\rm pair\; \natural\; 1}},
   \underbrace{{\bar z}_{j_3})(z_{j_3}}_{{\rm pair\; \natural\; 2}},{\bar z}_{j_4})\cdots (z_{j_{p-1}},{\bar
   z}_{j_p}) \cdots (z_{j_{\ell-1}},\underbrace{{\bar z}_{j_\ell})(z_{j_\ell}}_{{\rm pair\; \natural\; (\ell-1)}},
   {\bar z}_{j_1}) \nonumber
\end{eqnarray}
and
\begin{eqnarray}
   ({\bar z}_{j_1},\underbrace{{\bar z}_{j_2})(z_{j_2}}_{{\rm pair\; \natural\; 1}},
   \underbrace{{\bar z}_{j_3})(z_{j_3}}_{{\rm pair\; \natural\; 2}},{\bar z}_{j_4})\cdots (z_{j_{p-1}},{\bar
   z}_{j_p}) \cdots (z_{j_{\ell-1}},\underbrace{{\bar z}_{j_\ell})(z_{j_\ell}}_{{\rm pair\; \natural\; (\ell-1)}},
   z_{j_1}). \nonumber
\end{eqnarray}
Here, the mutually distinct $j_p$ take the values from $1$ to
$\ell$. The two strings are identical up to an exchange of the first
and the last arguments $z_{j_1} \rightleftarrows {\bar z}_{j_1}$.
The remaining $2(\ell-1)$ arguments are distributed between the
kernels in such a way that an adjacent string is formed in
accordance with the Definition 5.7; the $(\ell-1)$ underbraces
identifying $(\ell-1)$ pairs of complex conjugate arguments
highlight the structure of an adjacent string.

The total number ${\cal N}_A(\ell)$ of all adjacent loop-like
strings of the length $\ell$ equals the number of ways to generate
those strings from the two depicted above. As soon as there are (i)
$\ell$ ways to assign a number from $1$ to $\ell$ to the label
$j_1$, (ii) $(\ell-1)!$ ways to assign the remaining $(\ell-1)$
numbers to the $(\ell-1)$ pairs left (labelled by $j_2, \cdots,
j_{\ell}$), and (iii) $2^{\ell-1}$ ways to exchange the arguments
$z_{j_k} \rightleftarrows {\bar z}_{j_k}$ ($k=2,\cdots,\ell$) within
those $(\ell-1)$ pairs, we derive:
\begin{eqnarray}
    {\cal N}_A(\ell) = 2 \times \ell \times (\ell-1)! \times
    2^{\ell-1} = (2\ell)!!
\end{eqnarray}
End of proof.\footnote[2]{In particular, there should exist eight
adjacent strings in the ordered Pfaffian expansion for $\ell=2$.
This is in concert with the explicit counting in Example 5.7.}
$\;\;\;\blacksquare$
\newline\newline
{\bf Remark 5.1.} Since the above reasoning holds for loop-like
(sub)strings of any length $1\le p\le \ell$, one concludes that the
total number of adjacent (sub)strings of the length $p$ equals
${\cal N}_A(p)=(2p)!!$.
\newline\newline
{\bf Lemma 5.5.} {\it Any loop-like (sub)string of the length $p$
can be transformed into an adjacent (sub)string of the same length
by means of proper (i) permutation of kernels and/or (ii)
permutation of intra-kernel arguments.}
\newline\newline
{\bf Proof.} To be coherent with the notations used in the proof of
the Lemma 5.3, we deal below with a loop-like string of the length
$\ell$. However, the very same argument applies to any loop-like
(sub)string of the length $1\le p\le \ell$ so that our proof (based
on mathematical induction) holds generally.

(i) For $\ell=1$ and $\ell=2$, the Lemma's statement is obviously
true. Indeed, a loop-like string of the length $\ell=1$ is
automatically an adjacent one. For $\ell=2$, a loop-like string
composed of two kernels is reduced to an adjacent string by utmost
one intra-kernel permutation of arguments.

(ii) Now we assume the Lemma to hold for loop-like (sub)strings of
the length $\ell$ (that is, that any loop-like string of the length
$\ell$ can be reduced to an adjacent string by means of the two
types of allowed operations).

(iii) Given the previous assumption, we have to show that a
loop-like (sub)string of the length $(\ell+1)$ can also be reduced
to an adjacent (sub)string. It follows from the proof of the Lemma
5.3 (see the discussion around (\ref{pr-str-1})) that a loop-like
string of the length $(\ell+1)$ can be generated from a loop-like
string of a smaller length $\ell$ by adding an additional pair of
arguments $(z_{\ell+1}, {\bar z}_{\ell+1})$ followed by exchange of
either $z_{\ell+1}$ or ${\bar z}_{\ell+1}$ with one of the arguments
of the original loop-like string of the length $\ell$. Since, under
the induction assumption (ii), the latter can be made adjacent,
\begin{eqnarray}
\label{adj}
    (z_{\ell+1},{\bar z}_{\ell+1})\, \otimes\, \underbrace{[
    (z_{j_1},\dots)\, \cdots \, (\dots, {\bar z}_{j_q})\, (z_{j_q},\dots)\,
    \cdots \, (\dots, {\bar z}_{j_1})
    ]}_{{\rm adjacent\; string\; of \; the\; length \;}\ell},
\end{eqnarray}
one readily concludes that we are only two steps away from forming
an adjacent string of the length $(\ell+1)$ out of (\ref{adj}).
Indeed, an exchange of arguments ${\bar
z}_{\ell+1}\rightleftarrows{\bar z}_{j_q}$ brings (\ref{adj}) to the
form
\begin{eqnarray}
\label{adj-1}
    (z_{\ell+1},{\bar z}_{j_q})\,
    (z_{j_1},\dots)\, \cdots \, (\dots, {\bar z}_{\ell+1})\, (z_{j_q},\dots)\,
    \cdots \, (\dots, {\bar z}_{j_1})
\end{eqnarray}
which boils down to the required adjacent string
\begin{eqnarray}
\label{adj-2}
    \underbrace{(z_{j_1},\dots)\, \cdots \, (\dots, {\bar z}_{\ell+1})\, (z_{\ell+1},{\bar z}_{j_q})\,
    (z_{j_q},\dots)\,
    \cdots \, (\dots, {\bar z}_{j_1})}_{{\rm adjacent\; string\; of \; the\; length \;}(\ell+1)}
\end{eqnarray}
upon moving the pair $(z_{\ell+1},{\bar z}_{j_q})$ through $(q-1)$
pairs on the right. Exchanging  ${\bar z}_{\ell+1}\rightleftarrows{
z}_{j_q}$ instead can be done in the same way. $\;\;\;\blacksquare$
\newline\newline {\bf Remark
5.2.} The reduction of a loop-like (sub)string to an adjacent
(sub)string is {\it not} unique. For instance, the loop-like string
\begin{eqnarray}
+(1{\bar 3})({\bar 2}3)({\bar 1}2) \nonumber
\end{eqnarray}
(see the first term in (\ref{gr-3})) can be reduced, by permutation
of kernels and permutation of intra-kernel arguments, to one of the
following adjacent strings:
\begin{eqnarray}
\label{adj-set}
    +(1{\bar 3})(3{\bar 2})(2{\bar 1}), \;\; -({\bar 1}2)({\bar 2}3)({\bar
    3}1), \nonumber\\
    +(2{\bar 1})(1{\bar 3})(3{\bar 2}),\;\; -({\bar 2}3)({\bar 3}1)({\bar
    1}2), \nonumber \\
    +(3{\bar 2})(2{\bar 1})(1{\bar 3}), \;\; -({\bar 3}1)({\bar 1}2)({\bar
    2}3).
\end{eqnarray}
To handle the problem of the non-unique reduction of a loop-like
string to an adjacent string, the notion of string {\it handedness}
has to be introduced.

\subsubsection{Handedness of an adjacent substring\newline\newline}
\label{SubSec55}\hspace{-0.3cm}{\bf Definition 5.8.} {\it Close an
adjacent substring ${\cal S}$ into a loop by ``gluing'' the right
argument of the last kernel with the left argument of the first
kernel below the chain as in Figure 4.,
\begin{eqnarray} \fl
   {\cal S}:\;\;\rightsquigarrow \bullet \underbrace{(w_{j_1},{\bar w}_{j_2})}_{{\rm first\; kernel}}\cdot\,(w_{j_2},
   {\bar w}_{j_3})\cdots (w_{j_{q-1}},{\bar
   w}_{j_q}) \cdots (w_{j_{p-1}},{\bar w}_{j_p})\cdot\underbrace{(w_{j_p},
   {\bar w}_{j_1})}_{{\rm last\; kernel}}\bullet \rightsquigarrow
   \nonumber
\end{eqnarray}
(here, a set of the arguments $(w_1,\cdots, w_{2p})$ is specified by
(\ref{perm}) with $\ell$ set to $p$, and the symbol $\bullet$
denotes a gluing point). Read the arguments of a loop, one after the
other, in a clockwise direction (as depicted by the symbol
$\rightsquigarrow$), starting with ${\bar w}_{j_1}$ until you arrive
at $w_{j_p}$. If $\alpha_R$ is the number of times an argument
$z_{j_q}$ is followed by its conjugate ${\bar z}_{j_q}$ for all
$q\in (1,\cdots,p)$,
\begin{eqnarray}
    (\dots, z_{j_q})\cdot ({\bar z}_{j_q},\dots), \nonumber
\end{eqnarray}
an adjacent substring ${\cal S}$ is said to have the {\bf
handedness} ${\mathbb H}(\alpha_L, \alpha_R)$, where $\alpha_L =
p-\alpha_R$.}
\newline\newline
{\bf Example 5.8}. The handedness of eight adjacent strings
considered in the Example 5.7 with $p=l=2$ is listed below:
\begin{eqnarray}
\label{h58} {\mathbb H}(2,0),\;\; {\mathbb H}(0,2),\;\;{\mathbb
H}(0,2),\;\;
{\mathbb H}(2,0),\nonumber \\
{\mathbb H}(1,1),\;\; {\mathbb H}(1,1),\;\;{\mathbb H}(1,1),\;\;
{\mathbb H}(1,1).
\end{eqnarray}
\newline
{\bf Lemma 5.6}. {\it Out of $(2\ell)!!$ adjacent strings of the
length $\ell$ arising in an ordered Pfaffian expansion, there are
exactly
\begin{eqnarray}
\label{naL} {\cal N}_{\alpha}(\ell) = \ell! \, {\ell \choose \alpha}
\end{eqnarray}
strings with the handedness ${\mathbb H}(\alpha,\ell-\alpha)$.}
\newline\newline
{\bf Proof.} A string with the handedness ${\mathbb
H}(\alpha,\ell-\alpha)$ closed into a loop (see Definition 5.8)
contains $\alpha$ ``left'' fragments $(\dots,{\bar z}_{j_q})\cdot
(z_{j_q},\dots)$ and $(\ell-\alpha)$ ``right'' fragments $(\dots,
z_{j_q})\cdot({\bar z}_{j_q},\dots)$ with the opposite order of
complex conjugation in the nearest neighbouring kernels; each
fragment is labelled by an integer number $j_q\in (1,\cdots,\ell)$.
To count a total number of all strings with the handedness ${\mathbb
H}(\alpha,\ell-\alpha)$, we notice that there exist ${\ell \choose
\alpha}$ ways to distribute ``left'' and ``right'' fragments on the
loop, and $\ell!$ ways to assign $\ell$ integer numbers (from $1$ to
$\ell$) to the labels $j_1,\cdots,j_\ell$. Applying the
combinatorial multiplication rule completes the proof.
$\;\;\;\blacksquare$
\newline\newline
{\bf Remark 5.3.} It is instructive to realise that the Lemma 5.4
can be seen as a corollary to the Lemma 5.6. Indeed, the total
number of all adjacent strings of the length $\ell$ is nothing but
\begin{eqnarray}
    \sum_{\alpha=0}^\ell {\cal N}_\alpha(\ell)=
    \ell !\, \sum_{\alpha=0}^\ell {\ell \choose \alpha}
    = (2\ell)!! \nonumber
\end{eqnarray}
Not unexpectedly, this result is in concert with the Lemma 5.4.
\newline\newline
{\bf Corollary 5.2.} {\it The number of adjacent substrings of the
length $p$, $1 \le p < \ell$, with the handedness
$\mathbb{H}(\alpha, p-\alpha)$ equals
\begin{eqnarray}
    {\cal N}_\alpha (p) = p!\,{p \choose \alpha}.
\end{eqnarray}
The total number of all adjacent substrings of the length $p$ is
$(2p)!!$}
\newline\newline
{\bf Proof.} Follow the proof of the Lemma 5.6 and the Remark 5.3
with $\ell$ replaced by $p$. $\;\;\;\blacksquare$
\newline\newline
{\bf Example 5.9.} Out of $16$ longest loop-like strings arising in
the Pfaffian expansion for $\ell=2$, there are eight adjacent as
explicitly specified in Example 5.7. The handedness of those strings
was considered in Example 5.8. In accordance with the Lemma 5.6,
there must exist ${\cal N}_0(2)=2$ strings of the handedness
${\mathbb H}(0,2)$, ${\cal N}_2(2)=2$ strings of the handedness
${\mathbb H}(2,0)$, and ${\cal N}_1(2)=4$ strings of the handedness
${\mathbb H}(1,1)$. Direct counting (\ref{h58}) confirms that this
is indeed the case.

\subsubsection{Equivalence classes of adjacent
(sub)strings\newline\newline} \label{SubSec56}\hspace{-0.3cm}Having
defined a notion of the handedness of an adjacent string, we are
back to the issue of a {\it non-uniqueness} of reduction of a
loop-like string to an adjacent one. To deal with the indicated
non-uniqueness problem, we would like to define, and explicitly
identify, all distinct {\it equivalence classes} for $(2\ell)!!$
adjacent strings arising in the context of an ordered Pfaffian
expansion.
\newline\newline
{\bf Definition 5.9.} {\it Two adjacent strings, ${\cal S}_i$ and
${\cal S}_j$, are said to be {\bf equivalent adjacent strings},
${\cal S}_i \sim {\cal S}_j$, if they can be obtained from each
other by (i) permutation of kernels and/or (ii) intra-kernel
permutation of arguments.}
\newline\newline
{\bf Example 5.10.} The six adjacent strings (\ref{adj-set}) are
equivalent to each other.
\newline\newline
{\bf Definition 5.10.} {\it A group of equivalent adjacent strings
is called the {\bf equivalence class of adjacent strings}. The
$j$-th equivalence class to be denoted ${\cal AC}_j$ consists of
$\|{\cal AC}_j\|$ equivalent adjacent strings.}
\newline\newline
{\bf Lemma 5.7.} {\it All $(2\ell)!!$ adjacent strings arising in
the ordered Pfaffian expansion can be assigned to $(2\ell-2)!!$
equivalence classes $\{{\cal AC}_1,\cdots, {\cal
AC}_{(2\ell-2)!!}\}$ of adjacent strings, each class containing
$2\ell$ equivalent adjacent strings: $\| {\cal AC}_j\|=2\ell$ for
all $j\in 1,\cdots, (2\ell-2)!!$.}
\newline\newline
{\bf Proof.} Let us concentrate on a given adjacent string ${\cal
S}$ with the handedness ${\mathbb H}(\alpha,\ell-\alpha)$ that
belongs to an equivalence class ${\cal AC}_j$ and count a number of
ways to generate equivalent strings out of it by means of (i)
permutation of kernels and/or (ii) intra-kernel permutation of
arguments {\it without destroying the adjacency property}. Two
complementary generating mechanisms exist.

\begin{itemize}
  \item First way (M1): One starts with an adjacent string of the handedness ${\mathbb
H}(\alpha,\ell-\alpha)$,
\begin{eqnarray} \fl \label{as-1}
   \rightsquigarrow \bullet \underbrace{(w_{j_1},{\bar w}_{j_2})}_{{\rm first\; kernel}}\cdot\,(w_{j_2},
   {\bar w}_{j_3})\cdots (w_{j_{p-1}},{\bar
   w}_{j_p}) \cdots (w_{j_{\ell-1}},{\bar w}_{j_\ell})\cdot\underbrace{(w_{j_\ell},
   {\bar w}_{j_1})}_{{\rm last\; kernel}}\bullet \rightsquigarrow
\end{eqnarray}
(the reader is referred to the Definition 5.8 for the notation
used), simultaneously flips the intra-kernel arguments in all $\ell$
kernels,
\begin{eqnarray} \fl
\;\;\;\;\;\;\;
    \underbrace{({\bar w}_{j_2}, w_{j_1})}_{{\rm first\; kernel}}\cdot\,
   ({\bar w}_{j_3},w_{j_2})\cdots ({\bar
   w}_{j_p}, w_{j_{p-1}}) \cdots ({\bar w}_{j_\ell},
    w_{j_{\ell-1}})\cdot\underbrace{
   ({\bar w}_{j_1},w_{j_\ell})}_{{\rm last\; kernel}} \ ,
   \nonumber
\end{eqnarray}
and further permutes the kernels in a fan-like way so that the last
$\ell$-th kernel in the string becomes the first, the $(\ell-1)$-th
kernel in the string becomes the second, etc.:
\begin{eqnarray} \fl \label{as-2}
    \rightsquigarrow \bullet
   \underbrace{
   ({\bar w}_{j_1},w_{j_\ell})}_{{\rm last\; kernel}}
   \cdot\, ({\bar w}_{j_\ell}, w_{j_{\ell-1}})
   \cdots ({\bar
   w}_{j_p}, w_{j_{p-1}}) \cdots
   ({\bar w}_{j_3},w_{j_2})\cdot
   \underbrace{({\bar w}_{j_2}, w_{j_1})}_{{\rm first\; kernel}} \bullet
   \rightsquigarrow
\end{eqnarray}
The so obtained {\it adjacent} string is equivalent to the
initial one (\ref{as-1}) but possesses the complementary
handedness ${\mathbb H}(\ell-\alpha,\alpha)$.
\newline
  \item Second way (M2): One starts with an adjacent string of the handedness ${\mathbb
  H}(\alpha,\ell-\alpha)$, and permutes the $\ell$ kernels in a cyclic manner to
generate $(\ell-1)$ additional (but equivalent) adjacent strings
with the same handedness (to visualise the process, one may think of
moving a gluing point $\bullet$ through the kernels in
(\ref{as-1})).
\end{itemize}
The two mechanisms, M1 and M2, combined together bring up $2\ell$
equivalent adjacent strings due to the combinatorial multiplication
rule. As a result, we conclude that $\|{\cal AC}_j\|=2\ell$.
Consequently, the number of distinct equivalence classes equals
$(2\ell)!!/(2\ell)=(2\ell-2)!!$.  $\;\;\;\blacksquare$
\newline\newline
{\bf Example 5.11.} To illustrate the Lemma 5.7, we consider an
ordered Pfaffian expansion for $\ell=2$ as detailed in the Example
5.1. The eight adjacent strings of the length $\ell=2$ were
specified in the Example 5.7; their handedness was considered in the
Example 5.8. In accordance with the Lemma 5.7, there should exist
$2$ distinct equivalence classes, each containing $4$ adjacent
strings. Indeed, one readily verifies that those two equivalence
classes are
\begin{eqnarray}
    {\cal AC}_1:\;\;
   -(1{\bar 2})(2{\bar 1}),\;\; -({\bar 2}1)({\bar 1}2),\;\; -({\bar 1}2)({\bar
   2}1),\;\; -(2{\bar 1})(1{\bar 2}) \nonumber
\end{eqnarray}
and
\begin{eqnarray}
   {\cal AC}_2:\;\;+(12)({\bar 2}{\bar 1}),\;\; +(21)({\bar 1}{\bar 2}),\;\; +({\bar 1}{\bar
   2})(21),\;\; +({\bar 2}{\bar 1})(12). \nonumber
\end{eqnarray}
\newline
\begin{figure}[t]
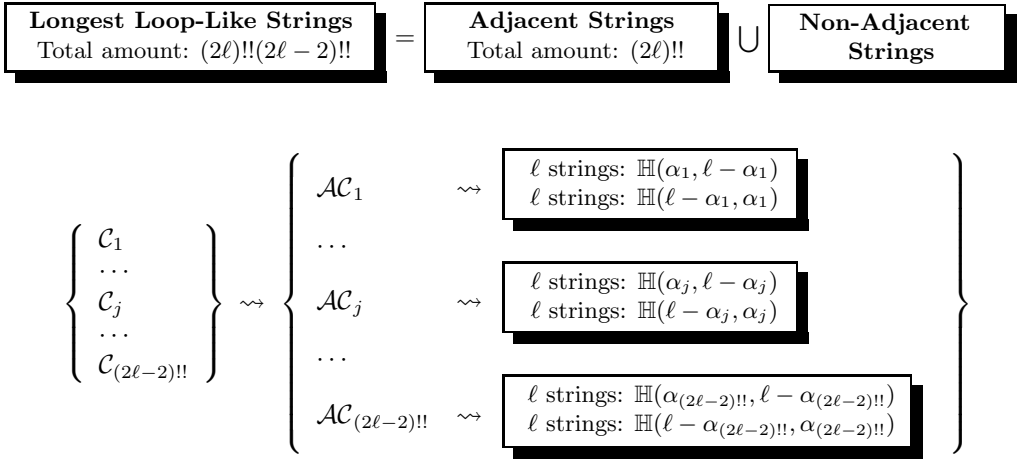

\vspace{0.5cm} $
\begin{array}{lll}
 {\begin{minipage}{5.4in} \raisebox{-0.5cm}
  {\shadowbox{
    \begin{minipage}{1.8in}
    \begin{center}
    \small{
   {\bf Longest Loop-Like Strings}
   \newline{Total amount: $(2\ell)!!(2\ell-2)!!$}}
   \end{center}
   \end{minipage}}} =
\raisebox{-0.5cm}
  {\shadowbox{
    \begin{minipage}{1.4in}
    \begin{center}
    \small{
   {\bf ~Adjacent Strings}
   \newline
   {Total amount: $(2\ell)!!$}}
   \end{center}
   \end{minipage}}} $\bigcup$
\raisebox{-0.5cm}
  {\shadowbox{
    \begin{minipage}{1.1in}
    \begin{center}
    \small{
   {\bf Non-Adjacent}}
   \newline
   \small{{\bf Strings}}
   \end{center}
   \end{minipage}}}
\end{minipage}}
& {} & {} \\
{} & {} & {} \\
 {} & {} & {}
\end{array}
\newline
\qquad \left\{ \begin{array}{l}
  {\cal C}_1 \\
  \cdots \\
    {\cal C}_j \\
  \cdots \\
  {\cal C}_{(2\ell-2)!!}
 \end{array}\right\}\
 \rightsquigarrow\;
\left\{ \begin{array}{ll}
  {\cal AC}_1 & \rightsquigarrow \begin{array}{l}
            \raisebox{-0.5cm}
  {\shadowbox{
    \begin{minipage}{1.4in}
    \begin{center}
    \small{
     $\ell$ strings: ${\mathbb H}(\alpha_1,\ell-\alpha_1)$\\
     $\ell$ strings: ${\mathbb H}(\ell-\alpha_1,\alpha_1)$}
   \end{center}
   \end{minipage}}}
    \end{array}
  \\
  \cdots \\
  {\cal AC}_j & \rightsquigarrow \begin{array}{l}
            \raisebox{-0.5cm}
  {\shadowbox{
    \begin{minipage}{1.4in}
    \begin{center}
    \small{
     $\ell$ strings: ${\mathbb H}(\alpha_j,\ell-\alpha_j)$\\
     $\ell$ strings: ${\mathbb H}(\ell-\alpha_j,\alpha_j)$}
   \end{center}
   \end{minipage}}}
    \end{array} \\
  \cdots \\
  {\cal AC}_{(2\ell-2)!!} & \rightsquigarrow \begin{array}{l}
            \raisebox{-0.5cm}
  {\shadowbox{
    \begin{minipage}{2.0in}
    \begin{center}
    \small{
     $\ell$ strings: ${\mathbb H}(\alpha_{(2\ell-2)!!},\ell-\alpha_{(2\ell-2)!!})$\\
     $\ell$ strings: ${\mathbb H}(\ell-\alpha_{(2\ell-2)!!},\alpha_{(2\ell-2)!!})$}
   \end{center}
   \end{minipage}}}
    \end{array}
 \end{array}\right\}
 \newline\newline\newline
$
 \caption{Any longest loop-like string from the equivalence classes
  $\{ {\cal C}_1, \cdots, {\cal C}_{(2\ell-2)!!} \}$ can be
 reduced to an adjacent string belonging to one of the equivalence classes
 $\{ {\cal AC}_1, \cdots, {\cal AC}_{(2\ell-2)!!} \}$ of adjacent strings
 by means of proper permutation of kernels and/or permutation of
 intra-kernel arguments (Lemma 5.5). This reduction is non-unique (Remark
 5.2). Each equivalence class ${\cal AC}_j$ of adjacent strings
 contains $\ell$ adjacent strings of the handedness ${\mathbb
 H}(\alpha_j,\ell-\alpha_j)$ and $\ell$ adjacent strings of the complementary handedness
${\mathbb H}(\ell-\alpha_j,\alpha_j)$ (Remark 5.4). Notice that two
distinct equivalence classes ${\cal AC}_i$ and ${\cal AC}_j$ (where
$i\neq j$) may have the same values for $\alpha$: $\alpha_i
=\alpha_j$.\newline} \label{Topology-2}
\end{figure}{\bf Remark 5.4.} In fact, a generic prescription can be given to
build $(2\ell-2)!!$ distinct equivalence classes $\{{\cal
AC}_1,\cdots, {\cal AC}_{(2\ell-2)!!}\}$ for $(2\ell)!!$ adjacent
strings arising in the ordered Pfaffian expansion. Because of the
``duality'' between equivalent adjacent strings with complementary
handedness ${\mathbb H}(\alpha,\ell-\alpha)$ and ${\mathbb
H}(\ell-\alpha,\alpha)$ discussed in the proof of the Lemma 5.6, the
adjacent strings whose handedness ${\mathbb H}(\alpha,\ell-\alpha)$
is restricted by the inequality $0\le \alpha \le \lfloor \ell/2
\rfloor$ will form a natural basis in the consideration to
follow.\newline
\begin{itemize}
  \item {\bf The case $\ell=2\lambda+1$ odd.}\newline\newline
  (i) For $0 \le \alpha_1 \le \lambda$, (i.a) pick up
  an adjacent
  string with the handedness ${\mathbb H}(\alpha_1, \ell-\alpha_1)$,
out of ${\cal N}_{\alpha_1}(\ell)$, and generate $\ell$
  equivalent adjacent strings with the same handedness through
  the mechanism M2 of the Lemma 5.7. (i.b) Apply the mechanism
  M1 of the same Lemma to each of the $\ell$ adjacent strings
  generated in (i.a) to create $\ell$ more equivalent adjacent
  strings of the handedness ${\mathbb H}(\ell-\alpha_1,
  \alpha_1)$ hereby raising their total amount to $2\ell$. The
  strings generated in (i.a) and (i.b) are said to belong to the
  {\it equivalence class} ${\cal AC}_1$.
  \newline\newline
  (ii) To generate the next equivalence class ${\cal AC}_2 \neq
  {\cal AC}_1$, pick up an adjacent string not belonging to
  ${\cal AC}_1$ with the handedness ${\mathbb H}(\alpha_2,
\ell-\alpha_2)$ out
  of $({\cal N}_{\alpha_1}(\ell) - \ell)$ left (again,
  $\alpha_2$ is restricted to $0\le \alpha_2\le \lambda$), and
  repeat the actions described in (i.a) and (i.b) to generate
  another set of $2\ell$ equivalent strings. These will belong
  to the equivalence class ${\cal AC}_2$ which is distinct from
  ${\cal AC}_1$.\newline\newline (iii) To generate the $j$-th
  equivalence class ${\cal AC}_j$, one picks up an adjacent
  string out of $({\cal N}_{\alpha_1}(\ell) - (j-1)\ell)$ left
  and repeats the actions sketched in (ii).\newline\newline (iv)
  The procedure stops once there are no adjacent strings left.
  Obviously, the total number of equivalence classes is
  \begin{eqnarray}\fl \label{count-odd}
    \frac{1}{\ell}\sum_{\alpha=0}^{\lfloor \ell/2\rfloor} {\cal N}_{\alpha}(\ell) =
    (2\lambda)! \sum_{\alpha=0}^\lambda {2\lambda+1 \choose \alpha}
    = (2\lambda)!\, 2^{2\lambda} = (2\ell-2)!!
  \end{eqnarray}
  This is in concert with the Lemma 5.7.\newline
  \item {\bf The case $\ell=2\lambda$ even.}\newline\newline
  In this case, special care should be exercised for the set of
adjacent strings with the
  handedness ${\mathbb H}(\lambda,\lambda)$ because these adjacent strings are {\it self}-complementary:
  the mechanism M1 applied to any of those adjacent strings generates a string with the {\it same}, not complementary,
  handedness. The latter circumstance can readily be accommodated
  when
  giving a prescription for building $(2\ell-2)!!$ distinct equivalence
  classes of adjacent strings. \newline\newline
  (i) First, we separate all adjacent strings with the handedness ${\mathbb
  H}(\alpha,\ell-\alpha)$ where $0\le \alpha \le \lambda-1$ and apply a procedure identically equivalent to that
  described for the case $\ell$ odd to generate distinct equivalence
  classes of adjacent strings. The total amount of distinct
  equivalence classes built in this way equals
    \begin{eqnarray}\fl \label{count-even-1}
    L_1=\frac{1}{\ell}\sum_{\alpha=0}^{\lfloor \ell/2\rfloor-1} {\cal N}_{\alpha}(\ell) =
    (2\lambda-1)! \sum_{\alpha=0}^{\lambda-1} {2\lambda \choose \alpha}
    = (2\ell-2)!! - \frac{(2\lambda-1)!}{2} \,{2\lambda \choose \lambda}.
  \end{eqnarray}
  \newline\newline
  (ii) Second, having generated in the previous step $L_1$ distinct
  equivalence classes $\{{\cal AC}_1,\cdots, {\cal AC}_{L_1}\}$ of adjacent strings,
  we concentrate on the adjacent strings with the
  handedness ${\mathbb H}(\lambda,\lambda)$ not treated so far.
  To this end, we
  (ii.a) pick up an adjacent string out of ${\cal
  N}_\lambda(2\lambda)$ with the handedness ${\mathbb
  H}(\lambda,\lambda)$ and perform the operations M1 and M2 to
  generate $2\ell=4\lambda$ equivalent strings with the {\it same}
  handedness. The $2\ell$ equivalent adjacent strings will belong to
  a certain equivalence class, say, ${\cal AC}_{L_1+1}$. (ii.b) In
  the next step, we pick up an adjacent string with the handedness
  ${\mathbb H}(\lambda,\lambda)$ out of $({\cal
  N}_\lambda(2\lambda)-2\ell)$ left, and perform the operations
  detailed in (ii.a) in order to generate yet another set of $2\ell$ equivalent
  adjacent strings belonging to an equivalence class ${\cal
  AC}_{L_1+2}$. (ii.c) We proceed further on until the last
  available  equivalence class composed of $2\ell$ adjacent strings is formed, ${\cal
  AC}_{L_1+L_2}$, where $L_2$ equals
\begin{eqnarray} \label{count-even-2}
    L_2 =\frac{1}{2\ell}\, {\cal N}_{\lambda}(\ell) =
    \frac{(2\lambda-1)!}{2} \,{2\lambda \choose \lambda}.
\end{eqnarray}
Hence, for $\ell=2\lambda$, the total number of equivalence classes
for adjacent strings equals
\begin{eqnarray}
L_1+L_2 = (2\ell-2)!!
\end{eqnarray}
as expected from the Lemma 5.7.
\end{itemize}

\subsection{Integrating out all longest loop-like strings of the length $\ell$} \label{SubSec57}
More spadework is needed to
prove the Pfaffian integration theorem. Below, we will be interested
in calculating the contribution $C_L(\ell)$ of longest loop-like
strings (of the length $\ell$) into the sought integral
(\ref{th-2}):
\begin{eqnarray}
    \label{LLS-1}\fl
    C_L(\ell) &=
    \int_{{\mathbb C}}\,  \prod_{j=1}^\ell \,  d\pi(z_j)\,\,
    \left({\rm pf\,} \left[
    \begin{array}{cc}
      Q_{n}(z_i, z_j) & Q_{n}(z_i, {\bar z}_j) \\
      Q_{n}({\bar z}_i, z_j) & Q_{n}({\bar z}_i, {\bar z}_j) \\
    \end{array}
    \right]_{2\ell \times 2\ell}\right)_{{\scriptsize \begin{array}{c}
                                               {\rm longest} \\
                                               {\rm loop-like\; strings}
                                             \end{array}}
    } \nonumber \\ \fl
    &= \frac{1}{2^\ell\ell!}\int_{{\mathbb C}}\,  \prod_{j=1}^\ell \,  d\pi(z_j)\,\,
    \sum_{\sigma\in S^\prime_{2\ell}}
    {\rm sgn} (\sigma) \, \prod_{j=0}^{\ell-1}
    Q_n(w_{\sigma(2j+1)},w_{\sigma(2j+2)}).
\end{eqnarray}
In the second line of (\ref{LLS-1}), only that part of the ordered
Pfaffian expansion (\ref{pf-ord}) appears which corresponds to a set
of all loop-like strings of the length $\ell$. They are accounted
for by picking up proper permutations $S^\prime_{2\ell} \subset
S_{2\ell}$ in the expansion (\ref{pf-ord}),
\begin{eqnarray} \nonumber
    S^{\prime}_{2\ell} \; \mapsto \; ( {\rm longest\; loop-like\; strings\; of\; the\; length \;\ell}).
\end{eqnarray}
Although, in accordance with the Lemma 5.3, the number of terms in
the expansion (\ref{LLS-1}) equals ${\cal N}_L(\ell)=(2\ell)!!
(2\ell-2)!!$, there is no need to integrate all of them out because
various loop-like strings belonging to the same equivalence class
yield identical contributions. The latter observation effectively
reduces the number of terms in (\ref{LLS-1}) so that
\begin{eqnarray}
    \label{LLS-2}\fl
    C_L(\ell) =
    \frac{(2\ell)!!}{2^\ell\ell!}\int_{{\mathbb C}}\,  \prod_{j=1}^\ell \,  d\pi(z_j)\,\,
    \sum_{\sigma\in S^{\prime\prime}_{2\ell}}
    {\rm sgn} (\sigma) \, \prod_{j=0}^{\ell-1}
    Q_n(w_{\sigma(2j+1)},w_{\sigma(2j+2)}).
\end{eqnarray}
Here, the prefactor $(2 \ell)!!= 2^\ell \ell!$ equals the number of
longest loop-like strings in each equivalence class; the
$\sigma$-series runs over the permutations
$S^{\prime\prime}_{2\ell}\subset S^\prime_{2\ell}$ corresponding to
$n_L(\ell)=(2\ell-2)!!$ longest loop-like strings, each of them
being {\it a} representative of one distinct equivalence class,
\begin{eqnarray} \label{LLS-3}
    S^{\prime\prime}_{2\ell}\; \mapsto \;  \{{\cal S}_1\in {\cal C}_1,\cdots,{\cal S}_{(2\ell-2)!!}\in {\cal
    C}_{(2\ell-2)!!}\},
\end{eqnarray}
see the Lemma 5.3. There are $(2\ell-2)!!$ terms in (\ref{LLS-2}).

To perform the integration explicitly, one has to reduce the longest
loop-like strings in (\ref{LLS-2}) to the form of adjacent strings
as discussed in the Lemma 5.5. In accordance with the Lemma 5.7,
there exist $(2\ell-2)!!$ equivalence classes $\{ {\cal
AC}_1,\cdots, {\cal AC}_{(2\ell-2)!!}\}$ of adjacent strings, each
of them containing $2\ell$ equivalent adjacent strings (see also
Fig. \ref{Topology-2}). This results in the representation
\begin{eqnarray}
    \label{LLS-2adj-000}\fl
    C_L(\ell) = \int_{{\mathbb C}}\,  \prod_{j=1}^\ell \,  d\pi(z_j)\,\,
    \sum_{\sigma\in {\tilde S}^{\prime\prime}_{2\ell}}
    {\rm sgn} (\sigma) \, \prod_{j=0}^{\ell-1}
    Q_n(w_{\sigma(2j+1)},w_{\sigma(2j+2)}),
\end{eqnarray}
where the $\sigma$-series runs over the permutations ${\tilde
S}^{\prime\prime}_{2\ell}\subset S^\prime_{2\ell}$ corresponding to
$(2\ell-2)!!$ adjacent loop-like strings of the length $\ell$, each
of them being {\it a} representative of each one of existing
equivalence classes of {\it adjacent strings},
\begin{eqnarray} \label{LLS-3-000}
    {\tilde S}^{\prime\prime}_{2\ell}\; \mapsto \;  \{\tilde{{\cal S}}_1\in
     {\cal AC}_1,\cdots,{\tilde {\cal S}}_{(2\ell-2)!!}\in {\cal
    AC}_{(2\ell-2)!!}\}.
\end{eqnarray}
The number of terms in (\ref{LLS-2adj-000}) is $(2\ell-2)!!$.

To proceed, we rewrite (\ref{LLS-2adj-000}) in a more symmetric form
that treats {\it all} adjacent strings on the same footing:
\begin{eqnarray}
    \label{LLS-2adj-111}\fl
    C_L(\ell) = \frac{1}{2\ell}\,\int_{{\mathbb C}}\,  \prod_{j=1}^\ell \,  d\pi(z_j)\,\,
    \sum_{\sigma\in S^{\prime\prime\prime}_{2\ell}}
    {\rm sgn} (\sigma) \, \prod_{j=0}^{\ell-1}
    Q_n(w_{\sigma(2j+1)},w_{\sigma(2j+2)}).
\end{eqnarray}
Here, the $\sigma$-series runs over the permutations
$S^{\prime\prime\prime}_{2\ell}\subset S^\prime_{2\ell}$
corresponding to {\it all} adjacent loop-like strings of the length
$\ell$:
\begin{eqnarray} \fl \label{LLS-3-222}
    S^{\prime\prime\prime}_{2\ell}\; \mapsto \;  \left\{
    \{\tilde{{\cal S}}^{(1)}_1,\cdots,  \tilde{{\cal S}}^{(1)}_{2\ell}\}    \in
     {\cal AC}_1,\cdots,\{{\tilde {\cal S}}_1^{((2\ell-2)!!)},\cdots,
            {\tilde {\cal S}}_{2\ell}^{((2\ell-2)!!)} \}
     \in {\cal
    AC}_{(2\ell-2)!!}
    \right\}.
\end{eqnarray}
As soon as there exist $2\ell$ equivalent adjacent strings in each
equivalent class ${\cal AC}_i$ of adjacent strings, the prefactor
$(2\ell)^{-1}$ was included into (\ref{LLS-2adj-111}) to avoid the
overcounting.

An advantage of the representation (\ref{LLS-2adj-111}) can be
appreciated with the help of the Lemma 5.6. According to it, the
summation over the permutations $\sigma \in
S_{2\ell}^{\prime\prime\prime}$ can be replaced with the summation
over all longest adjacent strings with a given handedness ${\mathbb
H}(\alpha,\ell-\alpha)$, for all $\alpha\in (0,\ell)$:
\begin{eqnarray}
\label{repsum}
    \sum_{\sigma\in S^{\prime\prime\prime}_{2\ell}}
    {\rm sgn} (\sigma) \, \prod_{j=0}^{\ell-1}
    Q_n(w_{\sigma(2j+1)},w_{\sigma(2j+2)}) =
    \sum_{\alpha=0}^{\ell} \sum_{i=1}^{{\cal N}_\alpha(\ell)} {\cal
    S}_i(\alpha).
\end{eqnarray}
Here, ${\cal S}_i(\alpha)$ denotes the $i$-th adjacent string of the
length $\| {\cal S}_i(\alpha) \|=\ell$ with the handedness ${\mathbb
H}(\alpha,\ell-\alpha)$. In accordance with the Lemma 5.6,
\begin{eqnarray}
\label{naL-rep} {\cal N}_{\alpha}(\ell) = \ell! \, {\ell \choose
\alpha}.
\end{eqnarray}
Given (\ref{repsum}), the integration in (\ref{LLS-2adj-111}) can be
performed explicitly. Due to the new representation
\begin{eqnarray}
    \label{LLS-2adj-555}
    C_L(\ell) = \frac{1}{2\ell}\,
    \sum_{\alpha=0}^{\ell} \sum_{i=1}^{{\cal N}_\alpha(\ell)}
    \int_{{\mathbb C}}\,  \prod_{j=1}^\ell \,  d\pi(z_j)\, {\cal
    S}_i(\alpha),
\end{eqnarray}
one has to calculate the contribution of a string ${\cal
S}_i(\alpha)$ with the handedness ${\mathbb H}(\alpha,\ell-\alpha)$
to the integral:
\begin{eqnarray}
    \label{LLS-2adj-666}
    I_\ell(\alpha)=\int_{{\mathbb C}}\,  \prod_{j=1}^\ell \,  d\pi(z_j)\, {\cal
    S}_i(\alpha).
\end{eqnarray}
(i) The case $\alpha=0$ is the simplest one. Having in mind the
definition (\ref{th-1}) and introducing an auxiliary matrix
${\boldsymbol {\hat \varsigma}}$ with the entries
\begin{eqnarray}
    \varsigma_{jk} = \frac{1}{2} \, \int_{\mathbb C} d\pi(z) \,
    q_j(z)\, q_k({\bar z}),
\end{eqnarray}
we straightforwardly derive:
\begin{equation} \fl \label{is-1}
    I_\ell(0)={\rm sgn}(\sigma_0) \int_{\mathbb C} \prod_{j=1}^\ell
    d\pi(z_j)
    \left[\prod_{j=1}^{\ell-1}
    Q_n({\bar z}_{j}, z_{j+1})\right]
    Q_n({\bar z}_\ell,z_1) = - {\rm tr}_{(0,n-1)} \left[({\boldsymbol {\hat \mu}}{\boldsymbol {\hat
    \varsigma}})^{\ell}\right].
\end{equation}
Here, the permutation sign, ${\rm sign}(\sigma_0)$, is ${\rm
sgn}(\sigma_0)=-1$ (see (\ref{perm})), while the trace ${\rm
tr}_{(0,n-1)}(\cdots)$ reflects the fact that the integrated
adjacent string is loop-like. Importantly, the result of the
integration (\ref{is-1}) does {\it not} depend on a particular
arrangement of the arguments $z_j$ and ${\bar z}_j$ as far as the
handedness ${\mathbb H}(0,\ell)$ is kept.
\newline\newline
(ii) The case $\alpha=\ell$ associated with the adjacent loop-like
strings of the handedness ${\mathbb H}(\ell,0)$ can be treated along
the same lines to bring
\begin{eqnarray} \label{is-L}
    I_\ell(\ell)= (-1)^{\ell-1} {\rm tr}_{(0,n-1)} \left[({\boldsymbol {\hat \mu}}{\boldsymbol {\hat
    \varsigma^*}})^{\ell}\right].
\end{eqnarray}
\newline
(iii) More care should be exercised for $0 < \alpha < \ell$. In this
case, two adjacent strings with the {\it same} handedness ${\mathbb
H}(\alpha,\ell-\alpha)$ may bring {\it different} contributions into
the integral (\ref{LLS-2adj-666}). For instance, the adjacent string
\footnote[3]{The empty products are interpreted to be $1$.}
\begin{eqnarray} \fl
    {\cal S}_1(\alpha) = {\rm sgn}(\sigma_\alpha)     \left[
    \prod_{j=1}^{\alpha-1}{Q_n}(z_{j},{\bar z}_{j+1})\right]
    Q_n(z_{\alpha}, z_{\alpha+1})\left[\prod_{j=\alpha+1}^{\ell-1}
    Q_n({\bar z}_{j}, z_{j+1})\right]
    Q_n({\bar z}_\ell,{\bar z}_1) \nonumber
\end{eqnarray}
yields the contribution
\begin{eqnarray} \label{is-L1}
    I^{(1)}_\ell(\alpha)= (-1)^{\alpha-1} {\rm tr}_{(0,n-1)} \left[({\boldsymbol {\hat \mu}}{\boldsymbol {\hat
    \varsigma}}^*)^{\alpha}
    ({\boldsymbol {\hat \mu}}{\boldsymbol {\hat
    \varsigma}})^{\ell-\alpha}
    \right].
\end{eqnarray}
At the same time, the adjacent string
\begin{eqnarray} \fl
    {\cal S}_{{\ell \choose
\alpha}}(\alpha) = {\rm sgn}(\sigma_\alpha) \nonumber \\ \fl \times
       \left[
    \prod_{j=1}^{\ell-\alpha-1}{Q_n}({\bar z}_{j},z_{j+1})\right]
    Q_n({\bar z}_{\ell-\alpha}, {\bar z}_{\ell-\alpha+1})
    \left[\prod_{j=\ell-\alpha+1}^{\ell-1}
    Q_n(z_{j}, {\bar z}_{j+1})\right]
    Q_n(z_\ell,z_1), \nonumber
\end{eqnarray}
possessing the very same handedness, yields
\begin{eqnarray} \label{is-LN}
    I^{({\ell \choose
\alpha})}_\ell(\alpha)= (-1)^{\alpha-1} {\rm tr}_{(0,n-1)}
\left[({\boldsymbol {\hat \mu}}{\boldsymbol {\hat
    \varsigma}})^{\ell-\alpha}
    ({\boldsymbol {\hat \mu}}{\boldsymbol {\hat
    \varsigma}}^*)^{\alpha}
    \right].
\end{eqnarray}
In deriving the results (\ref{is-L1}) and (\ref{is-LN}), we have
used the fact that the permutation sign ${\rm sign}(\sigma_\alpha)$
is ${\rm sign}(\sigma_\alpha)=(-1)^{\alpha-1}$.

It can readily be seen that the adjacent strings with a given
handedness ${\mathbb H}(\alpha,\ell-\alpha)$ bring all possible
${\ell \choose \alpha}$ contributions, or {\it words}, that can be
represented as a trace
\begin{eqnarray}
\label{word-j}
  W_j(\ell,\alpha) = - {\rm tr}_{(0,n-1)} \underbrace{\left[\cdots
  (-{\boldsymbol {\hat \mu}}{\boldsymbol {\hat
    \varsigma}}^*)\cdots
    ({\boldsymbol {\hat \mu}}{\boldsymbol {\hat
    \varsigma}}) \cdots (-{\boldsymbol {\hat \mu}}{\boldsymbol {\hat
    \varsigma}}^*)\cdots\right]}_{\alpha \; {\rm letters\;} (-{\boldsymbol {\hat \mu}}{\boldsymbol {\hat
    \varsigma}}^*)\; {\rm and\;} \ell-\alpha\; {\rm letters\;}({\boldsymbol {\hat \mu}}{\boldsymbol {\hat
    \varsigma}})}
\end{eqnarray}
of a product of $\alpha$ matrices, or {\it letters}, $(-{\boldsymbol
{\hat \mu}}{\boldsymbol {\hat \varsigma}}^*)$ and $\ell-\alpha$
matrices ({\it letters}) $({\boldsymbol {\hat \mu}}{\boldsymbol
{\hat \varsigma}})$ distributed in all possible ${\ell \choose
\alpha}$ ways. Hence, the index $j$ in (\ref{word-j}) takes the
values $1\le j \le {\ell \choose \alpha}$. Importantly, each word
$W_j(\ell,\alpha)$ appears exactly $\ell !$ times since there always
exist $\ell!$ adjacent strings ${\cal S}_{i_k}(\alpha)$
($k=1,\cdots,\ell!$) which are related to each other by a
permutation of the integration variables in (\ref{LLS-2adj-555}). As
a result, the latter is reduced to
\begin{eqnarray}
    \label{LLS-2adj-777}
    C_L(\ell) = \frac{1}{2\ell}\, \ell!\,
    \sum_{\alpha=0}^{\ell} \sum_{j=1}^{{\ell \choose \alpha}}
    W_j(\ell,\alpha).
\end{eqnarray}
Spotting that
\begin{eqnarray}
    \sum_{\alpha=0}^{\ell} \sum_{j=1}^{{\ell \choose \alpha}}
    W_j(\ell,\alpha) = - {\rm tr}_{(0,n-1)} \left[ \left(
    {\boldsymbol {\hat \mu}}{\boldsymbol {\hat
    \varsigma}} - {\boldsymbol {\hat \mu}}{\boldsymbol {\hat
    \varsigma}}^*
    \right)^\ell \right],
\end{eqnarray}
we obtain
\begin{eqnarray}
    C_L(\ell)
    =
    - \frac{1}{2}\, (\ell-1)!\,  {\rm tr}_{(0,n-1)} \left[
        ({\boldsymbol {\hat \mu}}{\boldsymbol {\hat
    \varsigma}}- {\boldsymbol {\hat \mu}}{\boldsymbol {\hat
    \varsigma^*}})^\ell
    \right].
\end{eqnarray}
Finally, noticing that the matrix $({\boldsymbol {\hat
\mu}}{\boldsymbol {\hat
    \varsigma}}- {\boldsymbol {\hat \mu}}{\boldsymbol {\hat
    \varsigma^*}})$ under the sign of trace is related to the matrix ${\boldsymbol {\hat \upsilon}}$
defined by (\ref{th-3}) as
\begin{eqnarray}
\label{ups}
    {\boldsymbol {\hat \upsilon}} = 2i\,
    ({\boldsymbol {\hat \mu}}{\boldsymbol {\hat
    \varsigma}}- {\boldsymbol {\hat \mu}}{\boldsymbol {\hat
    \varsigma^*}}),
\end{eqnarray}
we arrive at the remarkably compact result
\begin{eqnarray}
\label{CLL}
    C_L(\ell) = -\frac{1}{2}\, \frac{(\ell-1)!}{(2i)^\ell}\,
    {\rm tr}_{(0,n-1)} (
        {\boldsymbol {\hat \upsilon}}^\ell).
\end{eqnarray}
\newline
Hence, we have proven the following Theorem.
\newline\newline
{\bf Theorem 5.2.} {\it Let $d\pi(z)$ be any benign measure on $z
\in {\mathbb C}$, and the function $Q_n(x,y)$ be an antisymmetric
function of the form
\begin{eqnarray} \label{qk}
    Q_n(x,y) = \frac{1}{2} \sum_{j,k=0}^{n-1} q_j(x) \, {\hat \mu}_{jk} \,
    q_k(y) \nonumber
\end{eqnarray}
where $q_j$'s are arbitrary polynomials of $j$-th order, and
${\boldsymbol {\hat \mu}}$ is an antisymmetric matrix. Then the
integration formula
\begin{eqnarray} \fl
    \label{theorem-1}
    C_L(\ell) = \prod_{j=1}^\ell \, \int_{z_j\in {\mathbb C}} d\pi(z_j)\,\,
    &\left({\rm pf\,} \left[
    \begin{array}{cc}
      Q_{n}(z_i, z_j) & Q_{n}(z_i, {\bar z}_j) \\
      Q_{n}({\bar z}_i, z_j) & Q_{n}({\bar z}_i, {\bar z}_j) \\
    \end{array}
    \right]_{2\ell \times 2\ell}\right)_{{\scriptsize \begin{array}{c}
                                               {\rm longest} \\
                                               {\rm loop-like\; strings}
                                             \end{array}}
    } \nonumber \\ \fl
    &=
    -\frac{1}{2}\, \frac{(\ell-1)!}{(2i)^\ell}\,
    {\rm tr}_{(0,n-1)} (
        {\boldsymbol {\hat \upsilon}}^\ell)
\end{eqnarray}
holds, provided the integrals on the~l.h.s.~exist. Here, the matrix
${\boldsymbol {\hat \upsilon}}$ is determined by the entries
\begin{eqnarray}
    \label{sigma=def-rep}
    {\hat \upsilon}_{\alpha,\beta} = i \sum_{k=0}^{n-1} {\hat \mu}_{\alpha,k}
    \int_{z\in {\mathbb C}} d\pi(z) \left[
        q_k(z)\, q_\beta({\bar z}) - q_\beta(z) \, q_k({\bar z})
    \right]. \nonumber
\end{eqnarray}
}
\newline
The following corollary holds. \newline\newline{\bf Corollary 5.3.}
{\it Consider a set of $2p$ arguments
\begin{eqnarray}
\label{perm-p-long}
    \{ w_1=z_1,w_2={\bar z}_1,\cdots, w_{2p-1} = z_p, w_{2p}= {\bar
    z}_p \},
\end{eqnarray}
all $(2p)!$ permutations of which are denoted by $S_{2p}$. Take the
subset $S^\prime_{2p}\subset S_{2p}$ of $S_{2p}$ corresponding to
all loop-like strings ${\cal S}_i^{(p)}$ of the length $\| {\cal
S}_i^{(p)} \|=p$,
\begin{eqnarray}
{\cal S}_i^{(p)} = \prod_{j=0}^{p-1} Q_n(w_{\sigma_i
(2j+1)},w_{\sigma_i(2j+2)}). \nonumber
\end{eqnarray}
Here, $\sigma_i$ labels the $i$-th permutation $\sigma_i \in
S_{2p}^\prime$. The Theorem 5.2 implies:
\begin{eqnarray} \fl
\label{c53} C_L(p) = \frac{1}{2^p p!}\int_{{\mathbb C}}\,
\prod_{j=1}^p \, d\pi(z_j)\,\,
    \sum_{\sigma\in S^\prime_{2p}}
    {\rm sgn} (\sigma) \, \prod_{j=0}^{p-1}
    Q_n(w_{\sigma(2j+1)},w_{\sigma(2j+2)}) \nonumber \\=
    -\frac{1}{2}\frac{(p-1)!}{(2i)^p}\, {\rm tr}_{(0,n-1)}({\boldsymbol {\hat
    \upsilon}}^p).
\end{eqnarray}
}

\subsection{Integrating out compound strings}
\label{SubSec58} Having dealt with the longest loop-like strings
arising in the ordered Pfaffian expansion (see Theorem 5.2) and
having extended the result of this theorem to loop-like strings of a
smaller length (see Corollary 5.3), we now turn to the treatment of
the remaining $(2\ell)! - (2\ell)!! (2\ell-2)!!$ strings to be
referred to as {\it compound strings}. Our final goal here is to
calculate their contribution to the integral in the l.h.s. of
(\ref{th-2}).

\begin{figure}[t]
\hspace{0.5cm}
\includegraphics[scale=.90]{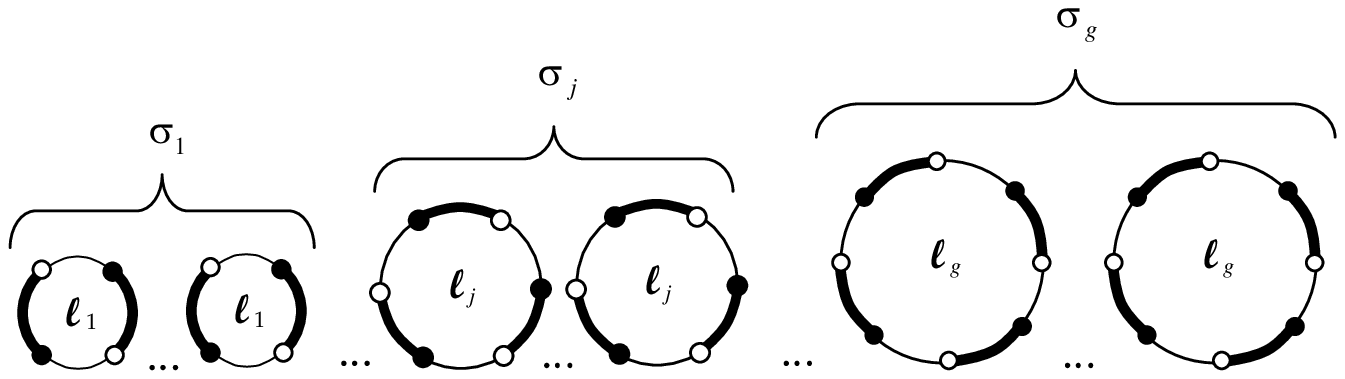}
 \caption{A diagrammatic representation of a compound string belonging
 to the topology class $
    {\boldsymbol \lambda} = (\ell_1^{\sigma_1},\cdots,
    \ell_g^{\sigma_g})
    $ with $\sum_{j=1}^g \ell_j \sigma_j = \ell$ and $\sum_{j=1}^g
    \sigma_j=r$. Here, $g$ denotes the number of inequivalent parts
    of the partition ${\boldsymbol \lambda}$ whilst $r$ equals the
    total number of loops.
 }\label{compound-string}
\end{figure}

To give a definition of a {\it compound string}, we remind that any
given string ${\cal S}_i$ of the length $\|{\cal S}_i\|=\ell$ from
the ordered Pfaffian expansion can be decomposed into a set of
loop-like substrings of smaller lengths (see Lemma 5.2). According
to the Lemma 5.5, each of the above loop-like substrings ${\cal
S}_i^{(\ell_j)}$ can further be reduced to the form of an adjacent
substring. This leads us to the following definition:
\newline\newline
{\bf Definition 5.11.} {\it A string ${\cal S}_i$ of the length
$\|{\cal S}_i\|=\ell$ from the ordered Pfaffian expansion is called
a compound string if it is composed of a set of adjacent substrings
${\cal S}_i^{(\ell_j)}$ of respective lengths $\| {\cal
S}_i^{(\ell_j)}\| =\ell_j$ such that $
    {\cal S}_i = \bigcup_{j} {\cal S}_i^{(\ell_j)}
   $ with $\sum_{j}\ell_j=\ell$.}
\newline\newline
{\bf Remark 5.5.} (i) This definition suggests that all compound
strings can be classified in accordance with all possible patterns
of {\it unordered partitions} ${\boldsymbol \lambda}$ of the size
$|{\boldsymbol \lambda}|=\ell$ of an integer $\ell$:
\begin{eqnarray}
\label{part-not}
    {\boldsymbol \lambda} = (\ell_1^{\sigma_1},\cdots,
    \ell_g^{\sigma_g}).
\end{eqnarray}
The frequency representation (\ref{part-not}) of the partition
${\boldsymbol \lambda}$ says that the part $\ell_j$ appears
$\sigma_j$ times so that
\begin{eqnarray}
\label{part-not-1}
    \ell = \sum_{j=1}^g \ell_j \sigma_j.
\end{eqnarray}
Here, $g$ is the number of inequivalent parts of the partition
${\boldsymbol \lambda}$. \newline\newline (ii) Alternatively, the
partition (\ref{part-not}) of an integer $\ell$ can be represented
as
\begin{eqnarray}
    {\boldsymbol \lambda} = ( {\bar \ell}_1,\cdots,{\bar \ell}_r),
\end{eqnarray}
where the order of $\bar{\ell}_j$'s is irrelevant, and some of them
can be equal to each other. Obviously,
\begin{eqnarray}
    \sum_{j=1}^r {\bar \ell}_j = \ell.
\end{eqnarray}
Here, $r$ is the length of unordered partition ${\boldsymbol
\lambda}$. \newline\newline (iii) The correspondence between
compound strings of the length $\ell$ and unordered partitions
${\boldsymbol \lambda}$ of the size $|{\boldsymbol \lambda}|=\ell$
gives rise to a topological interpretation of compound strings,
which can conveniently be represented in a diagrammatic form (see
Fig. \ref{compound-string}). The diagram for a generic compound
string
\begin{eqnarray}
    {\cal S}_i = \bigcup_{j=1}^r {\cal
    S}_i^{(\bar{\ell}_j)},\;\;\; \;\sum_{j=1}^r {\bar \ell}_j=\ell,
    \nonumber
\end{eqnarray}
consists of $r$ loops, the $j$-th loop depicting an adjacent
substring ${\cal S}_i^{({\bar \ell}_j)}$ of the length ${\bar
\ell}_j$. Such a diagram will be said to belong to a topology class
$\{{\bar \ell}_1,\cdots,{\bar \ell}_r\} =
\{\ell_1^{\sigma_1},\cdots, \ell_g^{\sigma_g}\}$.
\newline\newline
{\bf Remark 5.6.} The number of topologically distinct diagrams
equals the number $p(\ell)$ of unordered partitions of an integer
$\ell$ (see Remark 5.5). These are known to follow the sequence
\begin{eqnarray}
\label{nunop}
    p(\ell) = \{ 1,2,3,5,7,11,15,22,30,42,\cdots \}.
\end{eqnarray}
An exact evaluation of $p(\ell)$ can be performed with the help of
Euler's generating function (Andrews 1998)
\begin{eqnarray}
    \sum_{\ell=0}^\infty p(\ell)\, q^\ell =
    \prod_{m=1}^\infty \frac{1}{(1-q^m)} \equiv \frac{1}{(q)_\infty}.
\end{eqnarray}
The asymptotic behaviour of $p(\ell)$ for $\ell \gg 1$ was studied
by Hardy and Ramanujan (1918),
\begin{eqnarray}
    p(\ell) \sim \frac{1}{4\ell \sqrt{3}} \, \exp \left(
        \pi \sqrt{2\ell/3}
    \right).
\end{eqnarray}
\newline
To calculate the contribution of compound strings to the integral in
the l.h.s. of (\ref{th-2}), one has to determine (i) the
contribution of a diagram belonging to a given topology class to the
integral, and (ii) the number of diagrams within a given topology
class.
\newline\newline
{\bf Lemma 5.8.} {\it The number of diagrams belonging to the
topology class $
    {\boldsymbol \lambda} = (\ell_1^{\sigma_1},\cdots,
    \ell_g^{\sigma_g})
    $
equals
\begin{eqnarray}
\label{L58}
    {\cal N}_{\{\ell_1^{\sigma_1},\cdots,
    \ell_g^{\sigma_g}\}} = \ell! \prod_{j=1}^g\frac{1}{ (\ell_j!)^{\sigma_j}
    \,\sigma_j!}.
\end{eqnarray}}
\newline
{\bf Proof.} To determine the number ${\cal
N}_{\{\ell_1^{\sigma_1},\cdots, \ell_g^{\sigma_g}\}}$ of diagrams
belonging to a given topology class $\{{\bar \ell}_1,\cdots,{\bar
\ell}_r\} = \{\ell_1^{\sigma_1},\cdots, \ell_g^{\sigma_g}\}$, we use
the multiplication principle.
\begin{itemize}
  \item First, we distribute the pairs of arguments $\{{\bar \ell}_1,\cdots, {\bar
  \ell}_r\}$ between $r$ loops. This can be achieved by $m_1$ ways,
  \begin{eqnarray} \fl
    m_1 = {\ell \choose {\bar \ell}_1}\, {\ell-{\bar \ell}_1 \choose {\bar
    \ell}_2}\cdots {\ell-{\bar \ell}_1 -\cdots-{\bar \ell}_{r-2} \choose {\bar \ell}_{r-1}}
    \, {\ell-{\bar \ell}_1 -\cdots-{\bar \ell}_{r-1} \choose {\bar \ell}_{r}}
    \times \frac{1}{r!}. \nonumber
  \end{eqnarray}
  The factor $1/r!$ reflects the fact that the order of the $r$
  loops is irrelevant. (Indeed, a given topology class is
  associated with an {\it unordered} partition of an integer
  $\ell$). Simple rearrangements show that $m_1$ simplifies down
  to
  \begin{eqnarray}
    m_1 = \frac{\ell!}{{\bar \ell}_1!\cdots {\bar \ell}_r!}
    \frac{1}{r!}.
  \end{eqnarray}
  \item Next, we shuffle all {\it inequivalent} loops (those with distinct
  lengths). This can be achieved by $m_2$ ways,
    \begin{eqnarray}
        m_2 = \frac{(\sigma_1+\cdots+\sigma_g)!}{\sigma_1! \cdots
        \sigma_g!} = \frac{r!}{\sigma_1! \cdots
        \sigma_g!}.
    \end{eqnarray}
\end{itemize}
As a result, the total number of diagrams belonging to the topology
class $
    {\boldsymbol \lambda} = (\ell_1^{\sigma_1},\cdots,
    \ell_g^{\sigma_g})
    $
equals
\begin{eqnarray}
{\cal N}_{\{\ell_1^{\sigma_1},\cdots,
    \ell_g^{\sigma_g}\}} = m_1 m_2=\frac{\ell!}{{\bar \ell}_1!\cdots {\bar \ell}_r!}
    \frac{1}{\sigma_1! \cdots
        \sigma_g!}.
\end{eqnarray}
The observation $\prod_{j=1}^r{\bar \ell}_j! = \prod_{j=1}^g
(\ell_j!)^{\sigma_j}$ ends the proof. $\;\;\;\blacksquare$
\newline\newline\newline
{\bf Lemma 5.9.} {\it A diagram associated with a topology class $
    {\boldsymbol \lambda} = (\ell_1^{\sigma_1},\cdots,
    \ell_g^{\sigma_g})
    $ contributes
\begin{eqnarray}
\label{L59}
    C_{\{ \ell_1^{\sigma_1},\cdots,
    \ell_g^{\sigma_g}\}} = \prod_{j=1}^g C_L^{\sigma_j}(\ell_j)
\end{eqnarray}
to the integral in the l.h.s. of (\ref{th-2}). The function $C_L(p)$
is defined by (\ref{c53}) of the Corollary 5.3.
    }
    \newline\newline
    {\bf Proof.} The above claim is a direct consequence of the Lemma 5.2, Definition 5.11, and Remark 5.5.
    End of proof.  $\;\;\;\blacksquare$
\newline\newline
{\bf Theorem 5.3.} {\it All compound strings belonging to the
topology class $
    {\boldsymbol \lambda} = (\ell_1^{\sigma_1},\cdots,
    \ell_g^{\sigma_g})
    $ yield, after the integration in (\ref{th-2}), the contribution
    \begin{eqnarray}
    \label{c54}
    {\tilde C}_{{\boldsymbol \lambda}} = \frac{\ell!}{(2i)^\ell}
    \prod_{j=1}^g \left[ \frac{1}{\ell_j^{\sigma_j}\sigma_j!}
    \left(
        -\frac{1}{2}\, {\rm tr}_{(0,n-1)} ({\boldsymbol {\hat
        \upsilon}}^{\ell_j})
    \right)^{\sigma_j} \right].
    \end{eqnarray}
    }
\newline
{\bf Proof.} Observe that ${\tilde C}_{{\boldsymbol \lambda}} =
{\cal N}_{\{\ell_1^{\sigma_1},\cdots,
    \ell_g^{\sigma_g}\}} C_{\{\ell_1^{\sigma_1},\cdots,
    \ell_g^{\sigma_g}\}}$, and make use of (\ref{c53}), (\ref{L58}) and (\ref{L59}) to derive
    (\ref{c54}). End of proof. $\;\;\;\blacksquare$

\subsection{Proof of the Pfaffian integration theorem} \label{SubSec59} Now we
have all ingredients needed to complete the proof of the Pfaffian
integration theorem announced in Section 5.1. Indeed, in accordance
with the topological interpretation of the terms arising in the
Pfaffian expansion (\ref{pf-ord}), the integral on the l.h.s. of
(\ref{th-2}) is given by the sum of contributions of adjacent
strings of two types: the longest adjacent strings and the compound
strings. The contribution of the former, $C_L(\ell)$, is given by
the Theorem 5.2 while the contribution of the latter, ${\tilde
C}_{{\boldsymbol \lambda}}$, is determined by the Theorem 5.3. As
soon as
\begin{eqnarray}
    C_L(\ell) = {\tilde C}_{{\boldsymbol \lambda} = (\ell^1)},
    \nonumber
\end{eqnarray}
one immediately concludes that the integrated Pfaffian equals the
sum of ${\tilde C}_{\boldsymbol \lambda}$ over all unordered
partitions ${\boldsymbol \lambda}$ of the size $|{\boldsymbol
\lambda}|=\ell$:
\begin{eqnarray} \fl \label{pit-1}
    {\cal I} =
    \int_{{\mathbb C}}\,  \prod_{j=1}^\ell \,  d\pi(z_j)\,\,
    {\rm pf\,} \left[
    \begin{array}{cc}
      Q_{n}(z_i, z_j) & Q_{n}(z_i, {\bar z}_j) \\
      Q_{n}({\bar z}_i, z_j) & Q_{n}({\bar z}_i, {\bar z}_j) \\
    \end{array}
    \right]_{2\ell \times 2\ell} =\sum_{|{\boldsymbol
    \lambda}|=\ell} {\tilde C}_{{\boldsymbol
    \lambda}} \nonumber \\
    = \left( \frac{i}{2} \right)^\ell (-1)^\ell \ell! \,\sum_{|{\boldsymbol
    \lambda}|=\ell}
    \prod_{j=1}^g \left[ \frac{1}{\sigma_j!}
    \left(
        -\frac{1}{2\ell_j}\, {\rm tr}_{(0,n-1)} ({\boldsymbol {\hat
        \upsilon}}^{\ell_j})
    \right)^{\sigma_j} \right].
\end{eqnarray}
Quite remarkably, the r.h.s. of (\ref{pit-1}) can be recognised to
be a zonal polynomial (Macdonald 1998)
\begin{eqnarray}
\label{SymPol-5}
     Z_{(1^\ell)}(p_1,\cdots,p_\ell) =
    (-1)^\ell  \ell! \sum_{|\blambda|=\ell} \, \prod_{j=1}^g \frac{1}{\sigma_j!}
\left(
    - \frac{p_{\ell_j}}{\ell_j}
\right)^{\sigma_j} \nonumber
\end{eqnarray}
with the arguments
\begin{eqnarray}
    p_j = \frac{1}{2}\, {\rm tr}_{(0,n-1)} ({\boldsymbol {\hat
        \upsilon}}^{j}), \;\;\;\;\; j=1,2,\cdots, \ell.
\end{eqnarray}
As a result, we conclude that
\begin{eqnarray}  \label{pit-2}
    {\cal I} =
    \left( \frac{i}{2} \right)^\ell
    Z_{(1^\ell)}\left(
    \frac{1}{2} {\rm tr}_{(0,n-1)}
    {\boldsymbol {\hat \upsilon}}^1,\cdots,\frac{1}{2} {\rm tr}_{(0,n-1)}
    {\boldsymbol {\hat \upsilon}}^\ell
    \right).
\end{eqnarray}
This coincides with the statement (\ref{th-2}) of the Pfaffian
integration theorem. $\;\;\;\blacksquare$

\section{Probability function $p_{n,k}$: General solution and generating function}
\label{Sec6}
\subsection{General solution}
\label{SubSec61} The general solution for the probability function
$p_{n,k}$ of a fluctuating number of real eigenvalues in spectra of
GinOE is now straightforward to derive. Indeed, it was shown in
Section \ref{SubSec33} that the probability function $p_{n,k}$
admits the representation
\begin{eqnarray} \fl
    \label{pnk-comp-6}
    p_{n,k} = \frac{p_{n,n}}{\ell!} \, \left(\frac{2}{i}\right)^\ell
    \,
    \prod_{j=1}^\ell \int_{{\rm Im\,}z_j >0} d^2z_j\, \nonumber \\
    \times\; {\rm erfc}
    \left( \frac{z_j - {\bar z}_j}{i\sqrt{2} } \right)
    \,
    {\rm pf\,} \left[
    \begin{array}{cc}
      {\cal D}_{n}(z_i, z_j) & {\cal D}_{n}(z_i, {\bar z}_j) \\
      {\cal D}_{n}({\bar z}_i, z_j) & {\cal D}_{n}({\bar z}_i, {\bar z}_j) \\
    \end{array}
    \right]_{2\ell \times 2\ell}
\end{eqnarray}
with the kernel function ${\cal D}_n(x,y)$ given by (\ref{d2m}) and
(\ref{d2m+1}).

The $\ell$-fold integral in (\ref{pnk-comp-6}) can explicitly be
performed by virtue of the Pfaffian integration theorem after the
identification
\begin{eqnarray} \fl
    d\pi(z)= e^{-(z^2+{\bar z}^2)/2}{\rm erfc}
    \left( \frac{z - {\bar z}}{i\sqrt{2} } \right)\, \theta({\rm Im\,}z)\, d^2
    z, \;\;\;\;
    Q_n(x,y)= e^{(x^2+y^2)/2}{\cal D}_n(x,y). \nonumber
\end{eqnarray}
Straightforward calculations bring
\begin{eqnarray}
    p_{n,k} = \frac{p_{n,n}}{\ell!} \,
    Z_{(1^\ell)}\left(
    \frac{1}{2} {\rm tr}_{(0,n-1)}
    {\boldsymbol {\hat \upsilon}}^1,\cdots,\frac{1}{2} {\rm tr}_{(0,n-1)}
    {\boldsymbol {\hat \upsilon}}^\ell
    \right)
\end{eqnarray}
where the matrix ${\boldsymbol {\hat \upsilon}}$ is given by
(\ref{th-3}). Combining the definition (\ref{th-3}) with
(\ref{mu-even}), (\ref{mu-odd}), (\ref{cm}), (\ref{alpha-measure})
and (\ref{chi-matrix}), one concludes that
\begin{eqnarray}
    {\boldsymbol {\hat \upsilon}} = 2 i (\boldsymbol{\hat \mu} \boldsymbol{\hat \chi})
    =\boldsymbol{{\hat \sigma}},
\end{eqnarray}
see Appendices A and B. Finally, making use of the trace identity
\begin{eqnarray}
\label{trid-body}
    {\rm tr\,}_{(0,n-1)} {\boldsymbol {\hat \sigma}}^j = 2\,
    {\rm tr\,}_{(0,\lfloor n/2 \rfloor -1)} {\boldsymbol {\hat \varrho}}^j
\end{eqnarray}
proven in Appendix C, we end up with the exact formula
\begin{eqnarray}
    p_{n,k} = \frac{p_{n,n}}{\ell!} \,
    Z_{(1^\ell)}\left(
        {\rm tr\,}_{(0,\lfloor n/2 \rfloor -1)} {\boldsymbol {\hat
        \varrho}}^1, \cdots,
        {\rm tr\,}_{(0,\lfloor n/2 \rfloor -1)} {\boldsymbol {\hat
        \varrho}}^\ell
    \right).
\end{eqnarray}
The entries of the matrix $\boldsymbol{{\hat \varrho}}$, calculated
in Appendix C, are given by
\begin{eqnarray}
\label{rho-even-body}
    {{\hat \varrho}}_{\alpha,\beta}^{{\rm even}} &=&
    \int_0^\infty dy \,y^{2(\beta-\alpha)-1}\, e^{y^2}\, {\rm
    erfc\,}(y\sqrt{2}) \,
    \nonumber \\
    &\times& \left[ (2\alpha+1)\, L_{2\alpha+1}^{2(\beta-\alpha)-1}(-2y^2)
    + 2 y^2\, L_{2\alpha-1}^{2(\beta-\alpha)+1}(-2y^2) \right]
\end{eqnarray}
and
\begin{eqnarray}
   \label{rho-odd-body}
    {\hat \varrho}_{\alpha,\,\beta}^{\rm odd} =
    {\hat \varrho}_{\alpha,\,\beta}^{\rm even} - (-4)^{m-\beta}
    \frac{m!}{(2m)!}\frac{(2\beta)!}{\beta!}\,
    {\hat \varrho}_{\alpha,\,m}^{\rm even}
\end{eqnarray}
for $n=2m$ even and $n=2m+1$ odd, respectively.

\subsection{Generating function for $p_{n,k}$}
\label{SubSec62}
Interestingly, the entire generating function
\begin{eqnarray}
\label{gf-body}
    G_n(z) = \sum_{\ell=0}^{\lfloor n/2 \rfloor} z^\ell p_{n,n-2\ell}
\end{eqnarray}
for the probabilities $p_{n,k}$ can explicitly be determined. To
proceed, we make use of the summation formula
\begin{eqnarray}
    \sum_{r=0}^\infty \frac{z^r}{r!}\, Z_{(1^r)}(p_1,\cdots, p_r) =
    \exp \left(
        \sum_{r \ge 1} (-1)^{r-1} \frac{p_r z^r}{r}
    \right)
\end{eqnarray}
well known in the theory of symmetric functions (Macdonald 1998).
With $p_r={\rm tr\,}_{(0,\lfloor n/2 \rfloor -1)} {\boldsymbol {\hat
        \varrho}}^r$, the r.h.s. transforms to
\begin{eqnarray} \fl
    \exp \left( {\rm tr\,}_{(0,\lfloor n/2 \rfloor -1)}
        \sum_{r \ge 1} (-1)^{r-1} \frac{ (z {\boldsymbol {\hat
        \varrho}})^r }{r}
    \right) &=
    \exp \Big( {\rm tr\,}_{(0,\lfloor n/2 \rfloor -1)}
        \log (\boldsymbol{\hat 1}_{\lfloor n/2 \rfloor} +  z {\boldsymbol {\hat
        \varrho}})
    \Big),\nonumber \\
    &= \det\, \left[
        \boldsymbol{\hat 1} +  z {\boldsymbol {\hat
        \varrho}}
    \right]_{{\lfloor n/2 \rfloor} \times {\lfloor n/2 \rfloor}},
\end{eqnarray}
resulting in an amazingly simple answer:
\begin{eqnarray}
\label{gf-body-2}
    G_n(z) = \sum_{\ell=0}^{\lfloor n/2 \rfloor} z^\ell p_{n,n-2\ell}
    = p_{n,n}\,
    \det\, \left[
        \boldsymbol{\hat 1} +  z {\boldsymbol {\hat
        \varrho}}
    \right]_{{\lfloor n/2 \rfloor} \times {\lfloor n/2 \rfloor}}.
\end{eqnarray}

\subsection{Integer moments of the number of real eigenvalues}
\label{SubSec63} The result (\ref{gf-body-2}) allows us to formally
determine any integer moment ${\mathbb E}[{\cal N}_r^q]$ of the
fluctuating number ${\cal N}_r$ of real eigenvalues in the spectra
of GinOE. Denoting the fluctuating number of complex eigenvalues
through $2{\cal N}_c$, we derive:
\begin{eqnarray}
\label{nrq}
    {\mathbb E}[{\cal N}_r^q] = {\mathbb E}[(n-2{\cal N}_c)^q] =
    \sum_{j=0}^q {q \choose j} n^{q-j} (-2)^j {\mathbb E}[{\cal N}_c^j].
\end{eqnarray}
Since
\begin{eqnarray}
   {\mathbb E}[{\cal N}_c^j]  = \left( z \frac{\partial}{\partial z}
   \right)^j \, G_n(z) \Big|_{z=1},
\end{eqnarray}
the formula (\ref{nrq}) simplifies to
\begin{eqnarray}
\label{im}
   {\mathbb E}[{\cal N}_r^q] = p_{n,n} \,\left(n-2 z \frac{\partial}{\partial
   z}\right)^q \,  \det\, \left[
        \boldsymbol{\hat 1} +  z {\boldsymbol {\hat
        \varrho}}
    \right]_{{\lfloor n/2 \rfloor} \times {\lfloor n/2
    \rfloor}}\Big|_{z=1}.
\end{eqnarray}
Of course, to make the formulae (\ref{gf-body-2}) and (\ref{im})
explicit, the determinant
\begin{eqnarray}
d_n(z)= \det\, \left[
        \boldsymbol{\hat 1} +  z {\boldsymbol {\hat
        \varrho}}
    \right]_{{\lfloor n/2 \rfloor} \times {\lfloor n/2 \rfloor}}
\nonumber
\end{eqnarray}
has to be evaluated in a closed form. There are a few indications
that this is a formidable task, but we have not succeeded in the
calculation of $d_n(z)$ yet.

\section[Asymptotic analysis of the probability $p_{n,n-2}$]{Asymptotic analysis of the
 probability $p_{n,n-2}$ to find exactly one pair of complex conjugated eigenvalues}
\label{Sec7} To determine a qualitative behaviour of the probability
function $p_{n,k}$, an asymptotic analysis of the exact solution
(\ref{St-1}) is needed. In this section, the simplest probability
function $p_{n,n-2}$,
\begin{eqnarray}
    \label{prob-7}
    p_{n,n-2} = 2\, p_{n,n}\,
    \int_0^\infty dy\, y \, e^{y^2}\, {\rm
    erfc}(y \sqrt{2}) \,
     L_{n-2}^2(-2y^2),
\end{eqnarray}
is studied in the large-$n$ limit. Our consideration is based on an
alternative exact representation for $p_{n,n-2}$ (see the Theorem
7.1) which is more suitable for obtaining regular large-$n$
asymptotics.

\subsection{Alternative exact representation of $p_{n,n-2}$}
\label{SubSec71}
Let us define the sequence
\begin{eqnarray}
\label{s7-1}
    S_n = \int_0^\infty dy\, y \, e^{y^2}\, {\rm
    erfc}(y \sqrt{2}) \,
     L_{n}^2(-2y^2), \;\;\; n=0,1,\cdots,
\end{eqnarray}
such that
\begin{eqnarray} \label{newex}
p_{n,n-2}=2 p_{n,n}\, S_{n-2}.
\end{eqnarray}
To find an exact alternative representation for $S_n$ (and, hence,
for $p_{n,n-2}$), we (i) introduce a generating function $\tau(z)$
in the form
\begin{eqnarray}
\label{s7-2}
    \tau(z) = \sum_{n=0}^\infty S_n z^n
\end{eqnarray}
which is supposed to exist in some domain $\Omega_\tau \in {\mathbb
R}$ of the real line ${\mathbb R}$ (to be specified later on), (ii)
calculate $\tau(z)$ explicitly, and (iii) expand it back in $z \in
\Omega_\tau$.
\newline\newline
{\bf Lemma 7.1.} {\it The generating function $\tau(z)$ reads:
\begin{eqnarray}
\label{s7-0}
    \tau(z) = \frac{1}{2(1-z^2)(1-z)} \left(
        -1 + \sqrt{2} \sqrt{\frac{1-z}{1-3z}}
    \right),
\end{eqnarray}
where
\begin{eqnarray}
-1< z < \frac{1}{3}.
\end{eqnarray}}
\newline
{\bf Proof.}  The identity \footnote[4]{See Eq. (5.11.2.1) in
Prudnikov, Brychkov and Marichev (1986).}
\begin{eqnarray}
    \sum_{n=0}^\infty L_n^{\lambda} (x) \, z^n =
    (1-z)^{-\lambda-1}\, \exp \left(
        \frac{xz}{z-1}
    \right), \;\;\; |z| < 1, \nonumber
\end{eqnarray}
applied in the context of (\ref{s7-1}) and (\ref{s7-2}), gives rise
to the representation
\begin{eqnarray}
    \label{s7-3}
    \tau(z) = \frac{1}{(1-z)^{3}} \, \int_0^\infty dy\, y\, {\rm erfc\,} (y\sqrt{2})\,
    \exp \left(y^2
        \frac{1+z}{1-z}
    \right), \;\;\; |z| < 1.
\end{eqnarray}
A change of the integration variable $y$ to
\begin{eqnarray}
    \xi = y^2 \frac{1+z}{1-z}
\end{eqnarray}
followed by integration by parts results in
\begin{eqnarray} \fl
\label{s7-4}
    \tau(z) = \frac{1}{2(1-z^2)(1-z)} \left[
        e^{\xi}\, {\rm erfc\,}
    \left(a_z\, \sqrt{\xi}
    \right) \Big|_{\xi=0}^{\xi = \infty} - \int_0^\infty d\xi \,
    e^\xi\, \frac{d}{d\xi} \, {\rm erfc\,}
    \left(
               a_z\, \sqrt{\xi}
    \right)
    \right],
\end{eqnarray}
where
\begin{eqnarray}
    a_z = \sqrt{2}\sqrt{\frac{1-z}{1+z}}. \nonumber
\end{eqnarray}
For the boundary term in (\ref{s7-4}) to nullify at $\xi=\infty$,
the parameter $z$ has to belong to the domain
\begin{eqnarray}
\label{s7-5}
    \Omega_\tau:\; \; -1< z < \frac{1}{3}.
\end{eqnarray}
Performing the remaining integral, we end up with (\ref{s7-0}). End
of proof.  $\;\;\;\blacksquare$
\newline\newline
Having determined the generating function $\tau(z)$, we are going to
Taylor-expand it around $z=0$ in order to arrive at an alternative
formula for the sequence $S_n$. As $\tau(z)$ is a relatively simple
function, we may expect that $S_n$ obtained in this way will also
have a relatively simple form.
\newline\newline
{\bf Lemma 7.2.} {\it The following formula holds:
\begin{eqnarray}
\label{s7-6}
    S_n =  \frac{1}{\sqrt{2}}\, \sum_{j=0}^{\lfloor n/2
\rfloor}
    3^{j+\alpha_n/2}\, P_{2j+\alpha_n}\left( \frac{2}{\sqrt{3}}
    \right) - \frac{1}{2}\Big( \lfloor n/2 \rfloor+1\Big).
\end{eqnarray}
Here, $\alpha_n = \lceil n/2 \rceil - \lfloor n/2 \rfloor$.}
\newline\newline
{\bf Proof.} To expand the function $\tau(z)$ given by (\ref{s7-0})
around $z=0$, we represent it in the form
\begin{eqnarray}
    \tau(z) = \frac{1}{2}\,\tau_1(z)\left( - 1  +
    \sqrt{2}\,\tau_2(z)\right),
\end{eqnarray}
where
\begin{eqnarray}
    \tau_1(z) = \frac{1}{(1-z^2)(1-z)},\;\;\; \tau_2(z) =
    \sqrt{\frac{1-z}{1-3z}},
\end{eqnarray}
and constantly use a variant of the Cauchy formula
\begin{eqnarray}
    \label{cauchy-f}
    \left( \sum_{k=0}^\infty a_k \,z^k \right)
    \cdot
    \left( \sum_{k=0}^\infty b_k \, z^k \right) =
    \sum_{n=0}^\infty c_n \, z^n, \;\;\;\; c_n = \sum_{k=0}^n a_k \,
    b_{n-k},
\end{eqnarray}
where absolute convergence of the resulting series is assumed.
\newline\newline
{\it Expansion of $\tau_1(z)$}. To determine the coefficients
$c_k^{(1)}$ in the expansion
\begin{eqnarray}
\tau_1(z) = \sum_{k=0}^\infty c_k^{(1)} z^k,
\end{eqnarray}
we notice that
\begin{eqnarray}
    \frac{1}{1-z} = \sum_{k=0}^\infty z^k, \;\;\;
    \frac{1}{1-z^2} = \sum_{k=0}^\infty z^{2k}, \;\;\;
\end{eqnarray}
so that, in the notation of (\ref{cauchy-f}),
\begin{eqnarray}
a_k=1, \;\;\;  b_k = \frac{1+(-1)^k}{2}.
\end{eqnarray}
Straightforward application of the Cauchy formula yields
\begin{eqnarray}
\label{ck1fin}
    c_k^{(1)} = \sum_{j=0}^k a_j b_{k-j} = \frac{1}{2} \left[
        k+1 + \frac{1+(-1)^k}{2}
    \right] = \lfloor k/2 \rfloor +1.
\end{eqnarray}
\newline\newline
{\it Expansion of $\tau_2(z)$}. To determine the coefficients
$c_k^{(2)}$ in the expansion
\begin{eqnarray}
\tau_2(z) = \sum_{k=0}^\infty c_k^{(2)} z^k,
\end{eqnarray}
we notice that
\begin{eqnarray} \fl
    \sqrt{1-z} = \sum_{k=0}^\infty  {1/2 \choose k}\, (-z)^k, \;\;\;
    \frac{1}{\sqrt{1-3 z}} = \sum_{k=0}^\infty {-1/2 \choose k}\, (-3z)^{k}, \;\;\;
\end{eqnarray}
so that, in the notation of (\ref{cauchy-f}),
\begin{eqnarray}
a_k={1/2 \choose k}\, (-1)^k, \;\;\;  b_k = {-1/2 \choose k} \,
(-3)^k.
\end{eqnarray}
The Cauchy formula yields
\begin{eqnarray}
\label{ck2}
    c_k^{(2)} = \sum_{j=0}^k a_j b_{k-j} =
    3^k \frac{(1/2)_k}{k!}\,\, {}_2 F_1\left(
    -\frac{1}{2}, -k; \frac{1}{2}-k; \frac{1}{3}
    \right).
\end{eqnarray}
The latter can be expressed in terms of Legendre polynomials by
means of the identity~\footnote[5]{See Eq. (7.3.1.153) in Prudnikov,
Brychkov and Marichev (1990)}
\begin{eqnarray} \fl
    {}_2 F_1\left(
    -\frac{1}{2}, -k; \frac{1}{2}-k; w
    \right) = \frac{k!}{(1/2)_k} \, w^{k/2} \left[
        P_k\left( \frac{w+1}{2\sqrt{w}} \right) - \sqrt{w}\,
        P_{k-1}\left( \frac{w+1}{2\sqrt{w}} \right)
    \right]
\end{eqnarray}
that simplifies (\ref{ck2}) to
\begin{eqnarray}
\label{ck2fin}
    c_k^{(2)} = 3^{k/2} P_k\left( \frac{2}{\sqrt{3}} \right) - 3^{(k-1)/2}
        P_{k-1}\left( \frac{2}{\sqrt{3}} \right).
\end{eqnarray}
\newline\newline
{\it Expansion of the product $\tau_1(z)\tau_2(z)$.} To determine
the coefficients $c_k^{(3)}$ in the expansion
\begin{eqnarray}
    \tau_1(z)\tau_2(z) = \sum_{k=0}^\infty c_k^{(3)} \, z^k,
\end{eqnarray}
we again use the Cauchy formula
\begin{eqnarray}
    c_k^{(3)} = \sum_{j=0}^k c_{k-j}^{(1)} c_{j}^{(2)},
\end{eqnarray}
with $c_k^{(1)}$ and $c_k^{(2)}$ given by (\ref{ck1fin}) and
(\ref{ck2fin}), respectively. Lengthy but straightforward
calculations result in
\begin{eqnarray}
    c_k^{(3)} = \sum_{j=0}^{\lfloor k/2 \rfloor}
    3^{j+\alpha_k/2} \, P_{2j+\alpha_k}\left(
        \frac{2}{\sqrt{3}}
    \right),
\end{eqnarray}
where $\alpha_k = \lceil k/2 \rceil - \lfloor k/2 \rfloor$.
\newline\newline
The observation
\begin{eqnarray}
    S_n = \frac{1}{\sqrt{2}}\, c_n^{(3)} - \frac{1}{2}\, c_n^{(1)}
\end{eqnarray}
completes the proof.   $\;\;\;\blacksquare$
\newline\newline
{\bf Theorem 7.1.} {\it The probability $p_{n,n-2}$ to find exactly
one pair of complex conjugated eigenvalues in spectra of GinOE
admits the following exact representation:
\begin{eqnarray}
\label{th71eq}
    p_{n,n-2} = p_{n,n} \,\left[ \sqrt{2}\, \sum_{j=0}^{\lfloor n/2 \rfloor -1}
    3^{j+\alpha_n/2}\, P_{2j+\alpha_n}\left( \frac{2}{\sqrt{3}}
    \right) - \lfloor n/2 \rfloor\right].
\end{eqnarray}
Here, $\alpha_n = \lceil n/2 \rceil - \lfloor n/2 \rfloor$, and
$P_n$ stands for Legendre polynomials.}
\newline\newline
{\bf Proof.} Use the Lemma 7.2 and relation (\ref{newex}) to deduce
(\ref{th71eq}).  $\;\;\;\blacksquare$

\subsection{Asymptotic analysis of $p_{n,n-2}$}
\label{SubSec72}
The result (\ref{th71eq}), combined with the
integral representation of Legendre polynomials
\begin{eqnarray}
\label{LP-int}
    P_n(\phi) = \frac{1}{\pi} \int_{0}^\pi d\theta \left(
        \phi + \sqrt{\phi^2-1} \, \cos\theta
    \right)^n,
\end{eqnarray}
is particularly useful for carrying out an asymptotic analysis of
the probability $p_{n,n-2}$ in the large-$n$ limit. Indeed,
(\ref{LP-int}) facilitates performing a summation in (\ref{th71eq})
leading to
\begin{eqnarray} \fl
\label{pJ}
    p_{n,n-2} = p_{n,n} \left[
        \frac{\sqrt{2}}{\pi} \int_{-1}^{+1} \frac{dx}{\sqrt{1-x^2}}\, \frac{(x+2)^{\alpha_n} - (x+2)^n}{1-(x+2)^2}
        - \lfloor n/2 \rfloor
    \right].
\end{eqnarray}
\newline\newline
The large-$n$ behaviour of the integral
\begin{eqnarray}
\label{Jdef}
    J_n = \int_{-1}^{+1} \frac{dx}{\sqrt{1-x^2}}\, \frac{(x+2)^{\alpha_n} - (x+2)^n}{1-(x+2)^2}
\end{eqnarray}
is of our primary interest. A saddle-point analysis shows that the
saddle point $x_{\rm sp}\approx -2$ lies away from the integration
domain $x \in (-1,+1)$. As a result, the contribution of the end
points of the integration domain, $x_{\rm L}=-1$ and $x_{\rm R}=+1$,
should be examined. One can see that the close vicinity $x=1
-\epsilon$ of $x_{\rm R}=+1$ dominates {\it exponentially} in $n \gg
1$. Indeed, the vicinity $\epsilon\in (0,c_0)$ yields
\begin{eqnarray}  \nonumber
    J_n \approx \int_{0}^{c_0} \frac{d\epsilon}{\sqrt{\epsilon
    (2-\epsilon)}}\, \frac{3^n (1-\epsilon/3)^n -
    3^{\alpha_n}(1-\epsilon/3)^{\alpha_n}}{3^2 (1-\epsilon/3)^2-1},
\end{eqnarray}
where $c_0$ is a proper cut-off. In the large-$n$ limit, only a
region of order $n^{-1}$, $\epsilon = \tau/n$, effectively
contributes $J_n$ reducing it to
\begin{eqnarray}
    J_n \approx \frac{3^n}{8\sqrt{2n}} \int_{0}^{\infty} \frac{d\tau}{\sqrt{\tau}}\,
    e^{-\tau/3} = \frac{3^{n+1/2}}{8} \sqrt{\frac{\pi}{2n}}.
\end{eqnarray}
Combined with (\ref{pJ}) and (\ref{Jdef}), this estimate leads to
the following theorem.
\newline\newline
{\bf Theorem 7.2.} {\it The leading large-$n$ behaviour of the
probability $p_{n,n-2}$ is given by the formula
\begin{eqnarray}
    p_{n,n-2} \approx \frac{3^{n+1/2}}{8\sqrt{\pi\,n}}\, p_{n,n}
\end{eqnarray}
where $p_{n,n}=2^{-n(n-1)/4}$.}
\newline\newline
{\bf Remark 7.1.} The Theorem 7.2 implies the inequality $p_{n,n-2}
\gg p_{n,n}$.
\newline
\section{Correlations of complex eigenvalues of a matrix without real eigenvalues}
\label{Sec8}
\subsection{GOE correlations in GinOE spectra}
One of the earliest results on eigenlevel statistics in GinOE is due
to Ginibre (1965) who spotted that spectra of random real matrices
which happened to have {\it no complex eigenvalues} exhibit the
famous GOE behaviour. Indeed, for ${\boldsymbol {\cal H}}_n \in
{\mathbb T}(n/n)$, the j.p.d.f. (\ref{b=1}) reduces to
\footnote[6]{As a side remark, we notice that the explicit formula
(\ref{ed-res}) for the probability $p_{n,n}$ can easily be derived
by integrating $P_{{\boldsymbol {\cal H}}_n}$ over all of its
arguments. Due to Selberg's integral (Mehta 2004)
\begin{eqnarray}
    \prod_{j=1}^n \int_{\mathbb R} d\lambda_j \,
    e^{-\lambda_j^2/2}\,
    \prod_{i>j=1}^n |\lambda_i-\lambda_j| = 2^{n/2} n!\,
    \prod_{j=1}^n \Gamma(j/2), \nonumber
\end{eqnarray}
one obtains $p_{n,n}=2^{-n(n-1)/4}$. }
\begin{eqnarray} \label{gin-real}
\fl
    P_{{\boldsymbol {\cal H}}_n}
    (\lambda_1,\cdots,\lambda_n)
    &=
    \frac{2^{- n(n+1)/4}}{n! \,\prod_{j=1}^n \Gamma(j/2)}
    \prod_{i>j=1}^n |\lambda_i - \lambda_j|
    \, \prod_{j=1}^n \, \exp(-\lambda_j^2/2).
\end{eqnarray}
The GOE spectral correlations readily follow (Mehta 2004).

\subsection{GinSE-like correlations in GinOE spectra}
Below, we concentrate on just the opposite case of random real
matrices ${\boldsymbol {\cal H}}_0 \in {\mathbb T}(n/0)$ whose
spectrum occasionally contains {\it no real eigenvalues}. The
j.p.d.f. of all complex eigenvalues of ${\boldsymbol {\cal H}}_0$
can also be deduced from (\ref{b=1}), the result being
\begin{eqnarray} \label{gin-comp}
    \fl
    P_{{\boldsymbol {\cal H}}_0}
    (z_1,\cdots,z_\ell)
    &=
    \frac{2^{- n(n-1)/4} \, i^{n/2}}{(n/2)!\, \prod_{j=1}^n \Gamma(j/2)}\,
    \prod_{i>j=1}^{\ell} |z_i - z_j|^2 |z_i - {\bar
    z}_j|^2\,
     \nonumber \\
    &\times  \prod_{j=1}^\ell \, ({\bar z}_j-z_j)\,
     {\rm erfc} \left(
        \frac{z_j-{\bar z}_j}{i\sqrt{2}}
    \right)\,
    \exp\left(-\frac{z_j^2+{\bar z}_j^2}{2}\right)
\end{eqnarray}
with $n$ even, $n=2\ell$. Remarkably, while the above j.p.d.f.
resembles the j.p.d.f. of complex eigenvalues in GinSE
(\ref{jpdf=4}), it is manifestly different from the latter.

Is it possible to determine the correlation functions for the new
complex eigenvalue model (\ref{gin-comp})? The answer is positive.
\newline\newline
{\bf Lemma 8.1.} {\it Let ${\boldsymbol {\cal H}_0}$ be an $n\times
n$ random real matrix with no real eigenvalues such that its entries
are statistically independent random variables picked from a normal
distribution} $\textsf{N}(0,1)$. {\it Then, the $p$-point
correlation function ($1\le p\le \ell$) of its complex eigenvalues
equals}
\begin{eqnarray}\fl \label{rp-def}
R^{({\boldsymbol {\cal H}}_0)}_{0,p}(z_1,\cdots, z_p;n)
= p_{n,n}\, \frac{\prod_{j=0}^{\ell-1} r_j}
    {\prod_{j=1}^n \Gamma(j/2)}
    \prod_{j=1}^p {\rm erfc}\left( \frac{z_j-{\bar
    z}_j}{i\sqrt{2}}\right)\, \exp\left(
    -\frac{z_j^2+{\bar z}_j^2}{2}
    \right) \nonumber \\
    \times\, {\rm pf\,}
    \left[
    \begin{array}{cc}
      \kappa_{\ell}(z_i, z_j) & \kappa_{\ell}(z_i, {\bar z}_j) \\
      \kappa_{\ell}({\bar z}_i, z_j) & \kappa_{\ell}({\bar z}_i, {\bar z}_j) \\
    \end{array}
    \right]_{2p \times 2p}.
\end{eqnarray}
{\it Here, $n=2\ell$ and the `pre-kernel' $\kappa_\ell$ is
\begin{eqnarray}
\label{kzw-www}
    \kappa_\ell (z,z^\prime) = i \sum_{j=0}^{\ell-1} \frac{1}{r_j} \bigg[
        p_{2j}(z) p_{2j+1}(z^\prime) - p_{2j}(z^\prime)
    p_{2j+1}(z)
    \bigg].
\end{eqnarray}
The polynomials $p_j(z)$ in (\ref{kzw-www}) are skew orthogonal in
the complex half-plane ${\rm Im\,}z >0$,
\begin{eqnarray} \label{sp1}
    \langle p_{2j+1}, p_{2k} \rangle_{\rm c} = - \langle p_{2k}, p_{2j+1}
    \rangle_{\rm
    c} = i\,r_j \, \delta_{jk},  \\
    \label{sp2}
    \langle p_{2j+1}, p_{2k+1} \rangle_{\rm c} = \langle p_{2j}, p_{2k}
    \rangle_{\rm
    c}=0,
\end{eqnarray}
with respect to the skew product
\begin{eqnarray} \fl \label{csp}
    \langle f, g \rangle_{\rm c} = \int_{{\rm Im\,}z>0} d^2 z\,
    {\rm erfc} \left(
        \frac{z-{\bar z}}{i\sqrt{2}}
    \right)\,
    \exp\left(-\frac{z^2+{\bar z}^2}{2}\right)\,
    \left[
        f(z)g({\bar z}) - f({\bar z})g(z)
    \right].
\end{eqnarray}}
\newline\newline
{\bf Proof.} By definition (\ref{rpb1}), the $p$-point correlation
function is
\begin{eqnarray}\fl \label{rp}
R^{({\boldsymbol {\cal H}}_0)}_{0,p}(z_1,\cdots, z_p;n) =
    \frac{\ell!}{(\ell-p)!} \prod_{j=p+1}^\ell \int_{{\rm Im}z_j >0}
     d^2 z_j \, P_{{\boldsymbol {\cal H}}_0}
    (z_1,\cdots,z_\ell).
\end{eqnarray}
Since the $n \times n$ real matrix with no real eigenvalues has
$\ell$ pairs of complex conjugated eigenvalues, it holds that
$n=2\ell$.

Conceptually, the proof to be presented consists of three parts.
First, we concentrate on the j.p.d.f. $P_{{\boldsymbol {\cal H}}_0}
    (z_1,\cdots,z_\ell)$ and show that it
    can be represented in terms of a certain quaternion determinant.
    Second, we prove that the quaternion matrix under the quaternion determinant enjoys the projection property (see
Definition
    2.1). Third, we apply the Dyson integration theorem to carry out
    all $(\ell-p)$ integrations in (\ref{rp}).
\newline\newline
{\it Part 1.}---As the Vandermonde structure of (\ref{gin-comp})
mimics that of GinSE (\ref{jpdf=4}), it is tempting to employ the
identity
\begin{eqnarray} \fl
   \prod_{i>j=1}^{\ell} |z_i - z_j|^2 |z_i - {\bar
    z}_j|^2\, \prod_{j=1}^\ell ({\bar{z}_j - z_j}) = \det \big[
    z_j^{i-1}, \, {\bar z}_j^{i-1}
    \big]_{
    \matrix
    {
    \scriptstyle i=1,\cdots,2\ell\hfill \cr
    \scriptstyle j=1,\cdots,\ell\hfill\cr
    }
        }
\end{eqnarray}
that helps us reduce the j.p.d.f. $P_{{\boldsymbol {\cal H}}_0}$ to
the form
\begin{eqnarray} \label{pp1}
    \fl
    P_{{\boldsymbol {\cal H}}_0}
(z_1,\cdots,z_\ell)
    &= \frac{2^{- n(n-1)/4} \, i^{\ell}}{(n/2)!\, \prod_{j=1}^n \Gamma(j/2)}\,
    \det \big[
    p_{i-1}(z_j), \, p_{i-1}({\bar z}_j)
    \big]_{
    \matrix
    {
    \scriptstyle i=1,\cdots,2\ell\hfill \cr
    \scriptstyle j=1,\cdots,\ell\hfill\cr
    }
        }
    \nonumber \\
        &\times
        \prod_{j=1}^\ell
    {\rm erfc} \left(
        \frac{z_j-{\bar z}_j}{i\sqrt{2}}
    \right)\,
    \exp\left(-\frac{z_j^2+{\bar z}_j^2}{2}\right).
\end{eqnarray}
Here, $p_k(z)$ are {\it arbitrary} monic polynomials of degree $k$.

The very structure of the matrix under the determinant in
(\ref{pp1}),
\begin{eqnarray} \nonumber
    \det\,
    \left[
    \begin{array}{cccccc}
    p_0(z_1) & p_0({\bar z}_1) & \cdots & \cdots & p_0(z_\ell) & p_0({\bar
    z}_\ell) \\
    p_1(z_1) & p_1({\bar z}_1) & \cdots & \cdots & p_1(z_\ell) & p_1({\bar
    z}_\ell) \\
    \vdots & \vdots & {} & {} & \vdots & \vdots \\
    \vdots & \vdots & {} & {} & \vdots & \vdots \\
    p_{2\ell-2}(z_1) & p_{2\ell-2}({\bar z}_1) & \cdots & \cdots &
    p_{2\ell-2}(z_\ell) & p_{2\ell-2}({\bar z}_\ell) \\
    p_{2\ell-1}(z_1) & p_{2\ell-1}({\bar z}_1) & \cdots & \cdots &
    p_{2\ell-1}(z_\ell) & p_{2\ell-1}({\bar z}_\ell) \\
    \end{array}
    \right]
\end{eqnarray}
suggests that we introduce a set of quaternions $\{ {\boldsymbol
\psi}_0(z),\cdots, {\boldsymbol \psi}_{\ell-1}(z)\}$,
\begin{eqnarray}
    {\boldsymbol \psi}_j(z) &= \frac{p_{2j}(z)+p_{2j+1}(\bar{z})}{2}
    \, {\boldsymbol {\hat e}_0} +
    \frac{p_{2j}(z)-p_{2j+1}(\bar{z})}{2i}
    \, {\boldsymbol {\hat e}_1} \nonumber\\
    &+
    \frac{p_{2j}(\bar{z})-p_{2j+1}(z)}{2}
    \, {\boldsymbol {\hat e}_2}
    +
    \frac{p_{2j}(\bar{z})+p_{2j+1}(z)}{2i}
    \, {\boldsymbol {\hat e}_3}
\end{eqnarray}
whose $2\times 2$ matrix representation reads \footnote[7]{The
$2\times 2$ matrices ${\boldsymbol {\hat e}}_j$ are defined as
follows:
\begin{eqnarray}
    {\boldsymbol {\hat e}}_0 = \left(
                          \begin{array}{cc}
                            1 & 0 \\
                            0 & 1 \\
                          \end{array}
                        \right), \;\;\;
    {\boldsymbol {\hat e}}_1 = \left(
                          \begin{array}{cc}
                            i & 0 \\
                            0 & -i \\
                          \end{array} \right), \;\;\;
    {\boldsymbol {\hat e}}_2 = \left(
                          \begin{array}{cc}
                            0 & 1 \\
                            -1 & 0 \\
                          \end{array} \right), \;\;\;
    {\boldsymbol {\hat e}}_3 = \left(
                          \begin{array}{cc}
                            0 & i \\
                            i & 0 \\
                          \end{array} \right). \nonumber
\end{eqnarray}
}
\begin{eqnarray}
    \Theta[{\boldsymbol \psi}_j(z)] = \left[
                    \begin{array}{cc}
                      p_{2j}(z) & p_{2j}({\bar z}) \\
                      p_{2j+1}(z) & p_{2j+1}({\bar z}) \\
                    \end{array}
                  \right].
\end{eqnarray}
As a result, the above determinant can equivalently be written as
\begin{eqnarray} \label{det-3}
    \det\,
    \left[ p_{i-1}(z_j), \, p_{i-1}({\bar z}_j)\right]_{
    \matrix
    {
    \scriptstyle i=1,\cdots,2\ell\hfill \cr
    \scriptstyle j=1,\cdots,\ell\hfill\cr
    }
        } = \det\, \left[
            \Theta[{\boldsymbol \psi}_{i-1}(z_j)]
        \right]_{
    \matrix
    {
    \scriptstyle i=1,\cdots,\ell\hfill \cr
    \scriptstyle j=1,\cdots,\ell\hfill\cr
    }
        }.
\end{eqnarray}
The latter can be put into a quaternion determinant form (see
Corollary 5.1.3 in Mehta's book (2004)):
\begin{eqnarray} \fl \label{qd}
        \det\, \left[
            \Theta[{\boldsymbol \psi}_{i-1}(z_j)]
        \right]_{
    \matrix
    {
    \scriptstyle i=1,\cdots,\ell\hfill \cr
    \scriptstyle j=1,\cdots,\ell\hfill\cr
    }
        } = {\rm qdet\,}\left[
            \boldsymbol{{\cal A}\bar{{\cal A}}}\,
        \right]_{\ell\times \ell} = {\rm qdet\,}\left[
            \boldsymbol{\bar{{\cal A}}{\cal A}}\,
        \right]_{\ell\times \ell}.
\end{eqnarray}
Here, $\boldsymbol{{\cal A}}$ is an $\ell \times \ell$ quaternion
matrix with the entries
\begin{eqnarray}
    \boldsymbol{{\cal A}}_{ij} = {\boldsymbol \psi}_{i-1}(z_j)
\end{eqnarray}
and $\boldsymbol{{\bar{\cal A}}}$ is the dual quaternion matrix
whose entries
\begin{eqnarray}
    \boldsymbol{\bar{{\cal A}}}_{ij} = {\boldsymbol {\bar{\psi}}}_{j-1}(z_i)
\end{eqnarray}
are determined by the dual quaternion $\bar{{\boldsymbol
\psi}}_j(z)$,
\begin{eqnarray}
    {\bar {\boldsymbol \psi}}_j(z) &= \frac{p_{2j}(z)+p_{2j+1}(\bar{z})}{2}
    \, {\boldsymbol {\hat e}_0} -
    \frac{p_{2j}(z)-p_{2j+1}(\bar{z})}{2i}
    \, {\boldsymbol {\hat e}_1} \nonumber\\
    &-
    \frac{p_{2j}(\bar{z})-p_{2j+1}(z)}{2}
    \, {\boldsymbol {\hat e}_2}
    -
    \frac{p_{2j}(\bar{z})+p_{2j+1}(z)}{2i}
    \, {\boldsymbol {\hat e}_3}
\end{eqnarray}
such that
\begin{eqnarray}
    \Theta[{\boldsymbol {\bar \psi}}_j(z)] =
    \boldsymbol{{\hat e}}_2 \, \Theta[{\boldsymbol \psi}_j(z)]^{\rm T}\,
    \boldsymbol{{\hat e}}_2^{-1} =
    \left[
                    \begin{array}{cc}
                      p_{2j+1}(\bar{z}) & - p_{2j}({\bar z}) \\
                      - p_{2j+1}(z) & p_{2j}(z) \\
                    \end{array}
                  \right].
\end{eqnarray}
Combining (\ref{det-3}) and (\ref{qd}) into
\begin{eqnarray}
        \det\,
    \left[ p_{i-1}(z_j), \, p_{i-1}({\bar z}_j)\right]_{
    \matrix
    {
    \scriptstyle i=1,\cdots,2\ell\hfill \cr
    \scriptstyle j=1,\cdots,\ell\hfill\cr
    }
        } = {\rm qdet\,}\left[
            \boldsymbol{\bar{{\cal A}}{\cal A}}\,
        \right]_{\ell\times \ell},
\end{eqnarray}
we obtain
\begin{eqnarray}
    \det\,
    \left[ p_{i-1}(z_j), \, p_{i-1}({\bar z}_j)\right]_{
    \matrix
    {
    \scriptstyle i=1,\cdots,2\ell\hfill \cr
    \scriptstyle j=1,\cdots,\ell\hfill\cr
    }
        } = i^{-\ell} \prod_{j=0}^{\ell-1} r_j\; {\rm qdet} \left[
            {\boldsymbol {\hat \kappa}}_\ell(z_i,z_j)
        \right]_{\ell\times \ell}.
\end{eqnarray}
Here, the (self-dual) {\it quaternion kernel} ${\boldsymbol {\hat
\kappa}}_\ell$ admits a $2 \times 2$ matrix representation
\begin{eqnarray}
    \Theta \left[  {\boldsymbol {\hat \kappa}}_\ell(z_i,z_j) \right]
    =
    \left(
      \begin{array}{cc}
        -\kappa_\ell({\bar z}_i, z_j) & -\kappa_\ell({\bar z}_i, {\bar z}_j) \\
        \kappa_\ell(z_i, z_j) & \kappa_\ell(z_i, \bar{z}_j) \\
      \end{array}
    \right)
\end{eqnarray}
with
\begin{eqnarray}
\label{kzw}
    \kappa_\ell (z,w) = i \sum_{j=0}^{\ell-1} \frac{1}{r_j} \bigg[
        p_{2j}(z) p_{2j+1}(w) - p_{2j}(w)
    p_{2j+1}(z)
    \bigg].
\end{eqnarray}
The set of constants $\{r_j\}$ is not fixed so far.

The above consideration results in the following expression for the
j.p.d.f. $P_{{\boldsymbol {\cal H}}_0}$ [Eq. (\ref{pp1})]
\begin{eqnarray} \label{pp1-new}
    \fl
    P_{{\boldsymbol {\cal H}}_0}
(z_1,\cdots,z_\ell)
    &= \frac{2^{- n(n-1)/4}}{(n/2)!}\,\frac{\prod_{j=0}^{\ell-1} r_j}{\prod_{j=1}^n
    \Gamma(j/2)} \;
    {\rm qdet} \left[
            {\boldsymbol {\hat \kappa}}_\ell(z_i,z_j)
        \right]_{\ell\times \ell}
    \nonumber \\
        &\times
        \prod_{j=1}^\ell
    {\rm erfc} \left(
        \frac{z_j-{\bar z}_j}{i\sqrt{2}}
    \right)\,
    \exp\left(-\frac{z_j^2+{\bar z}_j^2}{2}\right),
\end{eqnarray}
that serves as a proper starting point for evaluating the $p$-point
correlation function $R^{{({\boldsymbol {\cal
H}}_0)}}_{0,p}(z_1,\cdots, z_p;n)$ specified in Eq. (\ref{rp-def}).
\newline\newline
{\it Part 2.}---Now, we are going to prove that the quaternion
${\boldsymbol {\hat \kappa}}_\ell$ satisfies the projection property
Definition 2.1. To simplify the consideration to follow, we set the
so far arbitrary monic polynomials $p_k(z)$ to be skew-orthogonal in
the complex half-plane, ${\rm Im\,} z >0$, as defined by (\ref{sp1})
-- (\ref{csp}).

Having imposed the skew-orthogonality on $p_k(z)$, with $r_j \in
{\mathbb R}$, we are in the position to verify whether or not the
projection property for the quaternion ${\boldsymbol {\hat
\kappa}}_\ell$ is fulfilled. In accordance with the Definition 2.1,
one has to consider the integral
\begin{eqnarray} \fl
    I = \int_{{\rm Im\,}w>0} d^2w\,
        {\rm erfc} \left(
        \frac{w-{\bar w}}{i\sqrt{2}}
    \right)\,
    \exp\left(-\frac{w^2+{\bar w}^2}{2}\right)\,
    \Theta \left[  {\boldsymbol {\hat \kappa}}_\ell(z_1,w) \right]
    \, \Theta \left[  {\boldsymbol {\hat \kappa}}_\ell(w,z_2)
    \right]
\end{eqnarray}
that equals
\begin{eqnarray}
    I
    =     \left(
      \begin{array}{cc}
        \delta_\ell({\bar z}_1, z_2) & \delta_\ell({\bar z}_1, {\bar z}_2) \\
        - \delta_\ell(z_1, z_2) & -\delta_\ell(z_1, \bar{z}_2) \\
      \end{array}
    \right).
\end{eqnarray}
Here, the function $\delta_\ell(z_1,z_2)$ is defined by the integral
\begin{eqnarray} \fl
    \delta_\ell(z_1, z_2) = \int_{{\rm Im\,}w>0} d^2w\,
        {\rm erfc} \left(
        \frac{w-{\bar w}}{i\sqrt{2}}
    \right)\,
    \exp\left(-\frac{w^2+{\bar w}^2}{2}\right)\, \nonumber \\
    \times
    \big[
        \kappa_\ell(z_1,w) \kappa_\ell({\bar w},z_2) - \kappa_\ell
        (z_2,w) \kappa_\ell({\bar w},z_1)
    \big].
\end{eqnarray}
Its evaluation, based on (\ref{kzw}), (\ref{csp}), (\ref{sp1}) and
(\ref{sp2}), is straightforward, the result being
\begin{eqnarray}
    \delta_\ell(z_1,z_2) = -\kappa_\ell(z_1, z_2)
\end{eqnarray}
so that
\begin{eqnarray}
I = \Theta\left[
    {\boldsymbol {\hat \kappa}}_\ell(z_1,z_2)
\right].
\end{eqnarray}
Put differently, the {\it quaternion kernel} ${\boldsymbol {\hat
\kappa}}_\ell$ satisfies the projection property in the form
\begin{eqnarray} \fl
\label{pp-gin}
    \int_{{\rm Im\,}w>0} d^2w\,
        {\rm erfc} \left(
        \frac{w-{\bar w}}{i\sqrt{2}}
    \right)\,
    \exp\left(-\frac{w^2+{\bar w}^2}{2}\right)\,
    {\boldsymbol {\hat \kappa}}_\ell(z_1,w)\, {\boldsymbol {\hat
    \kappa}}_\ell(w,z_2) = {\boldsymbol {\hat
    \kappa}}_\ell(z_1,z_2).
\end{eqnarray}
This is precisely (\ref{dt-2}) of the Dyson integration theorem with
a quaternion $\lambda=0$.
\newline\newline
{\it Part 3.}---The above proof of the projection property for the
quaternion kernel ${\boldsymbol {\hat \kappa}}_\ell$ in
(\ref{pp1-new}) paves the way for carrying out the $(\ell-p)$
integrations in (\ref{rp}). Indeed, the integrations therein can be
performed by virtue of the Dyson integration theorem (Theorem 2.1)
since
\begin{eqnarray}
    \Theta\left[{\bar {\boldsymbol {\hat \kappa}}}_\ell (z_1,z_2)\right] \stackrel{\rm def}{=}
    {\boldsymbol {\hat e}}_2 \, \Theta\left[
        {\boldsymbol {\hat \kappa}}_\ell (z_1,z_2)
    \right]^{\rm T} {\boldsymbol {\hat e}}_2^{-1} = \Theta\left[
        {\boldsymbol {\hat \kappa}}_\ell (z_2,z_1)
    \right].
\end{eqnarray}
To this end, one has to determine the constant $c$ defined by the
integral [see (\ref{dt-4})]
\begin{eqnarray} \fl
    c\, {\boldsymbol e_0} = \int_{{\rm Im\,}z>0} d^2z\,
        {\rm erfc} \left(
        \frac{z-{\bar z}}{i\sqrt{2}}
    \right)\,
    \exp\left(-\frac{z^2+{\bar z}^2}{2}\right)\,
    {\boldsymbol {\hat \kappa}}_\ell(z,z)
\end{eqnarray}
yielding
\begin{eqnarray} \fl
\label{c-value}
    c = \int_{{\rm Im\,}z>0} d^2z\,
        {\rm erfc} \left(
        \frac{z-{\bar z}}{i\sqrt{2}}
    \right)\,
    \exp\left(-\frac{z^2+{\bar z}^2}{2}\right)\,
    {\kappa}_\ell(z,\bar{z}) = \ell.
\end{eqnarray}
The projection property (\ref{pp-gin}) combined with the result
(\ref{c-value}) brings the key integration identity:
\begin{eqnarray} \fl
\label{ki}
    \frac{1}{(\ell-p)!}\prod_{j=p+1}^\ell \int_{{\rm Im\,}z_j>0} d^2z_j\,
        {\rm erfc} \left(
        \frac{z_j-{\bar z}_j}{i\sqrt{2}}
    \right)\,
    \exp\left(-\frac{z_j^2+{\bar z}_j^2}{2}\right)\,
    {\rm qdet} \left[
            {\boldsymbol {\hat \kappa}}_\ell(z_i,z_j)
        \right]_{\ell\times \ell}  \nonumber \\
        = {\rm qdet} \left[
            {\boldsymbol {\hat \kappa}}_\ell(z_i,z_j)
        \right]_{p \times p} =
        {\rm pf\,} \left[
       \begin{array}{cc}
      {\kappa}_{\ell}(z_i, z_j) & {\kappa}_{\ell}(z_i, {\bar z}_j) \\
      {\kappa}_{\ell}({\bar z}_i, z_j) & {\kappa}_{\ell}({\bar z}_i, {\bar
      z}_j) \\
    \end{array}
    \right]_{2p \times 2p}.
\end{eqnarray}
Applied to (\ref{rp}) and (\ref{pp1-new}), it results in
\begin{eqnarray} \fl
\label{rp-10}
R^{({\boldsymbol {\cal H}}_k)}_{0,p}(z_1, \cdots, z_p; n) =
    p_{n,n}\,\frac{\prod_{j=0}^{\ell-1} r_j}
    {\prod_{j=1}^n \Gamma(j/2)}
    \;     \prod_{j=1}^p {\rm erfc}\left( \frac{z_j-{\bar
    z}_j}{i\sqrt{2}}\right)\, \exp\left(
    -\frac{z_j^2+{\bar z}_j^2}{2}
    \right) \nonumber \\
    \times
        {\rm pf\,} \left[
       \begin{array}{cc}
      {\kappa}_{\ell}(z_i, z_j) & {\kappa}_{\ell}(z_i, {\bar z}_j) \\
      {\kappa}_{\ell}({\bar z}_i, z_j) & {\kappa}_{\ell}({\bar z}_i, {\bar
      z}_j) \\
    \end{array}
    \right]_{2p \times 2p},
\end{eqnarray}
where $n=2\ell$. This completes the proof of the Lemma 8.1.
$\;\;\;\blacksquare$

\subsection{Probability $p_{n,0}$ to find no real eigenvalues}
The technique exposed in the previous subsection allows us to
establish the following structural result for the probability to
find no real eigenvalues in GinOE spectra.
\newline\newline
{\bf Corollary 8.1.} {\it The probability $p_{n,0}$ to find no real
eigenvalues in spectra of GinOE equals}
\begin{eqnarray}
\label{pn0-res}
 p_{n,0} = \frac{p_{n,n}}{\prod_{j=1}^n \Gamma(j/2)}\, \prod_{j=0}^{\ell-1} r_j.
\end{eqnarray}
{\it Here, $n=2\ell$; the constants $r_j$ are defined by
(\ref{sp1}).}
\newline\newline
{\bf Proof.} For $n=2\ell$ even, the definition (\ref{pnk})
translates to
\begin{eqnarray}
    \label{pn0}
    p_{n,0} =
    \prod_{j=1}^\ell \int_{{\rm Im\,}z_j > 0} d^2 z_j\,
P_{{\boldsymbol {\cal H}_0}}(z_1,\cdots,z_\ell).
\end{eqnarray}
Given $P_{{\boldsymbol {\cal H}_0}}$ in the form
(\ref{pp1-new}), the above probability reduces to
\begin{eqnarray} \label{pn00}
    \fl
    p_{n,0} = \frac{p_{n,n}}{\ell!}\,\frac{ \prod_{j=0}^{\ell-1}
    r_j}{\prod_{j=1}^n
    \Gamma(j/2)}\,
    \prod_{j=1}^\ell \int_{{\rm Im\,}z_j > 0} d^2 z_j \;
    {\rm qdet} \left[
            {\boldsymbol {\hat \kappa}}_\ell(z_i,z_j)
        \right]_{\ell\times \ell}
    \nonumber \\
        \times
        \prod_{j=1}^\ell
    {\rm erfc} \left(
        \frac{z_j-{\bar z}_j}{i\sqrt{2}}
    \right)\,
    \exp\left(-\frac{z_j^2+{\bar z}_j^2}{2}\right).
\end{eqnarray}
Due to the projection property (\ref{pp-gin}) of the quaternion kernel ${\boldsymbol {\hat \kappa}}_\ell$,
the integration in (\ref{pn00}) can readily be performed [see, e.g., the identity (\ref{ki})] bringing
\begin{eqnarray} \fl
    \prod_{j=1}^\ell \int_{{\rm Im\,}z_j > 0} d^2 z_j \;
    {\rm qdet} \left[
            {\boldsymbol {\hat \kappa}}_\ell(z_i,z_j)
        \right]_{\ell\times \ell}
       \, \prod_{j=1}^\ell
    {\rm erfc} \left(
        \frac{z_j-{\bar z}_j}{i\sqrt{2}}
    \right)\,
    \exp\left(-\frac{z_j^2+{\bar z}_j^2}{2}\right) = \ell!
\end{eqnarray}
This establishes the result (\ref{pn0-res}) thus completing the
proof. $\;\;\;\blacksquare$
\newline\newline
{\bf Remark 8.1.} To make the formula (\ref{pn0-res}) explicit, one
has to know the normalisation constants $r_j$ which we have failed
to determine explicitly so far.

\section{Conclusions and open problems}
To summarise, an exact formula was derived for the probability
$p_{n,k}$ to find precisely $k$ real eigenvalues in the spectrum of
an $n\times n$ random matrix drawn from GinOE. Based on the Pfaffian
integration theorem (that can be seen as an extension of the Dyson
integration theorem to kernels that do not possess the projection
property), the solution found expresses the probability function
$p_{n,k}$ in terms of zonal polynomials \footnote[8]{To the best of
our knowledge, this is the first ever random-matrix-theory {\it
observable} admitting a representation in terms of symmetric
functions.}. This links the integrable structure of GinOE to the
theory of symmetric functions (Macdonald 1998). Undoubtedly, much
more effort is needed to accomplish the spectral theory of GinOE.
Below, we list some of the open problems that have to be addressed.

\textsc{\newline (I)~Probability Function $p_{n,k}$ and Associated
Generating Function $G_n(z)$.} The exact solution for the
probability $p_{n,k}$ expresses the probability in terms of zonal
polynomials of some complicated arguments. Does a more explicit
representation (like the one for $p_{n,n-2}$, see (\ref{prob-7}) and
(\ref{th71eq})) exist for $p_{n,k}$? A similar question arises in
the context of the solution for the determinantal generating
function $G_n(z)$ given by (\ref{gf-body-2}). Can the determinant
(\ref{gf-body-2}) be calculated in a closed form as a function of
$k$, $n$ and $z$? We have a few indications that this is a
formidable task.

\textsc{\newline (II)~Asymptotic Analysis of $p_{n,k}$ in the
Large-$n$ Limit.} The problem we wish to pose here concerns the
large-$n$ behaviour of $p_{n,k}$ in various scaling limits: (a) $k
\sim n^0$, (b) $n-k \sim n^0$, and (c) $k \sim n^{1/2}$ (this
scaling is prompted by the result (\ref{EnAs}) for an average number
of real eigenvalues). If available, the large-$n$ formulae of this
kind would facilitate a comparison of our exact theory with existing
numerical and experimental data, as reported \footnote[1]{The
numerics by Halasz {\it et al} (1997) refers to the chiral
counterparts of the Ginibre ensembles.} by Halasz {\it et al} (1997)
and Kwapie\'{n} {\it et al} (2000).

\textsc{\newline (III)~Correlation Functions of Complex Eigenvalues
and $p_{n,0}$.} The solution presented for these two spectral
characteristics involves specific polynomials which are skew
orthogonal in the complex plane with respect to a somewhat unusual
weight function containing the complementary error function. Can
those be determined explicitly to eventually bring closed formulae
for both the correlation functions and the probability $p_{n,0}$?
Their large-$n$ analysis would be of great interest, too.

\textsc{\newline (IV)~Correlation Functions of Both Real and Complex
Eigenvalues and a Generalised Pfaffian Integration Theorem.} The
calculation of all partial $(p,q)$-point correlation functions
$R_{p,q}^{({\boldsymbol {\cal H}}_k)}$ for GinOE matrices with a
given number $(k)$ of real eigenvalues (defined as an integral of
the j.p.d.f. over all but $p$ real and $q$ complex eigenvalues, see
(\ref{rpb1})) is yet another important problem to tackle. Also, can
the unconditional $(p,q)$-point correlation functions $R_{p,q} =
\sum_{k} R_{p,q}^{({\boldsymbol {\cal H}}_k)}$ be explicitly
determined?
 We believe that progress in this
direction can be achieved through a proper extension of the Pfaffian
integration theorem:
\begin{eqnarray} \fl
    \prod_{j=p+1}^\ell \, \int_{z_j\in {\mathbb C}} d\pi(z_j)\,\,
    &{\rm pf\,} \left[
    \begin{array}{cc}
      Q_{n}(z_i, z_j) & Q_{n}(z_i, {\bar z}_j) \\
      Q_{n}({\bar z}_i, z_j) & Q_{n}({\bar z}_i, {\bar z}_j) \\
    \end{array}
    \right]_{2\ell \times 2\ell}
=\; ?
\end{eqnarray}
Here, the notation of the Theorem 5.1 was used. Notice that one
should not assume the projection property for the kernel $Q_n$.
\newline\newline
The above list of open problems calls for further research of GinOE that will eventually unveil
the rich mathematical structures underlying this classical but still largely unexplored non-Hermitean random matrix model.

\section*{Acknowledgements}
\addcontentsline{toc}{section}{Acknowledgements} The authors
appreciate a clarifying correspondence with A. Borodin (Caltech), V.
B. Kuznetsov (Leeds) and G. Olshanski (IITP, Moscow) and useful
discussions with V.~Al.~Osipov (H.I.T.). A part of this work was
done during the visits to Brunel University West London (E.K.),
University of Warwick (E.K. and G.A.), and H.I.T. -- Holon Institute
of Technology (G.A.) supported by a BRIEF grant from Brunel
University. The work of G.A. is supported by EU network ENRAGE
MRTN-CT-2004-005616 and by EPSRC grant EP/D031613/1. The research of
E.K. is supported by the Israel Science Foundation through the grant
No 286/04.
\newpage \fancyhead{} \fancyhead[RE,LO]{Appendix A}
\fancyhead[LE,RO]{\thepage}

\appendix
\section*{Appendix A. The $n \times n$ matrix ${\boldsymbol {\hat \chi}}$}
\label{ap-a}
\renewcommand{\theequation}{A.\arabic{equation}}
\addcontentsline{toc}{section}{Appendix A. The $n\times n$ matrix
${\boldsymbol {\hat \chi}}$} {\it A1. Symmetries of ${\boldsymbol
{\hat \chi}}$ and useful expansions.}---The definition
(\ref{chi-matrix}) of the matrix ${\boldsymbol{\hat \chi}}$ suggests
that it is antisymmetric,
\begin{eqnarray}
\label{sr}
    {\hat \chi}_{jk} = - {\hat \chi}_{kj},
\end{eqnarray}
and purely imaginary
\begin{eqnarray}
{\bar {\hat
    \chi}}_{jk} = - {\hat \chi}_{jk}
\end{eqnarray}
as confirmed by the formula
\begin{eqnarray} \fl
    \label{chi-red}
    {\hat \chi}_{jk} = i  \int d\alpha(w) \, e^{-(w^2+{\bar w}^2)/2}
    \, \left[
       \,{\rm Re\,} q_j(w)\, {\rm Im\,} q_k({\bar w})
       + {\rm Im\,} q_j(w)\, {\rm Re\,} q_k({\bar w})
    \right].
\end{eqnarray}
Due to (\ref{skew-hermite}) that relates the skew orthogonal
polynomials $q_j(w)$ to Hermite polynomials $H_j(w)$, the following
expansions\footnote[2]{\texttt{http://functions.wolfram.com/05.01.19.0001.01}
and \texttt{.../05.01.19.0002.01}} are useful for even/odd order
Hermite polynomials taken at $w=x+iy$:
\begin{eqnarray} \fl
    \label{Hermite-complex-01}
    {\rm Re\,}
    H_{2m}(w) = \sum_{j=0}^{m}
    \frac{(-1)^j 2^{2j}}{(2j)!}\, (-2m)_{2j} \, y^{2j} H_{2m-2j}(x),
    \\ \fl \label{Hermite-complex-02}
    {\rm Im\,}
    H_{2m}(w) =
    \sum_{j=0}^{m-1}
    \frac{(-1)^{j-1} 2^{2j+1}}{(2j+1)!} \, (-2m)_{2j+1} \, y^{2j+1}
    H_{2m-2j-1}(x),
\end{eqnarray}
and
\begin{eqnarray}
\fl \label{Hermite-complex-03}
    {\rm Re\,}
    H_{2m+1}(w) =
    \sum_{j=0}^{m}
    \frac{(-1)^j 2^{2j}}{(2j)!}\, (-2m-1)_{2j} \, y^{2j}
    H_{2m+1-2j}(x), \\ \fl
    \label{Hermite-complex-04}
    {\rm Im \,}  H_{2m+1}(w) = \sum_{j=0}^{m}
    \frac{(-1)^{j-1} 2^{2j+1}}{(2j+1)!} \, (-2m-1)_{2j+1} \, y^{2j+1}
    H_{2m-2j}(x).
\end{eqnarray}
Here, $(-a)_n$ is the Pochhammer symbol ($a > 0$),
\begin{eqnarray}
    (-a)_n = (-1)^n \frac{\Gamma(a+1)}{\Gamma(a-n+1)}. \nonumber
\end{eqnarray}
\newline
{\it A2. Calculation of the matrix elements ${\hat
\chi}_{2\alpha,2\beta}$ and ${\hat
\chi}_{2\alpha+1,2\beta+1}$.}---We claim that
\begin{eqnarray}
\label{chi-even-0}
    {\hat \chi}_{2\alpha,2\beta}={\hat \chi}_{2\alpha+1,2\beta+1}=0
\end{eqnarray}
for all $\alpha=0,1,\cdots$ and $\beta=0,1,\cdots$. To prove this
statement, let us consider ${\hat \chi}_{2\alpha,2\beta}$. In
accordance with (\ref{chi-red}), this matrix element is related to
the integral containing
\begin{eqnarray}
e^{-(w^2+{\bar w}^2)/2}\,\Big[ {\rm Re\,} q_{2\alpha}(w)\, {\rm
Im\,} q_{2\beta}({\bar w})+ {\rm Im\,} q_{2\alpha}(w)\, {\rm Re\,}
q_{2\beta}({\bar w})\Big]. \nonumber
\end{eqnarray}
Having in mind (\ref{skew-hermite}), (\ref{Hermite-complex-01}) and
(\ref{Hermite-complex-02}), we observe that the above expression,
being integrated over the $x$-part of the integration measure
\begin{eqnarray} \label{meas}
    d\alpha(w) = dx\, \cdot{\rm erfc}(y\sqrt{2}) \,\theta(y) \, dy,
\end{eqnarray}
nullifies due to the products of Hermite polynomials $H_{{\rm
even}\,\natural}(x) \, H_{{\rm odd}\,\natural}(x)$ of even and odd
orders. Thus, we conclude that ${\hat \chi}_{2\alpha,2\beta}=0$. By
the same token, the matrix element ${\hat
\chi}_{2\alpha+1,2\beta+1}$ is zero as well.
\newline\newline
{\it A3. Calculation of the matrix elements ${\hat
\chi}_{2\alpha+1,2\beta}$ and ${\hat
\chi}_{2\alpha,2\beta+1}$.}---We claim that
\begin{eqnarray}
    \label{chi-ans} \fl
    {\hat \chi}_{2\alpha+1,2\beta} =
    \frac{i}{2}\, (-1)^{\beta-\alpha} h_\alpha \,
    \int_0^\infty dy \,y^{2(\beta-\alpha)-1}\, e^{y^2}\, {\rm
    erfc\,}(y\sqrt{2}) \,
    \nonumber \\
    \times \left[ (2\alpha+1)\, L_{2\alpha+1}^{2(\beta-\alpha)-1}(-2y^2)
    + 2 y^2\, L_{2\alpha-1}^{2(\beta-\alpha)+1}(-2y^2) \right].
\end{eqnarray}
In accordance with the symmetry relation (\ref{sr}), it holds
\begin{eqnarray}
\label{chi-sym}
    {\hat \chi}_{2\alpha,2\beta+1} = - {\hat
    \chi}_{2\beta+1,2\alpha}.
\end{eqnarray}
To prove the result (\ref{chi-ans}), we start with the definition
(\ref{chi-red}),
\begin{eqnarray} \fl
    \label{chi-red-1}
    {\hat \chi}_{2\alpha+1,2\beta} = i  \int d\alpha(w) \, e^{-(w^2+{\bar w}^2)/2}
    \, \nonumber \\
    \times \left[
       \,{\rm Re\,} q_{2\alpha+1}(w)\, {\rm Im\,} q_{2\beta}({\bar w})
       + {\rm Im\,} q_{2\alpha+1}(w)\, {\rm Re\,} q_{2\beta}({\bar w})
    \right],
\end{eqnarray}
where the skew orthogonal polynomials $q_j(w)$ are given by
(\ref{skew-hermite}). Substituting them into (\ref{chi-red-1}),
representing $w$ as $w=x+iy$ and making use of the expansions
(\ref{Hermite-complex-01}) to (\ref{Hermite-complex-04}), one can
carry out the integration over $x \in {\mathbb R}$
straightforwardly, due to the factorised integration measure
(\ref{meas}) and the orthogonality of Hermite polynomials on
${\mathbb R}$ with respect to the weight $\exp(-x^2)$. Lengthy but
straightforward calculations result in
\begin{eqnarray}
    \label{chi-red-2}
    {\hat \chi}_{2\alpha+1,2\beta} =
    \frac{i\, \alpha!}{2} \left( {\tilde \gamma}_{\alpha,\beta} - {\tilde \gamma}_{\alpha-1,\beta}
    \right)
\end{eqnarray}
where ${\tilde \gamma_{\alpha,\beta}}$ is
\begin{eqnarray} \fl \label{a14}
    {\tilde \gamma}_{\alpha,\beta} = (-1)^{\beta-\alpha} \,
    (2\alpha+1)\,
    \frac{h_\alpha}{\alpha!}\,
    \int_0^\infty dy \, y^{2(\beta-\alpha)-1}\, e^{y^2}\, {\rm erfc\,}(y\sqrt{2})
     \nonumber
    \\
    \times (2\beta)!\, \sum_{j=0}^{2\alpha+1} \frac{(2y^2)^j}{j!\, (2\alpha+1-j)! \, (j+2(\beta-\alpha)
    -1)!}.
\end{eqnarray}
Notice that for $\alpha\le -1$, the sum in (\ref{a14}) is void so
that ${\tilde \gamma}_{\alpha\le-1,\beta}=0$. Otherwise, the series
can be summed up in terms of Laguerre polynomials,
\begin{eqnarray}
    \label{sumup} \fl
    (2\beta)!\, \sum_{j=0}^{2\alpha+1} \frac{(2y^2)^j}{j!\, (2\alpha+1-j)! \, (j+2(\beta-\alpha)
    -1)!} = L_{2\alpha+1}^{2(\beta-\alpha)-1}(-2y^2),
\end{eqnarray}
to yield
\begin{eqnarray} \fl
\label{gamma-ab}
{\tilde \gamma}_{\alpha,\beta} =
(-1)^{\beta-\alpha} \,
    (2\alpha+1)\,
    \frac{h_\alpha}{\alpha!}\,
    \int_0^\infty dy \,y^{2(\beta-\alpha)-1}\, e^{y^2}\, {\rm erfc\,}(y\sqrt{2}) \,
    L_{2\alpha+1}^{2(\beta-\alpha)-1}(-2y^2).
\end{eqnarray}
From now on, the Laguerre polynomials of ``negative order'' are
interpreted to be zeros. Substituting it back to (\ref{chi-red-2}),
we end up with (\ref{chi-ans}).
\newline\newline Finally, the formula for ${\hat {\chi}}_{2\alpha,
2\beta+1}$ follows from the symmetry relation (\ref{chi-sym}).
\newline\newline
{\it A4. The structure of ${\boldsymbol {\hat \chi}}$.}---To
summarise the calculations of Appendix A, the structure of the $n
\times n$ matrix ${\boldsymbol {\hat \chi}}$ is presented below. For
$n=2m$ even, the matrix ${\boldsymbol {\hat \chi}}$ is
\begin{eqnarray} \fl
\label{chi-even-str}
    {\hat \chi}_{ij}^{\rm even} =  \left(
      \begin{array}{ccccc}
        0 & {\hat \chi}_{01} &  \cdots & 0 & {\hat \chi}_{0,2m-1}\\
        -{\hat \chi}_{01} & 0 &  \cdots & {\hat \chi}_{1,2m-2} & 0\\
        \vdots & \vdots & \ddots & \vdots & \vdots\\
        0 & -{\hat \chi}_{1,2m-2} & \cdots & 0 & {\hat \chi}_{2m-2,2m-1}\\
        -{\hat \chi}_{0,2m-1} & 0  & \cdots & -{\hat \chi}_{2m-2,2m-1} & 0\\
        \end{array}
    \right).
\end{eqnarray}
For $n=2m+1$ odd, the structure of ${\boldsymbol {\hat \chi}}$
follows the pattern
\begin{eqnarray} \fl
\label{chi-odd-str}
      {\hat \chi}_{ij}^{\rm odd} =  \left(
      \begin{array}{ccccc}
        0 & {\hat \chi}_{01} &  \cdots & {\hat \chi}_{0,2m-1} & 0\\
        -{\hat \chi}_{01} & 0 &  \cdots & 0 & {\hat \chi}_{1,2m}\\
        \vdots & \vdots & \ddots & \vdots & \vdots\\
        -{\hat \chi}_{0,2m-1} & 0 & \cdots & 0 & {\hat \chi}_{2m-1,2m}\\
        0 & -{\hat \chi}_{1,2m}  & \cdots & -{\hat \chi}_{2m-1,2m} & 0\\
        \end{array}
    \right).
\end{eqnarray}
\newline
\fancyhead{} \fancyhead[RE,LO]{Appendix B}
\fancyhead[LE,RO]{\thepage}
\section*{Appendix B. The $n \times n$ matrix $\boldsymbol{{\hat
      \sigma}}=2i\,\boldsymbol{ ({\hat \mu}{\hat \chi})}$}
\label{ap-b} \setcounter{equation}{0}
\renewcommand{\theequation}{B.\arabic{equation}}
\addcontentsline{toc}{section}{Appendix B. The $n\times n$ matrix
$\boldsymbol{{\hat \sigma}}=2i\,\boldsymbol{ ({\hat \mu}{\hat
\chi})}$} {\it B1. Calculation of the entries ${\hat
\sigma}_{\alpha,\beta}=2i\,({\hat \mu}{\hat \chi})_{\alpha,\beta}$
for $n$ even.}---The matrices ${\boldsymbol {\hat \mu}}$,
${\boldsymbol {\hat \chi}}$ and thus $\boldsymbol{ ({\hat \mu}{\hat
\chi})}$ are sensitive to the parity of $n$, the matrix
dimensionality. For $n=2m$, we make use of (\ref{mu-even}),
(\ref{chi-even-0}), (\ref{chi-sym}) and (\ref{chi-even-str}) to
derive
\begin{eqnarray} \fl
    \label{chi-mu-even}
        ({\hat \mu}{\hat \chi})_{2\alpha,2\beta+1}^{\rm even} =
        ({\hat \mu}{\hat \chi})_{2\alpha+1,2\beta}^{\rm even} =
        0, \\
        \fl \label{b2}
        ({\hat \mu}{\hat \chi})_{2\alpha, 2\beta}^{\rm even} =
        -\frac{1}{h_\alpha}\, {\hat \chi}_{2\alpha+1,2\beta}=
        \frac{1}{h_\alpha}\, {\hat \chi}_{2\beta,2\alpha+1}, \\
        \fl \label{b3}
        ({\hat \mu}{\hat \chi})_{2\alpha+1, 2\beta+1}^{\rm even} =
        \frac{1}{h_\alpha}\, {\hat \chi}_{2\alpha,2\beta+1} =
        - \frac{1}{h_\alpha}\, {\hat \chi}_{2\beta+1,2\alpha} =
        \frac{h_\beta}{h_\alpha}\, ({\hat \mu}{\hat
        \chi})_{2\beta,2\alpha}^{\rm even}.
\end{eqnarray}
Here, $\alpha=0,1,\cdots$ and $\beta=0,1,\cdots$.

An integral representation for the entries ${\hat \sigma}^{\rm
even}_{2\alpha,2\beta}$ and ${\hat \sigma}^{\rm
even}_{2\alpha+1,2\beta+1}$ can be read off from (\ref{chi-ans}).
Explicitly, we have
\begin{eqnarray}
    \label{mu-chi-ans-even} \fl
    {\hat \sigma}_{2\alpha,2\beta}^{\rm even} =  (-1)^{\beta-\alpha}
    \int_0^\infty dy \,y^{2(\beta-\alpha)-1}\, e^{y^2}\, {\rm
    erfc\,}(y\sqrt{2}) \,
    \nonumber \\
    \times \left[ (2\alpha+1)\, L_{2\alpha+1}^{2(\beta-\alpha)-1}(-2y^2)
    + 2 y^2\, L_{2\alpha-1}^{2(\beta-\alpha)+1}(-2y^2) \right]
\end{eqnarray}
and
\begin{eqnarray}
    \label{mu-chi-and-odd}
        {\hat \sigma}_{2\alpha+1, 2\beta+1}^{\rm even} =
        \frac{h_\beta}{h_\alpha}\, {\hat \sigma}_{2\beta,2\alpha}^{\rm even}.
\end{eqnarray}
All other entries are zeros.

Yet another representation in terms of
${\tilde\gamma}_{\alpha,\beta}$ given by (\ref{gamma-ab}) is useful.
Equations (\ref{chi-red-2}), (\ref{b2}) and (\ref{b3}) yield:
\begin{eqnarray}
\label{g1}
      {\hat \sigma}_{2\alpha, 2\beta}^{\rm even} &=
      \frac{\alpha!}{h_\alpha} \, \left(
        {\tilde \gamma}_{\alpha,\beta} - {\tilde \gamma}_{\alpha-1,\beta}
      \right), \\
\label{g2}
      {\hat \sigma}_{2\alpha+1, 2\beta+1}^{\rm even} &=
        \frac{\beta!}{h_\alpha} \, \left(
        {\tilde \gamma}_{\beta,\alpha} - {\tilde \gamma}_{\beta-1,\alpha}
      \right).
\end{eqnarray}
All other entries nullify in accordance with (\ref{chi-mu-even}).
\newline\newline
{\it B2. Calculation of the entries ${\hat
\sigma}_{\alpha,\beta}=2i\,({\hat \mu}{\hat \chi})_{\alpha,\beta}$
for $n$ odd.}---For $n=2m+1$, we make use of (\ref{mu-odd}),
(\ref{chi-even-0}), (\ref{chi-sym}), (\ref{chi-odd-str}), (\ref{b2})
and (\ref{b3}) to derive
\begin{eqnarray} \fl
    \label{chi-mu-odd}
        ({\hat \mu}{\hat \chi})_{2\alpha,2\beta+1}^{\rm odd} =
        ({\hat \mu}{\hat \chi})_{2\alpha+1,2\beta}^{\rm odd} =
        0, \\
        \fl \label{bbb2}
        ({\hat \mu}{\hat \chi})_{2\alpha, 2\beta}^{\rm odd} =
        (1-\delta_{\alpha,m}) ({\hat \mu}{\hat \chi})_{2\alpha, 2\beta}^{\rm
        even}
        -
        \frac{m!}{h_m} \, \delta_{\alpha,m} \, \sum_{j=0}^{m-1}
        \frac{h_j}{j!} \, ({\hat \mu}{\hat \chi})_{2j, 2\beta}^{\rm
        even},\\
        \fl \label{bbb3}
        ({\hat \mu}{\hat \chi})_{2\alpha+1, 2\beta+1}^{\rm odd} =
        \frac{h_\beta}{h_\alpha}\,
        \left[
            ({\hat \mu}{\hat \chi})_{2\beta, 2\alpha}^{\rm even} -
            \frac{m!}{h_m}\,\frac{h_\alpha}{\alpha!}
            ({\hat \mu}{\hat \chi})_{2\beta, 2m}^{\rm even}
        \right].
\end{eqnarray}
Here, $\alpha=0,1,\cdots$ and $\beta=0,1,\cdots$.

The above formulae simplify if expressed in terms of ${\tilde
\gamma}_{\alpha,\beta}$ akin to (\ref{g1}) and (\ref{g2}).
Straightforward calculations lead to the following result for the
entries of ${\boldsymbol {\hat \sigma}}^{\rm odd}$:
\begin{eqnarray} \fl
    \label{g3}
    {\hat \sigma}_{2\alpha,2\beta}^{\rm odd} =
     \frac{\alpha!}{h_\alpha} \Big[
    (1-\delta_{\alpha,m}) ({\tilde \gamma}_{\alpha,\beta} - {\tilde \gamma}_{\alpha-1,\beta})
    - \delta_{\alpha,m} \, {\tilde \gamma}_{m-1,\beta} \Big], \\
    \label{g4} \fl
    {\hat \sigma}_{2\alpha+1,2\beta+1}^{\rm odd} =
     \frac{\beta !}{\alpha !}
    \Big[
        \frac{\alpha!}{h_\alpha} ({\tilde \gamma}_{\beta,\alpha} - {\tilde \gamma}_{\beta-1,\alpha})
        -
        \frac{m!}{h_m} ({\tilde \gamma}_{\beta,m} - {\tilde \gamma}_{\beta-1,m})
    \Big].
\end{eqnarray}
Finally, the integral representations for (\ref{g3}) and (\ref{g4})
can be obtained from (\ref{gamma-ab}).

\section*{Appendix C. The $\lfloor n/2 \rfloor\times \lfloor n/2 \rfloor$
matrix ${\boldsymbol{\hat{\varrho}}}$ and the trace identity}
\label{ap-c} \fancyhead{} \fancyhead[RE,LO]{Appendix C}
\fancyhead[LE,RO]{\thepage} \setcounter{equation}{0}
\renewcommand{\theequation}{C.\arabic{equation}}
\addcontentsline{toc}{section}{Appendix C. The $\lfloor n/2
\rfloor\times \lfloor n/2 \rfloor$ matrix
${\boldsymbol{\hat{\varrho}}}$ and the trace identity} {\it C1. The
definition of ${\boldsymbol {\hat \varrho}}$.}---Since the $n\times
n$ matrix $\boldsymbol{{\hat \sigma}}$ has half of its entries
vanishing, it is useful to define a reduced, $\lfloor n/2 \rfloor
\times \lfloor n/2 \rfloor$ matrix ${\boldsymbol {\hat \varrho}}$,
such that
\begin{eqnarray}
\label{d1}
    {{\hat \varrho}}_{\alpha,\beta}^{{\rm even}} =
    (-1)^{\beta-\alpha}\,{\hat \sigma}_{2\alpha, 2\beta}^{\rm even}
    = (-1)^{\beta-\alpha} \frac{\alpha !}{h_\alpha}\, \left(
    {\tilde \gamma}_{\alpha,\beta} - {\tilde \gamma}_{\alpha-1,\beta}
    \right)
\end{eqnarray}
for $n=2m$ even, and
\begin{eqnarray} \fl
\label{d2}
    {{\hat \varrho}}_{\alpha,\beta}^{{\rm odd}} = (-1)^{\beta-\alpha}
    \frac{h_\beta}{h_\alpha}\,{\hat \sigma}_{2\beta+1, 2\alpha+1}^{\rm odd}
    = (-1)^{\beta-\alpha}
    \left[
            {\hat \sigma}_{2\alpha, 2\beta}^{\rm even} -
            \frac{m!}{h_m}\,\frac{h_\beta}{\beta!}
            {\hat \sigma}_{2\alpha, 2m}^{\rm even}
        \right] \nonumber \\
        =
           {\hat \varrho}_{\alpha, \beta}^{\rm even} -
            (-1)^{\beta-m}\, \frac{m!}{h_m}\,\frac{h_\beta}{\beta!}
            {\hat \varrho}_{\alpha, m}^{\rm even}
\end{eqnarray}
for $n=2m+1$ odd. Explicitly,
\begin{eqnarray}
\label{rho-even-app}
    {{\hat \varrho}}_{\alpha,\beta}^{{\rm even}} &=&
    \int_0^\infty dy \,y^{2(\beta-\alpha)-1}\, e^{y^2}\, {\rm
    erfc\,}(y\sqrt{2}) \,
    \nonumber \\
    &\times& \left[ (2\alpha+1)\, L_{2\alpha+1}^{2(\beta-\alpha)-1}(-2y^2)
    + 2 y^2\, L_{2\alpha-1}^{2(\beta-\alpha)+1}(-2y^2) \right]
\end{eqnarray}
and
\begin{eqnarray}
   \label{rho-odd-app}
    {\hat \varrho}_{\alpha,\,\beta}^{\rm odd} =
    {\hat \varrho}_{\alpha,\,\beta}^{\rm even} - (-4)^{m-\beta}
    \frac{m!}{(2m)!}\frac{(2\beta)!}{\beta!}\,
    {\hat \varrho}_{\alpha,\,m}^{\rm even}.
\end{eqnarray}
The above formulae have been obtained from the definitions
(\ref{d1}) and (\ref{d2}) with the help of (\ref{mu-chi-ans-even})
and (\ref{hj}).
\newline\newline
{\it C2. The trace identity.}---It turns out that the probability
function $p_{n,k}$ is most economically expressed in terms of
${\boldsymbol {\hat \varrho}}$. To realise this, it is instructive
to prove the following trace identity.
\newline\newline
{\bf Lemma C1.} {\it Let ${\boldsymbol {\hat \sigma}}$ be an $n
\times n$ matrix with the entries given by either (\ref{g1}),
(\ref{g2}) or (\ref{g3}), (\ref{g4}) depending on the parity of $n$.
Then, the trace identity
\begin{eqnarray}
\label{trid}
    {\rm tr\,}_{(0,n-1)} {\boldsymbol {\hat \sigma}}^j = 2\,
    {\rm tr\,}_{(0,\lfloor n/2 \rfloor -1)} {\boldsymbol {\hat \varrho}}^j
\end{eqnarray}
holds for all $j=1,2,\cdots$ and the matrix ${\boldsymbol {\hat
\varrho}}$ defined by (\ref{d1}) and (\ref{d2}).}
\newline\newline
{\bf Proof.} Since the matrix elements ${\hat
\sigma}_{2\alpha,2\beta+1}$ and ${\hat \sigma}_{2\alpha+1,2\beta}$
are zeros, the trace of ${\boldsymbol {\hat \sigma}}^j$
($j=1,2,\cdots$) can always be separated into two pieces:
\begin{eqnarray}
\label{sab}
    {\rm tr\,}_{(0,n-1)} {\boldsymbol {\hat \sigma}}^j =
    {\rm tr\,}_{(0,\lceil n/2 \rceil -1)} {\boldsymbol {\hat a}\,}^j
    +
    {\rm tr\,}_{(0,\lfloor n/2 \rfloor -1)} {\boldsymbol {\hat
    b}\,}^j.
\end{eqnarray}
The matrices ${\boldsymbol {\hat a}}$ and ${\boldsymbol {\hat b}}$
are defined through their entries,
\begin{eqnarray}
    \label{af}
    {\hat a}_{\alpha,\beta} = {\hat \sigma}_{2\alpha,2\beta},
    \\
    \label{bf}
    {\hat b}_{\alpha,\beta} = {\hat \sigma}_{2\alpha+1,2\beta+1},
\end{eqnarray}
and explicitly depend on the parity of $n$ in accordance with the
discussion in Appendix B.
\newline\newline
(i) {\it The case $n=2m$ even} is the simplest one because of the
relation (\ref{mu-chi-and-odd}). Indeed, its straightforward use
results in
\begin{eqnarray}
\label{c9} {\rm tr\,}_{(0,m -1)} {\boldsymbol {\hat
    b}\,}^j =
 {\rm tr\,}_{(0,m -1)} {\boldsymbol {\hat
    a}\,}^j.
\end{eqnarray}
This can further be simplified due to(\ref{c9}), (\ref{af}),
(\ref{sab}) and (\ref{d1}),
\begin{eqnarray}
    {\rm tr\,}_{(0,2m - 1)} {\boldsymbol {\hat \sigma}}^j =
    2\,
    {\rm tr\,}_{(0,m -1)} {\boldsymbol {\hat
    a}\,}^j = 2\,
    {\rm tr\,}_{(0,m -1)} {{\boldsymbol {\hat
    \varrho}\,}^{\rm even}}^j.
\end{eqnarray}
This completes the proof of (\ref{trid}) for $n=2m$ even.
\newline\newline
(ii) {\it The case $n=2m+1$ odd} is a bit more complicated since a
simple analogue of (\ref{mu-chi-and-odd}) does not exist. Instead,
we have [(\ref{af}) and (\ref{bf})]
\begin{eqnarray}
\label{aodd}
    {\hat a}_{\alpha,\beta} =
     c_\alpha \,\Big[
    (1-\delta_{\alpha,m}) ({\tilde \gamma}_{\alpha,\beta} - {\tilde \gamma}_{\alpha-1,\beta})
    - \delta_{\alpha,m} \, {\tilde \gamma}_{m-1,\beta} \Big], \\
    \label{bodd}
    {\hat b}_{\alpha,\beta} =
        c_\alpha\, ({\tilde \gamma}_{\beta,\alpha} - {\tilde \gamma}_{\beta-1,\alpha})
        -
        c_m\, ({\tilde \gamma}_{\beta,m} - {\tilde
        \gamma}_{\beta-1,m}),
\end{eqnarray}
where $c_\alpha=\alpha!/h_\alpha$ has been defined in (\ref{cm}). In
writing (\ref{bodd}), we dropped the prefactor $\beta !/\alpha !$
appearing in (\ref{g4}) since it does not affect the value of the
second trace in (\ref{sab}). To prove (\ref{trid}), we will start
with (\ref{sab}) in order to demonstrate that
\begin{eqnarray}
\label{c91} {\rm tr\,}_{(0,m -1)} {\boldsymbol {\hat
    b}\,}^j =
 {\rm tr\,}_{(0,m)} {\boldsymbol {\hat
    a}\,}^j.
\end{eqnarray}
This will be followed by a proof that either of these traces reduces
to $ {\rm tr\,}_{(0, m -1)} {\boldsymbol {\hat \varrho}}^{{\rm
odd\;}j}$.

Let us prove (\ref{c91}) by focusing on the eigenvalues
$\{\lambda\}$ of the matrices ${\boldsymbol {\hat a}}$ and
${\boldsymbol {\hat b}}$ which are the roots of the secular
equations
\begin{eqnarray}
    \label{secular-a}
    \det\, \big[{\boldsymbol {\hat a}} - \lambda\, {\boldsymbol {\hat 1}}
    \big]_{(m+1)\times (m+1)} = 0, \\
    \label{secular-b}
    \det\, \big[{\boldsymbol {\hat b}} - \lambda\, {\boldsymbol {\hat 1}}
    \big]_{m \times m}=0.
\end{eqnarray}
We claim that (\ref{c91}) holds because exactly {\it one} out of
$(m+1)$ eigenvalues of the matrix ${\boldsymbol {\hat a}}$ is {\it
zero} whilst the remaining $m$ eigenvalues of ${\boldsymbol {\hat
a}}$ coincide with $m$ eigenvalues of ${\boldsymbol {\hat b}}$. Put
differently, we are going to prove that
\begin{eqnarray}
    \det\, \big[{\boldsymbol {\hat a}} - \lambda\, {\boldsymbol {\hat 1}}
    \big]_{(m+1)\times (m+1)} = - \lambda\,
    \det\, \big[{\boldsymbol {\hat b}} - \lambda\, {\boldsymbol {\hat 1}}
    \big]_{m \times m}.
\end{eqnarray}
The proof consists of four steps.
\newline\newline
Step $\natural\, 1$. Consider the $(m+1)\times (m+1)$ matrix
${\boldsymbol {\hat a}}-\lambda {\boldsymbol {\hat 1}}$ under the
determinant in the secular equation (\ref{secular-a}),
\begin{eqnarray} \fl
    \left(
      \begin{array}{ccccc}
        c_0\, {\tilde \gamma}_{0,0} -\lambda & c_0\, {\tilde \gamma}_{0,1}
        &\cdots & c_0\, {\tilde \gamma}_{0,m-1}&
        c_0\, {\tilde \gamma}_{0,m} \\
        c_1\, \Gamma_{1,0} & c_1\, \Gamma_{1,1}-\lambda & \cdots
        & c_1 \, \Gamma_{1,m-1}&
        c_1\, \Gamma_{1,m} \\
        \vdots & \vdots & \ddots & \vdots & \vdots \\
        c_{m-1}\, \Gamma_{m-1,0} & c_{m-1}\, \Gamma_{m-1,1} &\cdots
        & c_{m-1}\, \Gamma_{m-1,m-1} - \lambda &
        c_{m-1}\, \Gamma_{m-1,m} \\
        - c_{m}\, {\tilde \gamma}_{m-1,0} & -c_m\, {\tilde \gamma}_{m-1,1}& \cdots &
        -c_m\, {\tilde \gamma}_{m-1,m-1}&
        - c_{m}\, {\tilde \gamma}_{m-1,m}-\lambda \\
      \end{array}
    \right), \nonumber \\ {} \label{det1}
\end{eqnarray}
where $\Gamma_{\alpha,\beta}={\tilde \gamma}_{\alpha,\beta}-{\tilde
\gamma}_{\alpha-1,\beta}$. Let us perform a number of operations
with rows and columns that will leave the value of the secular
determinant intact. First, we multiply the content of the first row
by $c_1/c_0$ and add it to the second row. Having done this, we
multiply a new content of the second row by $c_2/c_1$ and add it to
the content of the third row. We go on with this procedure until we
arrive at the modified $m$th row whose content is multiplied by
$c_m/c_{m-1}$ and further added to the last, $(m+1)$th, row. While
not affecting a value of the determinant in (\ref{secular-a}), the
above sequence of transformations brings (\ref{det1}) to the form
\begin{eqnarray} \fl
    \left(
      \begin{array}{cccc}
        c_0\, {\tilde \gamma}_{0,0} -\lambda & c_0\, {\tilde \gamma}_{0,1} & \cdots & c_0\, {\tilde \gamma}_{0,m} \\
        c_1\, {\tilde \gamma}_{1,0}-(c_1/c_0)\lambda  & c_1\, {\tilde \gamma}_{1,1}-\lambda & \cdots
        &
        c_1\, {\tilde \gamma}_{1,m} \\
        \vdots & \vdots & \ddots  & \vdots \\
        c_{m-1}\, {\tilde \gamma}_{m-1,0}-(c_{m-1}/c_0)\lambda & c_{m-1}\, {\tilde \gamma}_{m-1,1}-(c_{m-1}/c_1)\lambda &\cdots
        & c_{m-1}\, {\tilde \gamma}_{m-1,m} \\
        -(c_{m}/c_0)\lambda & -(c_m/c_1)\lambda &\cdots &
        -(c_{m}/c_m)\lambda \\
      \end{array} \right). \nonumber \\ {} \label{det2}
\end{eqnarray}
The last row of this equation suggests that a factor $\lambda$ can
be taken out of the secular determinant. In other words, $\lambda=0$
is always an eigenvalue of the $(m+1)\times (m+1)$ matrix
${\boldsymbol {\hat a}}$.
\newline\newline
Step $\natural \,2$. Next, we multiply the content of the first row
in (\ref{det2}) by $c_1/c_0$ and subtract it from the content of the
second row; having done that, we multiply the content of the
modified second row by $c_2/c_1$ and subtract it from the third row;
going on with this procedure, we arrive at the $(m-1)$th row,
multiply it by $c_{m-1}/c_{m-2}$ to subtract this from the content
of the $m$th row. We do not touch the last, $(m+1)$th row. This set
of transformations yields
\begin{eqnarray} \fl
    \left(
      \begin{array}{ccccc}
        c_0\, {\tilde \gamma}_{0,0} -\lambda & c_0\, {\tilde \gamma}_{0,1}
        &\cdots & c_0\, {\tilde \gamma}_{0,m-1}&
        c_0\, {\tilde \gamma}_{0,m} \\
        c_1\, \Gamma_{1,0} & c_1\, \Gamma_{1,1}-\lambda & \cdots
        & c_1 \, \Gamma_{1,m-1}&
        c_1\, \Gamma_{1,m} \\
        \vdots & \vdots & \ddots & \vdots & \vdots \\
        c_{m-1}\, \Gamma_{m-1,0} & c_{m-1}\, \Gamma_{m-1,1} &\cdots
        & c_{m-1}\, \Gamma_{m-1,m-1} - \lambda &
        c_{m-1}\, \Gamma_{m-1,m} \\
        -(c_{m}/c_0)\lambda & -(c_{m}/c_1)\lambda &\cdots &
        -(c_{m}/c_{m-1})\lambda &-(c_{m}/c_m)\lambda \\
      \end{array}
    \right). \nonumber \\ {} \label{det3}
\end{eqnarray}
Note, that all but the last row of the matrix (\ref{det3}) coincide
with those in (\ref{det1}).
\newline\newline
Step $\natural \,3$. Now, let us factor out $c_0$ from the first
row, $c_1$ from the second row, ..., $c_{m-1}$ from the $m$th row to
obtain
\begin{eqnarray}
    \prod_{j=0}^{m-1} c_j \nonumber
\end{eqnarray}
times
\begin{eqnarray} \fl
    \left(
      \begin{array}{ccccc}
        {\tilde \gamma}_{0,0} -\lambda/c_0 & {\tilde \gamma}_{0,1} &\cdots & {\tilde \gamma}_{0,m-1}&
        {\tilde \gamma}_{0,m} \\
        \Gamma_{1,0} & \Gamma_{1,1}-\lambda/c_1 & \cdots
        & \Gamma_{1,m-1}&
        \Gamma_{1,m} \\
        \vdots & \vdots & \ddots & \vdots & \vdots \\
        \Gamma_{m-1,0} & \Gamma_{m-1,1} &\cdots
        & \Gamma_{m-1,m-1} - \lambda/c_{m-1} &
        \Gamma_{m-1,m} \\
        -(c_{m}/c_0)\lambda & -(c_{m}/c_1)\lambda &\cdots &
        -(c_{m}/c_{m-1})\lambda &-(c_{m}/c_m)\lambda \\
      \end{array}
    \right). \label{det4}
\end{eqnarray}
Next, we multiply the first {\it column} by $c_0$, the second column
by $c_1$, ..., the $m$th column by $c_{m-1}$, and do not alter the
last, $(m+1)$th column. This leads us to
\begin{eqnarray} \fl
    \left(
      \begin{array}{ccccc}
        c_0\,{\tilde \gamma}_{0,0} -\lambda & c_1\, {\tilde \gamma}_{0,1} &\cdots & c_{m-1}\,{\tilde \gamma}_{0,m-1}&
        {\tilde \gamma}_{0,m} \\
        c_0\,\Gamma_{1,0} & c_1\, \Gamma_{1,1}-\lambda & \cdots
        & c_{m-1}\,\Gamma_{1,m-1}&
        \Gamma_{1,m} \\
        \vdots & \vdots & \ddots & \vdots & \vdots \\
        c_0\, \Gamma_{m-1,0} & c_1\,\Gamma_{m-1,1} &\cdots
        & c_{m-1}\,\Gamma_{m-1,m-1} - \lambda &
        \Gamma_{m-1,m} \\
        -c_{m}\lambda & -c_{m}\lambda &\cdots &
        -c_{m}\lambda &-\lambda \\
      \end{array}
    \right).  \label{det5}
\end{eqnarray}
\newline\newline
Step $\natural \,4$. Now, we multiply the last, $(m+1)$th column by
$c_m$ and subtract it from the first, second, ..., the $m$th column
to reduce (\ref{det5}) to
\begin{eqnarray} \fl
\label{det6} \left(
      \begin{array}{cccc}
        c_0\,{\tilde \gamma}_{0,0} - c_m{\tilde \gamma}_{0,m}-\lambda &
        \cdots & \cdots &
         {\tilde \gamma}_{0,m} \\
        c_0\, \Gamma_{1,0}
        -c_m\,\Gamma_{1,m} & \cdots
        & \cdots  &
        \Gamma_{1,m} \\
        \vdots & \ddots & \vdots &  \vdots \\
        c_0\,\Gamma_{m-1,0}
        -c_m\,\Gamma_{m-1,m} & \cdots & \cdots
        &\Gamma_{m-1,m}  \\
        0 & \cdots &0 &
        -\lambda \\
      \end{array}
    \right).
\end{eqnarray}
The determinant of the latter matrix can be calculated via expanding
with respect to its last row. As a result, the secular equation
(\ref{secular-a}) is reduced to
\begin{eqnarray} \fl
\label{det7}
    - \lambda \cdot \det
    \left[
        c_\beta \left({\tilde \gamma}_{\alpha,\beta} -
            {\tilde \gamma}_{\alpha-1,\beta}
        \right)
        -
        c_m \left({\tilde \gamma}_{\alpha,m} -
        {\tilde \gamma}_{\alpha-1,m}
        \right) -\lambda \,\delta_{\alpha,\beta}
    \right]_{(\alpha,\beta) = 0,\cdots,m-1}.
\end{eqnarray}
A comparison with (\ref{bodd}) allows us to rewrite (\ref{det7}) in
the form
\begin{eqnarray}
    - \lambda \, \det
    \left[
        {\boldsymbol {\hat b}}^{\rm T}
        - \lambda \, {\boldsymbol {\hat 1}}
    \right]_{m\times m}.
\end{eqnarray}
This establishes (\ref{c91}).
\newline\newline
Finally, it remains to show that, for all $j=1,2,\cdots$, the
identity
\begin{eqnarray}
\label{sodd-rho} {\rm tr}_{(0,m-1)}{\boldsymbol {\hat b}\,}^j = {\rm
tr}_{(0,m-1)}{\boldsymbol {\hat \varrho}\,}^{{\rm odd\;}j},
\end{eqnarray}
holds. That (\ref{sodd-rho}) is indeed true, follows from (\ref{bf})
and (\ref{d2}). This completes our proof of the Lemma.
$\;\;\;\blacksquare$

\section*{Appendix D. Calculation of the trace ${\rm tr}_{(0,\lfloor n/2
    \rfloor-1)}{\boldsymbol {\hat \varrho}}$}
\label{ap-d} \setcounter{equation}{0} \fancyhead{}
\fancyhead[RE,LO]{Appendix D} \fancyhead[LE,RO]{\thepage}
\renewcommand{\theequation}{D.\arabic{equation}}
\addcontentsline{toc}{section}{Appendix D. Calculation of the trace
${\rm tr}_{(0,\lfloor n/2 \rfloor-1)}{\boldsymbol {\hat \varrho}}$}

Since the matrix ${\boldsymbol {\hat \varrho}}$ is sensitive to the
parity of $n$, two separate calculations are needed.
\newline\newline
(i) {\it The case $n=2m$ even.}---To calculate the trace, we make
use of (\ref{rho-even-app}) to write down
\begin{eqnarray} \fl
    \label{for-1}
    {\rm tr}_{(0,m-1)}
    {\boldsymbol {\hat \varrho}}^{\rm even} = \sum_{\alpha=0}^{m-1} {\hat
    \varrho}_{\alpha,\alpha} = \int_0^\infty dy\, y^{-1} \, e^{y^2}\, {\rm
    erfc}(y \sqrt{2}) \,
    \nonumber \\
    \times
        \sum_{\alpha=0}^{m-1}
    \Big[
        (2\alpha+1) \, L_{2\alpha+1}^{-1}(-2y^2)
        + 2y^2 \, L_{2\alpha-1}^{1}(-2y^2)
    \Big].
\end{eqnarray}
First, to put (\ref{for-1}) into a more tractable form, we apply the
identity
\footnote[3]{\texttt{http://functions.wolfram.com/05.08.17.0009.01}}
\begin{eqnarray}
    \label{tr-1}
    L_n^{-m}(w) = \frac{w^m}{(-n)_m} \, L_{n-m}^{m}(w)
\end{eqnarray}
which, in the context of (\ref{for-1}), reads
\begin{eqnarray}
    \label{for-2}
    L_{2\alpha+1}^{-1} (w) = - \frac{w}{2\alpha+1} \,
    L_{2\alpha}^1 (w).
\end{eqnarray}
The use of (\ref{for-2}) reduces the integrand of (\ref{for-1}) to
\begin{eqnarray} \fl
    \label{for-3}
        (2\alpha+1) \, L_{2\alpha+1}^{-1}(-2y^2)
        + 2y^2 \, L_{2\alpha-1}^{1}(-2y^2)
        =
        2y^2 \Big[
            L_{2\alpha}^1 (-2y^2) + L_{2\alpha-1}^1 (-2y^2)
        \Big].
\end{eqnarray}
Second, we spot that the transformation
\begin{eqnarray}
    \label{tr-2}
    L_\nu^{\lambda-1}(w) = L_\nu^{\lambda} (w) - L_{\nu-1}^{\lambda}
    (w)
\end{eqnarray}
applied to (\ref{for-3}), yields
\begin{eqnarray}
    \label{for-4}
        \underbrace{L_{2\alpha}^1 (-2y^2)}_{L_{2\alpha}^2 - L_{2\alpha-1}^2} +
        \underbrace{L_{2\alpha-1}^1
        (-2y^2)}_{L_{2\alpha-1}^2 - L_{2\alpha-2}^2}
        = L_{2\alpha}^2(-2y^2) - L_{2\alpha-2}^2(-2y^2).
\end{eqnarray}
As a consequence, the summation in (\ref{for-1}) can be performed
explicitly,
\begin{eqnarray} \fl
    \label{for-5}
            \sum_{\alpha=0}^{m-1}
    \Big[
        (2\alpha+1) \, L_{2\alpha+1}^{-1}(-2y^2)
        + 2y^2 \, L_{2\alpha-1}^{1}(-2y^2)
    \Big] \nonumber \\= 2y^2 \sum_{\alpha=0}^{m-1}
    \Big[L_{2\alpha}^{2}(-2y^2) -
        L_{2\alpha-2}^{2}(-2y^2)
    \Big] \nonumber \\ = 2y^2\, L_{2m-2}^2(-2y^2),
\end{eqnarray}
resulting in a remarkably simple formula
\begin{eqnarray}
    \label{for-6}
    {\rm tr}_{(0,m-1)}
    {\boldsymbol {\hat \varrho}}^{\rm even} = 2
    \int_0^\infty dy\, y \, e^{y^2}\, {\rm
    erfc}(y \sqrt{2}) \,
    L_{2m-2}^2(-2y^2).
\end{eqnarray}
\newline\newline
(ii) {\it The case $n=2m+1$ odd.}---To calculate the trace, we
combine (\ref{d1}) and (\ref{rho-odd-app}) into
\begin{eqnarray}
    {\hat \varrho}_{\alpha,\alpha}^{\rm odd} =
    {\hat \varrho}_{\alpha,\alpha}^{\rm even} - \frac{m!}{h_m} \,
    \left(
        {\tilde \gamma}_{\alpha,m} - {\tilde \gamma}_{\alpha-1,m}
    \right).
\end{eqnarray}
Summing it up, we derive
\begin{eqnarray}
    {\rm tr}_{(0,m-1)}{\boldsymbol {\hat \varrho}}^{\rm odd}=
    {\rm tr}_{(0,m-1)}{\boldsymbol {\hat \varrho}}^{\rm even}
    - \frac{m!}{h_m} \,
    {\tilde \gamma}_{m-1,m}.
\end{eqnarray}
Further use of (\ref{for-6}) and (\ref{gamma-ab}) yields
\begin{eqnarray} \fl
    \label{odd-id-1}
    {\rm tr}_{(0,m-1)}
    {\boldsymbol {\hat \varrho}}^{\rm odd} = 2
    \int_0^\infty dy\, y \, e^{y^2}\, {\rm
    erfc}(y \sqrt{2}) \,
    \left[ L_{2m-2}^2(-2y^2)+ L_{2m-1}^1(-2y^2)\right].
\end{eqnarray}
With the help of the identity (\ref{tr-2}), this eventually
simplifies to
\begin{eqnarray}
    \label{odd-id-2}
    {\rm tr}_{(0,m-1)}
    {\boldsymbol {\hat \varrho}}^{\rm odd} = 2
    \int_0^\infty dy\, y \, e^{y^2}\, {\rm
    erfc}(y \sqrt{2}) \,
     L_{2m-1}^2(-2y^2).
\end{eqnarray}
\newline\newline
The formulae (\ref{for-6}) and (\ref{odd-id-2}) can be unified into
a single equation, holding for $n$ of arbitrary parity:
\begin{eqnarray}
    \label{d-final}
    {\rm tr}_{(0,\lfloor n/2 \rfloor -1)}
    {\boldsymbol {\hat \varrho}} = 2
    \int_0^\infty dy\, y \, e^{y^2}\, {\rm
    erfc}(y \sqrt{2}) \,
     L_{n-2}^2(-2y^2).
\end{eqnarray}
\smallskip
\section*{References}
\fancyhead{} \fancyhead[RE,LO]{References}
\fancyhead[LE,RO]{\thepage}
\addcontentsline{toc}{section}{\protect\enlargethispage*{100pt}References}
\begin{harvard}

\item[] Adler M, Forrester P J, Nagao T, van Moerbeke P 2000
        Classical skew orthogonal polynomials and random matrices
        {\it J. Stat. Phys.} {\bf 99} 141

\item[] Agam O, Bettelheim E, Wiegmann P B, and Zabrodin A 2002
        Viscous fingering and the shape of an electronic droplet in
        the quantum Hall regime
        \PRL {\bf 88} 236801

\item[] Akemann 2005
    The Complex Laguerre Symplectic Ensemble of Non-Hermitean Matrices
    \NP B {\bf 730} 253

\item[] Akemann G and Basile F 2007
        Massive partition functions and complex eigenvalue correlations
        in matrix models with symplectic symmetry
        {\it Nucl. Phys.} B {\bf 766} 150

\item[] Akemann G 2007
        Matrix Models and QCD with Chemical Potential
        {\it arXiv:~hep-th/0701175 [Int. J. Mod. Phys. A (in press)]}

\item[] Andrews G E 1998
        {\it The Theory of Partitions} (Cambridge: Cambridge University Press)

\item[] Bai Z D 1997
        Circular law
        {\it Ann. Probab.} {\bf 25} 494

\item[] Borodin A and Strahov E 2005
        Averages of characteristic polynomials in random matrix theory
        {\it Commun. Pure and Appl. Math.} {\bf LVIII} 0001

\item[] Chalker J T and Mehlig B 1998
        Eigenvector statistics in non-Hermitean random matrix
        ensembles
        \PRL {\bf 81} 3367

\item[] Dyson F J 1970
        Correlations between eigenvalues of a random matrix
        {\it Commun. Math. Phys.} {\bf 19} 235

\item[] Dyson F J 1972
        Quaternion determinants
        {\it Helv. Phys. Acta} {\bf 49} 289

\item[] Edelman A, Kostlan E, and Shub M 1994
        How many eigenvalues of a random matrix are real?
        {\it J. Amer. Math. Soc.} {\bf 7} 247

\item[] Edelman A 1997 The probability that a random real Gaussian
        matrix has $k$ real eigenvalues, related distributions, and
        the circular law
        {\it J. Mult. Analysis} {\bf 60} 203

\item[] Efetov K B 1997a
        Directed quantum chaos
        \PRL {\bf 79} 491

\item[] Efetov K B 1997b
        Quantum disordered systems with a direction
        \PR B {\bf 56} 9630

\item[] Eynard B 2001
        Asymptotics of skew orthogonal polynomials
        \JPA {\bf 34} 7591

\item[] Forrester P J 2005
        {\it Log-Gases and Random Matrices (web-book)}

\item[] Fyodorov Y V, Khoruzhenko B, and Sommers H-J 1997
        Almost Hermitean random matrices: Crossover from
        Wigner-Dyson  to Ginibre eigenvalue statistics
        \PRL {\bf 79} 557

\item[] Fyodorov Y V and Sommers H-J 2003
        Random matrices close to Hermitean or unitary: Overview of
        methods and results
        \JPA {\bf 36} 3303

\item[] Ginibre J 1965
        Statistical ensembles of complex, quaternion, and real
        matrices
        \JMP {\bf 19} 133

\item[] Girko V L 1984
        Circle law
        {\it Theory Probab. Appl.} {\bf 29} 694

\item[] Girko V L 1986
        Elliptic law
        {\it Theory Probab. Appl.} {\bf 30} 677

\item[] Grobe R, Haake F, and Sommers H-J 1988
        Quantum distinction of regular and chaotic dissipative
        motion
        \PRL {\bf 61} 1899

\item[] Grobe R and Haake F 1989
        Universality of cubic-level repulsion for dissipative
        quantum chaos
        \PRL {\bf 62} 2893

\item[] Guhr T, M\"uller-Groeling A, and Weidenm\"uller H A 1998
        Random matrix theories in quantum physics: Common concepts
        {\it Phys. Reports} {\bf 299} 189

\item[] Halasz M A, Osborn J C, and Verbaarschot J J M 1997
        Random matrix triality at nonzero chemical potential
        \PR D {\bf 56} 7059

\item[] Hardy G H, Ramanujan S 1918
        Asymptotic formulae in combinatory analysis
        {\it Proc. London Math. Soc.} B {\bf 17} 75

\item[] Jack H 1976 A class of polynomials in search of a definition,
        or the symmetric group parameterized {\it Jack,
        Hall-Littlewood and Macdonald polynomials (Contemporary Mathematics
        Series)}
        ed V B Kuznetsov and S Sahi (Providence: AMS 2006)

\item[] Janik R A, N\"orenberg W, Nowak M A, Papp G, and Zahed I
        1999
        Correlations of eigenvectors for non-Hermitean random-matrix
        models
        \PR E {\bf 60} 2699

\item[] Kanzieper E 2002a
        Eigenvalue correlations in non-Hermitean
        symplectic random matrices
        \JPA {\bf 35} 6631

\item[] Kanzieper E 2002b
        Replica field theories, Painlev\'e transcendents, and exact
        correlation functions
        \PRL {\bf 89} 250201

\item[] Kanzieper E 2005
        Exact replica treatment of non-Hermitean
        complex random matrices
        {\it Frontiers in Field Theory} ed O Kovras (New York: Nova Science
        Publishers) p 23

\item[] Kanzieper E and Akemann G 2005
        Statistics of real eigenvalues in Ginibre's ensemble of random real
        matrices
        \PRL {\bf 95} 230201

\item[] Khoruzhenko B A and Mezzadri F 2005
        {\it Private communication}

\item[] Kolesnikov A V and Efetov K B 1999
        Distribution of complex eigenvalues for symplectic ensembles
        of non-Hermitean matrices
        \WRM {\bf 9} 71

\item[] Kwapie\'{n} J, Dro\.{z}d\.{z} S, and Ioannides A A 2000
        Temporal correlations versus noise in the correlation matrix
        formalism: An example of the brain auditory response
        \PR E {\bf 62} 5557

\item[] Kwapie\'{n} J, Dro\.{z}d\.{z} S, G\'{o}rski A Z, and
        O\'{s}wi\c{e}cimka P
        2006
        Asymmetric matrices in an analysis of financial correlations
        {\it Acta Phys. Polonica} {\bf B37}, 3039

\item[] Le Ca\"er G and Ho J S 1990
        The Voronoi tessellation generated from eigenvalues of
        complex random matrices
        \JPA {\bf 23} 3279

\item[] Le Ca\"er G and Delannay R 1993
        Topological models of 2D fractal cellular structures
        {\it J. Phys. I France} {\bf 3} 1777

\item[] Lehmann N and Sommers H J 1991
        Eigenvalue statistics of random real matrices
        \PRL {\bf 67} 941

\item[] Macdonald I G 1998
        {\it Symmetric Functions and Hall Polynomials} (Oxford: Oxford
        University Press)

\item[] Mahoux G and Mehta M L 1991
        A method of integration over matrix variables
        {\it J. Phys. I (France)} {\bf 1} 1093

\item[] Markum H, Pullirsch R, and Wettig T 1999
        Non-Hermitean random matrix theory and lattice QCD with
        chemical potential
        \PRL {\bf 83} 484

\item[] Mehlig B and Chalker J T 2000
        Statistical properties of eigenvectors in non-Hermitean
        Gaussian random matrix ensembles
        \JMP {\bf 41} 3233

\item[] Mehta M L and Srivastava P K 1966
        Correlation functions for eigenvalues of real quaternion
        matrices
        \JMP {\bf 7} 341

\item[] Mehta M L 1976
        A note on certain multiple integrals
        \JMP {\bf 17} 2198

\item[] Mehta M L 2004
        {\it Random Matrices} (Amsterdam: Elsevier)

\item[] Mineev-Weinstein M, Wiegmann P B, and Zabrodin A 2000
        Integrable structure of interface dynamics
        \PRL {\bf 84} 5106

\item[] Muirhead R J 1982
        {\it Aspects of Multivariate Statistical Theory} (New York: John
        Wiley \& Sons)

\item[] Nagao T and Nishigaki S M 2001
        Massive random matrix ensembles at $\beta=1$ and $4$: QCD in
        three dimensions
        \PR D {\bf 63} 045011

\item[] Nishigaki S M and Kamenev A 2002
        Replica treatment of
        non-Hermitean disordered Hamiltonians
        \JPA {\bf 35} 4571

\item[] Osborn J C 2004
Universal results from an alternate random matrix model for QCD with a baryon
chemical potential
\PRL {\bf 93} (2004) 222001

\item[] Prudnikov A P, Brychkov Yu A, and Marichev O I 1986
        {\it Integrals and Series} vol 2 (New York: Gordon and Breach)

\item[] Prudnikov A P, Brychkov Yu A, and Marichev O I 1990
        {\it Integrals and Series} vol 3 (New York: Gordon and Breach)

\item[] Sinclair C D 2006
        Averages over Ginibre's ensemble of random real matrices
        {\it arXiv:~math-ph/0605006}

\item[] Sommers H J, Crisanti A, Sompolinsky H, and Stein Y 1988
        Spectrum of large random asymmetric matrices
        \PRL {\bf 60} 1895

\item[] Sompolinsky H, Crisanti A, Sommers H J 1988
        Chaos in random neural networks
        \PRL {\bf 61} 259

\item[] Splittorff K and Verbaarschot J J M 2004
        Factorization of Correlation Functions and the Replica Limit of the Toda
        Lattice Equation
        \NP B {\bf 683} 467

\item[] Stephanov M 1996
        Random matrix model of QCD at finite density and the nature
        of the quenched limit
        \PRL {\bf 76} 4472

\item[] Timme M, Wolf F, and Geisel T 2002
        Coexistence of regular and irregular dynamics in complex
        networks of pulse-coupled oscillators
        \PRL {\bf 89} 258701

\item[] Timme M, Wolf F, and Geisel T 2004
        Topological speed limits to network synchronization
        \PRL {\bf 92} 074101

\item[] Tracy C A and Widom H 1998
        Correlation functions, cluster functions, and spacing
        distributions for random matrices
        {\it J. Stat. Phys.} {\bf 92} 809

\item[] Wigner E P 1957
        Statistical properties of real symmetric matrices with many
        dimensions
        {\it Proc. 4th Can. Math. Cong. (Toronto)} p~174

\item[] Wigner E P 1960
        The unreasonable effectiveness of mathematics in natural
        sciences
        {\it Commun. Pure and Applied Mathematics} {\bf 13} 1

\item[] Zabrodin A 2003
        New applications of non-Hermitean random matrices
        {\it Proceedings of the International Conference
        on Theoretical Physics (TH-2002)} ed D Iagolnitzer, V Rivasseau, and
        J Zinn-Justin (Basel: Birkh\"auser Verlag)

\smallskip
\end{harvard}

\end{document}